\documentclass[11pt,twoside,letterpaper]{article} 
\usepackage{times,fancyhdr}
\usepackage[dvips]{graphicx}
\usepackage{color}

\usepackage[pdfborder={0 0 0},colorlinks=true]{hyperref}


\makeatletter
\def\cleardoublepage{\clearpage\if@twoside \ifodd\c@page\else%
    \hbox{}%
    \thispagestyle{empty}%
    \newpage%
    \if@twocolumn\hbox{}\newpage\fi\fi\fi}
\makeatother

\renewcommand{\thesection}{\arabic{section}.}
\renewcommand{\thesubsection}{\thesection\arabic{subsection}.}

\def\figurename{Figure}
\makeatletter
\renewcommand{\fnum@figure}[1]{\figurename~\thefigure.}
\makeatother

\def\tablename{Table}
\makeatletter
\renewcommand{\fnum@table}[1]{\tablename~\thetable.}
\makeatother

\def\q{\mathbf{q}}
\def\k{\mathbf{k}}

\def\K{\mathbf{K}}

\def\Q{\mathbf{Q}}

\newcommand{\bm}[1]{\mathbf{#1}}

\begin{document}
\title{
{\begin{flushleft}
\vskip 0.45in
{\normalsize\bfseries\textit{Chapter~6}}
\end{flushleft}
\vskip 0.45in
\bfseries\scshape Magnetic mechanisms of pairing in strongly correlated electron system of copper oxides}}
\author{\bfseries\itshape S.G. Ovchinnikov$^{1,2}$\thanks{E-mail address: sgo@iph.krasn.ru}, I.A. Makarov$^{1}$, E.I. Shneyder$^{1,3}$, Yu.N. Togushova$^{2}$,\\ \bfseries\itshape V.A. Gavrichkov$^{1,2}$, M.M. Korshunov$^{1,2}$\\
$^1$Kirensky Institute of Physics SBRAS,   Krasnoyarsk, Russia \\
$^2$Siberian Federal University,   Krasnoyarsk, Russia \\
$^3$Siberian State Aerospace University,  Krasnoyarsk, Russia}

\date{}
\maketitle
\thispagestyle{empty}
\setcounter{page}{1}
\thispagestyle{fancy}
\fancyhead{}
\fancyhead[L]{In: Recent Advances in Superconductivity Research \\
Editor: Christopher B. Taylor, pp. 93-143} 
\fancyhead[R]{ISBN 978-1-62618-406-0 \\
\copyright~2007 Nova Science Publishers, Inc.}
\fancyfoot{}
\renewcommand{\headrulewidth}{0pt}


\pagestyle{fancy}
\fancyhead{}
\fancyhead[EC]{S.G. Ovchinnikov, I.A. Makarov, E.I. Shneyder \textit{et al.}}
\fancyhead[EL,OR]{\thepage}
\fancyhead[OC]{Magnetic mechanisms of pairing in strongly correlated cuprates}
\fancyfoot{}
\renewcommand\headrulewidth{0.5pt}
\begin{abstract}
The multielectron LDA+GTB approach has been developed to calculate electronic structure of strongly correlated cuprates. At low energies the effective Hamiltonian of the $t - t' - t'' - {t_ \bot } - {J^ * } - {J_ \bot }$-model has been derived with parameters coming from the ab initio calculation for LSCO. The electronic structure of LSCO has been calculated self-consistently  with the short-range antiferromagnetic order for various doping level. Two Lifshitz-type quantum phase transitions with Fermi surface topology changes have been found at dopings $x_{c1}=0.15$ and $x_{c2}=0.24$. Its effect on normal and superconducting properties has been calculated. The interatomic exchange parameter and its pressure dependence has been calculated within LDA+GTB scheme. The magnetic mechanisms of d-wave pairing induced by static and dynamical spin correlations are discussed. Simultaneous treatment of magnetic and phonon pairing results in the conclusion that both contributions are of the same order. For two layer cuprates like YBCO the interlayer hopping and exchange effects on the electronic structure and doping dependence of $T_c$ is discussed as well as the Coulomb interaction induced mechanism of pairing.
\end{abstract}

\vspace{.08in}

\section{Introduction}

High-$T_c$ copper oxides, or cuprates, are known for their high critical temperature, unconventional superconducting state, and unusual normal state properties.
The phase diagram of cuprates is quite complicated. Doping is accomplished by replacing one of the spacer ions with another one with different valence or adding extra out-of-plane oxygen, e.g. La$_{2-x}$Sr$_x$CuO$_2$, Nd$_{2-x}$Ce$_x$CuO$_2$, and YBa$_2$Cu$_3$O$_{6+\delta}$. The additional electron or hole is then assumed to dope the CuO$_2$ plane. The undoped compounds exhibit antiferromagnetism, which vanishes with doping; superconductivity appears at some nonzero doping and then disappears, such that $T_c(x)$ forms a ``dome''.
The undoped materials are Mott insulators, thus the Mott-Hubbard physics of a half-filled Hubbard model may be a good starting point. Underdoped cuprates manifest pseudogap behavior,
as well as a variety of competing orders. At least for hole-doped cuprates, a strange metal phase near optimal doping is characterized by the linear-$T$ resistivity over a wide range of temperatures. One of the most widespread point of view is that these peculiarities are due to the Mott-Hubbard physics.

Although high-$T_c$ superconducting cuprates have long been studied, two global problems: (i) the nature of the ``anomalous'' normal state and (ii) mechanism of superconductivity with the $d_{x^2-y^2}$ pairing symmetry, are not completely solved yet. What is the role of the electron (magnetic, charge, current) Bose-like fluctuations along with phonons in the formation of the superconducting state? How does a transition from a Mott-Hubbard insulator upon doping with holes or electrons to a Fermi-liquid normal metal occurs in highly doped cuprates through a
pseudogap state in the low-doping region? These and many other fundamental physical problems of high-$T_c$ superconductivity are currently attract considerable attention of the condensed matter community.

It is natural to expect that unusual properties of the normal state in cuprates affect the superconductivity. Considering the complicated nature of the normal state we first discuss it in the following Section and then proceed to the magnetic mechanisms of pairing. Here we mainly focus on the limit of the strong electronic correlations and, in particular, discuss the results of the LDA+GTB method (local density approximation+generalized tight-binding method) while also discussing some results of the Fermi-liquid approach.

\section{Normal State of Cuprates}

One of the most attractive feature of the condensed matter physics is the ability to explain and even predict properties of the materials which comprise our world. Significant leap in this direction was made in mid 1960 when Hohenberg et al.
\cite{Hohenberg_1964,Kohn_1965} formulated a density functional theory (DFT). Because its starting point is the Shr\"odinger equation for the particular atomic arrangement and orbital and spin configurations, this theory is often referred to as the ``first principles'' or the \textit{ab initio} calculations. Augmented with the local density approximation (LDA) or the generalized gradient approximation (GGA) for the initially unknown quantity, exchange-correlation energy, DFT provides quantitative description of the ground state energy and the band structure of various atoms, molecules, and crystalline solids (see e.g., \cite{Jones_1989}).
Despite its success for s and p atoms in solids, LDA failed to describe transition metal oxides with partially filled 3d- and 4f-orbitals. The most pronounced failure is that LDA predicts La$_2$CuO$_4$ to be a metal whereas experimentally it is an insulator. The root of the problem is the unscreened on-site Coulomb interaction (Hubbard repulsion)~\cite{r7LDAGTB}. In a single-band system on the mean-field level if Hubbard repulsion $U$ is larger than the bandwidth $W$ , it splits the band into two Hubbard subbands with a gap $\propto U$. The spectral weight of a quasiparticle is redistributed between these subbands. At a half-filling the Fermi level is inside the gap and the system is an insulator. In a multiorbital system, along with the Hubbard repulsion other local interactions like the Hund's exchange $J_H$ and the interorbital Hubbard repulsion $U'$ are present and provide a rich set of physical properties. The opening of the Hubbard gap and moreover the major role played by the local interactions near the half-filling are beyond the scope of LDA and GGA.

There are several extensions to LDA which includes or simulates the effects of the electron correlations. One of them is LDA+U~\cite{AnZaOKA} and another one is SIC-LSDA (self-interaction--corrected local spin density approximation)~\cite{SvGun}. Both methods consider local interactions in the Hartree-Fock sense and result in the antiferromagnetic insulator as the ground state for La$_{2-x}$Sr$_x$CuO$_4$ contrary to LDA, but the origin of the insulating gap is incorrect. In both LDA+U and SIC-LSDA, it is formes by the local single-electron states splitted by the spin or orbital polarization. Therefore, the paramagnetic phase above the N\'eel temperature $T_N$ of the undoped La$_{2}$CuO$_4$ will be metallic in spite of strong correlation regime $U \gg W$. There is one more significant drawback in these approximations, namely, they disregard the redistribution of the spectral weight between the Hubbard subbands. Latter effect is incorporated in a different approach to \textit{ab initio} calculations for strongly correlated systems - LDA+DMFT (LDA+dynamical mean field theory)~\cite{AnPotKor,LichKats,HeNekBlu,KotSavHau}. The method is based on the self-consistent procedure where the LDA band structure is used to calculate the electron self-energy in DMFT. DMFT utilizes the fact that in the infinite dimensional limit of the Hubbard model, $D \to \infty $, the self-energy is momentum independent, $\Sigma \left( {\bf{k},\omega } \right) \to \Sigma \left( \omega  \right)$~\cite{MetVol,Vol,GeoKotKra}. The remaining frequency dependence is exact in $D \to \infty $ limit and carries very important information about dynamical correlations and Mott-Hubbard transition. On the other hand, the spatial correlations become crucial in low-dimensional systems like layered high-$T_c$ cuprates. That is why the correct band dispersion and spectral intensities for these systems cannot be obtained within LDA+DMFT. Natural extension of this method, LDA+cluster or cellular DMFT ~\cite{HetTahzad,KotSavPal,Potthoff,MaiJarPru}, LDA+DMFT+$\Sigma_{\bf{k}}$ \cite{KuNePchSad,NeKoKuSaKa,NeKoKuPchSa,KoKuNe,NePaKu,NeKuSa} and SDFT (spectral density functional theory)~\cite{SavKot} provides momentum dependent self-energy and thus allow for the non-local correlations.

Here we focus on the results obtained within the combined local density approximation (LDA) and generalized tight-binding (GTB) method. This combined method involves the \textit{ab initio} LDA calculation of the parameters of the multielectron tight-binding Hamiltonian with the Coulomb interactions and cluster perturbation theory within the generalized tight-binding method~\cite{KorGavOvch,OvchGavKor}. From the very beginning the GTB method has been suggested to extend the microscopic band structure calculations to take into account the strong electronic correlations (SEC) in the Mott-Hubbard insulators like the transition metal oxides~\cite{OvchSan}. Similar to conventional tight binding (TB) approach we start with a particular local electron states (with all multiorbital effects, symmetry and chemistry) and then by a Fourier transform move to the momentum space and obtain a band structure. Because of SEC we can not use free electron local states, our local fermion in a $d$-orbital system is a quasiparticle given by the excitations between multielectron ${d^n}$ and ${d^{n \pm 1}}$ terms contrary to the conventional TB. In other words, GTB is the strongly correlated version of the TB method. The first computer codes and successful application of GTB has been developed for cuprates~\cite{GavOvchBor}.
There are a lot of papers where the Hubbard model is used to describe properties of HTSC cuprates in different approximations. In LDA+GTB method we use the Hubbard ideas and generalized them to describe the electronic structure and properties of a real compound with its all multiorbital states of different atoms. We obtain a set of effective Hamiltonians that are different for a particular energy scale of interest.

\subsection{LDA+GTB Method}
\label{LDA_GTB:2}
LDA+GTB procedure consists of the following steps~\cite{KOGNP}:

\textbf{Step I: LDA.} Calculation of the LDA band structure, construction of Wannier functions with the given symmetry, and computation of the one- and two-electron matrix elements of the TB Hamiltonian with the local and nearest-neighbor Coulomb interactions.

\textbf{Step II: Exact diagonalization (ED).} Separation of the total Hamiltonian $H$ into the intra- and inter-cell parts, $H = {H_c} + {H_{cc}}$, where ${H_c}$ represents the sum of the orthogonal unit cells, ${H_c} = \sum\limits_f {{H_f}} $. ED of a single unit cell term, ${{H_f}}$, and construction of the Hubbard $X$-operators $X_f^{pq} = \left| p \right\rangle \left\langle q \right|$ using the complete orthogonal set of eigenstates $\left\{ {\left| p \right\rangle } \right\}$ of ${{H_f}}$.

\textbf{Step III: Perturbation theory.} Within the $X$-representation, local interactions are diagonal and all intercell hoppings and long-range Coulomb interaction terms have the bilinear form in the $X$-operators. Various perturbation approaches known for the Hubbard model in the $X$-representation can be used. The most general one includes treatment within the generalized Dyson equation obtained by the diagram technique~\cite{OvchVal}.

Below we discuss each step in detail.

\subsubsection{Step I: LDA}
\label{stepILDA:2.1}

LDA provides us a set of Bloch functions $\left| {{\Psi _{\lambda \bf{k}}}} \right\rangle $ ($\lambda $ is band index) and band energies ${\varepsilon _\lambda }\left( {\bf{k}} \right)$. For example, LDA band structure calculation for La$_{2}$CuO$_4$ and Nd$_{2}$CuO$_4$ was done within the TB-LMTO-ASA (linear muffin-tin orbitals using atomic sphere approximation in the tight-binding) method~\cite{OKAPaw}. Using the Wannier functions (WFs) formalism~\cite{AnisKond} or the NMTO method~\cite{OKASaha}, we obtain single electron energies ${\varepsilon _\lambda }$ and hopping integrals $T_{fg}^{\lambda \lambda '}$ of the TB model~\cite{KOGNP,GavOvchNek}

\begin{eqnarray}
 H &=& \sum\limits_{f,\lambda,\sigma}(\epsilon_ {\lambda}-\mu) n_ {f \lambda \sigma} + \sum\limits_{f \neq g} \sum\limits_ {\lambda,\lambda',\sigma} T_{fg}^{\lambda \lambda'} c_{f \lambda \sigma}^\dag c_{g \lambda' \sigma} \nonumber \\
 &+& \frac {1}{2}\sum\limits_{f,g,\lambda,\lambda'} \sum\limits_{\sigma_1,\sigma_2,\sigma_3,\sigma_4} V_{fg}^{\lambda \lambda'} c_{f \lambda \sigma_1}^\dag c_{f \lambda \sigma_3} c_{g \lambda' \sigma_2}^\dag c_{g \lambda' \sigma_4},
 \label{eq1}
\end{eqnarray}
where ${{c_{f\lambda \sigma }}}$ is the annihilation operator in the Wannier representation of the hole at the site $f$ on the orbital $\lambda $ and with the spin $\sigma $, ${n_{f\lambda \sigma }} = c_{f\lambda \sigma }^ + {c_{f\lambda \sigma }}$. Note that a number and a symmetry of chosen WFs are determined by the energy window that we are interested in.
The values of Coulomb parameters ${V_{fg}^{\lambda \lambda '}}$ are obtained by LDA supercell calculations. For Cu in La$_{2}$CuO$_4$, Hubbard parameter $U$ and Hund's exchange $J_H$ are equal to 10 eV and 1 eV, respectively.

\subsubsection{Step II: Exact Diagonalization}
\label{stepIIED:2.2}

In transition metal (Me) oxides the unit cell may be chosen as the MeO$_n$ (n=6,5,4) cluster and usually there is a common oxygen shared by two adjacent cells. All other ions provide the electroneutrality and contribute to the high energy electronic structure. In the low energy sector, they are inactive.
After the orthogonalization where we solve the problem of nonorthogonality of the oxygen molecular orbitals of adjacent cells, the Hamiltonian (\ref{eq1}) can be written as a sum of intracell and intercell contributions
\begin{equation}
\label{eq2}
H = H_c + H_{cc}, \;\;\; H_c  = \sum\limits_f {H_f}, \;\;\; H_{cc}  = \sum\limits_{f,g} {H_{fg} }
\end{equation}
with orthogonal states in different cells described by ${H_f}$.
The ED of ${H_f}$ gives us the set of eigenstates $\left| p \right\rangle  = \left| {{n_p},{i_p}} \right\rangle $ with the energy ${E_p}$. Here, ${n_p}$ is the number of electrons per unit cell and ${i_p}$ denotes all other quantum numbers like spin, orbital moment, etc. We perform the ED with all possible excited eigenstates, not the Lancoz procedure.

From the local electroneutrality we determine which configurations are relevant. For stoichiometric $3d$ metal the three relevant subspaces of the Hilbert space are ${d^{n - 1}}$, ${d^{n}}$ and ${d^{n + 1}}$. For each subspace, the ED provides a set of multielectron states $\left| {n,i} \right\rangle $ with energy ${E_i}\left( n \right)$, $i = 0,1,2,...,{N_n}$. Within this set of multielectron terms, there are charge-neutral Bose-type excitations with the energy ${\omega _i} = {E_i}\left( {n} \right) - {E_0}\left( n \right)$. Electron addition excitations (local Fermi-type quasiparticles) have energies ${\Omega _{c,i}} = {E_i}\left( {n + 1} \right) - {E_0}\left( n \right)$. Here, index ``$c$'' means that quasiparticles (QPs) form the empty conductivity band. Similarly, the valence band is formed by the electron removal Fermi-type QPs with energies ${\Omega _{v ,i}} = {E_0}\left( n \right) - {E_i}\left( {n - 1} \right)$. This multielectron language have been used in the spectroscopy, see for example~\cite{ZaanSaw}. The proper mathematical tool to study both the local QP and their intercell hoppings is given by the Hubbard $X$-operators~\cite{r7LDAGTB},
\begin{equation}
\label{eq3}
X_f^{pq}  = \left| p \right\rangle \left\langle q \right|.
\end{equation}
with algebra given by the multiplication rule $X_f^{pq}X_g^{rs} = {\delta _{fg}}{\delta _{qr}}X_f^{ps}$, and by the completeness condition $\sum\limits_p {X_f^{pp}}  = 1$.
The operator $X_f^{pq}$ describes the transition from initial state $\left| q \right\rangle $ to the final state $\left| p \right\rangle $, ${X^{pq}}\left| q \right\rangle  = \left| p \right\rangle $. Since $X$-operators are the projective operators, any local operator can be represented as the linear combination of $X$-operators.
If ${n_p} - {n_q}$ is odd (${ \pm 1,\, \pm 3}$, etc.) the ${X^{pq}}$ is called quasifermionic operator; if ${n_q} - {n_p}$ is even ($0, \pm 2$, etc.) the ${X^{pq}}$ is a quasibosonic one.
We use a simplified notation
%
$X_f^{pq} \to X_f^{\vec \alpha _m } \to X_f^m,$
%
where the number $m$ enumerate the excitation and plays the role of the QP band index; each pair $\left( {p,q} \right)$ corresponds to some vector $\vec \alpha \left( {p,q} \right)$ that is called ``root vector'' in the diagram technique. In this notation, a single-electron (hole) creation operator is given by a linear combination in the $X$-representation, ${c_{f\lambda \sigma }} = \sum\limits_m {{\gamma _{\lambda \sigma }}\left( m \right)X_f^m} $, ${\gamma _{\lambda \sigma }}\left( {p,q} \right) = \left\langle p \right|{c_{f\lambda \sigma }}\left| q \right\rangle $.
In the $X$-representation, the intracell part of the Hamiltonian is diagonal,
\begin{equation}
\label{eq5}
H_c  = \sum\limits_{f,p} \left(E_p - n_p \mu \right) X_f^{pp}.
\end{equation}

The intercell hopping is given by
\begin{equation}
\label{eq6}
{H_{cc}} = \sum\limits_{f \ne g} {\sum\limits_{m,m'} {t_{fg}^{mm'}} } \mathop {X_f^m}\limits^ +  X_g^{m'},
\end{equation}
where the matrix elements are
$t_{fg}^{mm'} = \sum\limits_{\sigma,\lambda,\lambda'} T_{fg}^{\lambda \lambda'} \gamma_{\lambda\sigma}^*(m) \gamma_{\lambda'\sigma}(m').$
%

All intraatomic $d - d$ Coulomb interactions are included in ${H_c}$ and since ${H_c}$ is diagonal in $X$-operators, they treated exactly. The dominant part of the $p - d$ and $p - p$ Coulomb interactions is also included in ${H_c}$ and gives contribution to energies ${E_p}$, while a small part of it ($\sim$10 \%) provides the intercell Coulomb interaction that is also bilinear in the $X$-operators, $H_{cc}^{Coul.} = \sum\limits_{f \ne g} {\sum\limits_{p,q,p',q'} {V_{fg}^{pq,p'q'}X_f^{pq}X_g^{p'q'}} } $.

\subsubsection{Step III: Perturbation Theory}
\label{stepIIIPT:2.3}

The characteristic local energy scale is given by the effective Hubbard parameter ${U_{eff}} = {E_0}\left( {n + 1} \right) - {E_0}\left( {n - 1} \right) - 2{E_0}\left( n \right)$ that is given by difference of the initial ${d^n} + {d^n}$ and the excited ${d^{n - 1}} + {d^{n + 1}}$ configurations. The same energy can be obtained as the local gap between the conductivity and valence bands, ${U_{eff}} = {\Omega _{c,0}} - {\Omega _{\upsilon ,0}}$. Depending on the ratio of the bare Hubbard $U$ and the charge excitation energy ${\Delta _{pd}} = {\varepsilon _p} - {\varepsilon _d}$, the ${U_{eff}}$ may represent the Mott-Hubbard gap for $U < {\Delta _{pd}}$ or the charge transfer (CT) gap ${E_{CT}}$ for $U > {\Delta _{pd}}$. The intercell hoppings and interactions result in the dispersion and decrease the energy gap, ${E_{gap}} < {U_{eff}}$. The intercell hoppings, Eq.~(\ref{eq6}), and the non-local Coulomb interactions can be treated by a perturbation theory. We'd like to emphasize that in the $X$-representation the perturbation, Eq.~(\ref{eq6}), has exactly the same structure as the hopping Hamiltonian in the conventional Hubbard model. That is why the accumulated experience of the Hubbard model study in the $X$-representation can be used here.
Single-electron Green function for a particle with momenta $\mathbf{k}$, energy $E$, spin $\sigma $, and orbital indices $\lambda $ and $\lambda '$, $G_{\mathbf{k}\sigma}^{\lambda \lambda'}(E) \equiv \left\langle\left\langle c_{\mathbf{k}\lambda\sigma} \left| c_{\mathbf{k}\lambda'\sigma}^\dag \right. \right\rangle\right\rangle_E$, is given by a linear combination of the Hubbard operator's Green functions $D_{{\bf{k}}\sigma }^{mn}\left( E \right) = {\left\langle {\left\langle {X_{{\bf{k}}\sigma }^m\left| {\mathop {X_{{\bf{k}}\sigma }^n}\limits^ +  } \right.} \right\rangle } \right\rangle _E}$,
\begin{equation}
\label{eq8}
 G_{\mathbf{k}\sigma}^{\lambda\lambda'}(E) = \sum\limits_{mm'} \gamma_{\lambda\sigma}(m) \gamma_{\lambda'\sigma }^*(m') D_{\mathbf{k}\sigma}^{mm'}(E).
\end{equation}

To determine $X$-operators's Green function we use a generalized Dyson equation obtained in the diagram technique for the $X$-operators \cite{OvchVal} assuming a small kinetic energy compared to the local Coulomb energy \cite{WestPaw,Zaitsev},
\begin{equation}
\label{eq9}
 \hat D_{\bf{k}\sigma}(E) = \left\{ \left(E - \Omega_m\right) \delta_{mm'} - \hat P_{\bf{k}\sigma}(E) \hat t(\bf{k}) - \hat \Sigma_{\mathbf{k}\sigma}(E) \right\}^{-1} \hat P_{\bf{k}\sigma}(E).
\end{equation}
Here ${\Omega _m} = {E_p}\left( {n + 1} \right) - {E_q}\left( n \right)$ is the $m$-th QP local energy. Besides the self-energy matrix ${{{\hat \Sigma }_{\bf{k}\sigma }}\left( E \right)}$, the unconventional term ${{{\hat P}_{\bf{k}\sigma }}\left( E \right)}$  called ``the strength operator'' appears due to the $X$-operators algebra and determines the QP spectral weight as well as the renormalized bandwidth.
The Hubbard I solution~\cite{r7LDAGTB} is obtained by setting ${{\hat \Sigma }_{\bf{k}\sigma }}\left( E \right) = 0$ and ${\left( {{{\hat P}_{\bf{k}\sigma }}\left( E \right)} \right)_{mm'}} = {F_m}{\delta _{mm'}}$, where ${F_m} = \left\langle {{X^{pp}}} \right\rangle  + \left\langle {{X^{qq}}} \right\rangle $. We call ${F_m}$ ``occupation factor''; it provides a spectral weight for the QP excitations between empty states. The dispersion equation for the QP band structure of the Hubbard fermions in this case is given by

\begin{equation}
\label{eq10}
\det \left\| \delta_{mn} \left(E - \Omega_m\right)/F(m) - t^{mn}(\bf{k}) \right\| = 0.
\end{equation}
The dispersion equation, Eq.~(\ref{eq10}), is similar to the conventional TB dispersion equation, but instead of a single electron local energy ${\varepsilon _\lambda }$ we have a local QP energy ${\Omega _m}$. That is why we call this approach the ``generalized TB method''.
The occupation numbers $\left\langle {{X^{pp}}} \right\rangle $ are calculated self-consistently via the chemical potential equation, ${n_e} = \frac{{\left\langle {{N_e}} \right\rangle }}{N} = \frac{1}{N}\sum\limits_{f,n,i} {n\left\langle {X_f^{n,i;n,i}} \right\rangle}$. The change of the concentration ${n_e}$ redistributes the occupation numbers and due to the occupation factors, ${F_m}$, changes the QP band structure.
In the GTB method, the intracell Green function can be found exactly
\begin{equation}
\label{eq11}
 G_{0 \sigma}^{\lambda\lambda'}(E) = \sum\limits_m \left| \gamma_{\lambda\sigma}(m) \right|^2 \delta_{\lambda\lambda'} D_{0 \sigma}^m(E),
\end{equation}
where $D_{0\sigma }^m\left( E \right) = {{{F_m}} \mathord{\left/ {\vphantom {{{F_m}} {\left( {E - {\Omega _m} + i\delta } \right)}}} \right. \kern-\nulldelimiterspace} {\left( {E - {\Omega _m} + i\delta } \right)}}$. Comparing this exact local Green function with the Lehmann representation, we can say that electron here is a linear combination of local (Hubbard) fermions with QP energy ${\Omega _m}$ and a spectral weight ${\left| {{\gamma _{\lambda \sigma }}\left( m \right)} \right|^2}{F_m}$.
It should be stressed that the LDA+GTB bands are not the single electron conventional bands. There is no any single particle Schr\"odinger equation with the effective potential that gives the LDA+GTB band structure. These QP are excitations between different multielectron terms. The LDA+GTB bands depends on the multielectron term occupation numbers through ${{{\hat P}_{\bf{k}\sigma }}\left( E \right)}$ and ${{{\hat \Sigma }_{\bf{k}\sigma }}\left( E \right)}$ that should be calculated via the chemical potential equation. There is no rigid band behavior from the very beginning; the band structure depends on doping, temperature, pressure, and external fields.

\subsection{Electronic Structure of Cuprates}
\label{ES:3}
In this section we present the results for cuprates with one and two CuO$_2$ layer in the unit cell: hole doped cuprate La$_{2-x}$Sr$_x$CuO$_4$ and YBa$_2$Cu$_3$O$_7$. Both electronic and magnetic structure should be calculated self-consistently. Thus, we consider separately the antiferromagnetic (AFM) phase with the long-range AFM order and the spin-liquid phase with the short-range AFM correlations.

\subsubsection{Band Structure of the Undoped La$_2$CuO$_4$}
\label{ES:3.1}

The ED of the multiband $p - d$ Hamiltonian Eq.~(\ref{eq1}) for the CuO$_6$ cluster which includes the apical oxygens, results in the following local eigenstates (in the hole representation): 1) ${n_h} = 0$, the vacuum state $\left| 0 \right\rangle $ formed by ${d^{10}}{p^6}$ orbital configuration; 2) ${n_h} = 1$, the spin doublets $\left| {\sigma ,\lambda } \right\rangle $ with different orbital symmetries. The lowest one is ${b_{1g}}$, $\left| \sigma  \right\rangle $, and the first excited is ${a_{1g}}$ molecular orbital; 3) ${n_h} = 2$. A set of two-hole singlets and triplets, spread in the energy region of about ${U_d} \sim 10$ eV. The lowest one is the ${}^1{A_1}$ singlet $\left| S \right\rangle $ that includes the Zhang-Rice singlet among other two-hole singlets, nevertheless we will call this state as Zhang-Rice one in the rest of the paper. The first excited triplet $\left| {TM} \right\rangle $ ($M =  + 1,0, - 1$) has the ${}^3{B_{1g}}$ symmetry. The total number of eigenstates is about 100.

\begin{figure}[t]
\begin{center}
 \includegraphics*[width=0.7\textwidth]{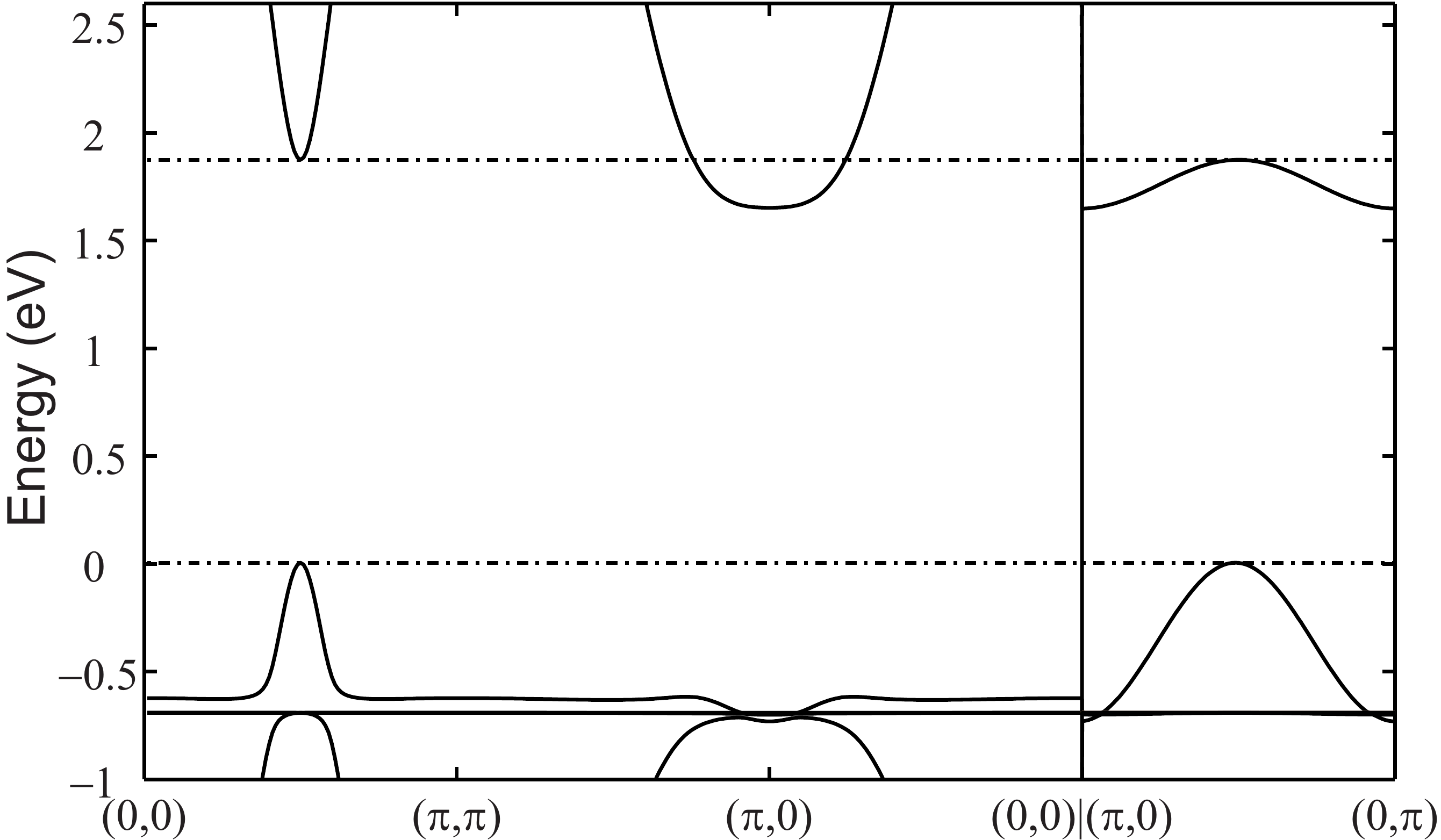}
 \caption{The LDA+GTB band structure of the AFM La$_2$CuO$_4$ along the principal cuts of the Brillouin zone \protect\cite{KOGNP}.
\label{fig3}}
\end{center}
\end{figure}

The next practical step is the calculation of the matrix elements, $\left\langle 0 \right|{c_{f\lambda \sigma }}\left| {1,\sigma '\lambda } \right\rangle $, $\left\langle {1,\sigma '\lambda } \right|{c_{f\lambda \sigma }}\left| {2,i} \right\rangle $, and the construction of $X$-operators for all single electron orbitals. Here, $\lambda $ stands for Cu-${d_{{x^2} - {y^2}}}$, Cu-${d_{z^2}}$, O-$b$, O-$a$, or O-$p_z$ orbital. For the AFM-ordered La$_2$CuO$_4$, we use the two-sublattice ($A$ and $B$) version of the Hubbard I solution with the two occupation factors, ${F_{m,A}}$ and ${F_{m,B}}$. Due to the effective molecular field, the local ${b_{1g}}$ spin doublet is splitted so that at $T = 0$ : $\left\langle {X_A^{ \uparrow  \uparrow }} \right\rangle  = \left\langle {X_B^{ \downarrow  \downarrow }} \right\rangle  = 1$, $\left\langle {X_A^{ \downarrow  \downarrow }} \right\rangle  = \left\langle {X_B^{ \uparrow  \uparrow }} \right\rangle  = 0$. The GTB band structure and the DOS in the wide energy region with all excited two-hole states $\left| {2,i} \right\rangle $ have been calculated in Ref.~\cite{Ovchinnikov}. The empty conductivity band is formed by only one Hubbard fermion, $X_f^{0,\sigma }$. It is separated by the CT gap ${U_{eff}} = {E_{CT}} \approx 2$ eV from the filled valence band. The valence band is formed by a large number of Hubbard fermions $X_f^{\sigma ,2i}$ and consists of a set of narrow bands with the total bandwidth about 6 eV. If we are interested in a smaller energy window around the ${E_{CT}}$ (for example, to study ARPES), it is possible to simplify the calculation by neglecting the high-energy states from both $\left| {2,i} \right\rangle $ and $\left| {1,\sigma '\lambda } \right\rangle $ sets. Then the minimal realistic basis is $\left\{ {\left| 0 \right\rangle ,\left| \sigma  \right\rangle ,\left| S \right\rangle ,\left| {TM} \right\rangle } \right\}$.
The $X$-representation for the fermionic operators in this basis is ${c_{f{d_{{x^2} - {y^2}}}\sigma }} = uX_f^{0\sigma } + 2\sigma {\gamma _x}X_f^{\bar \sigma S}$, ${c_{f{p_b}\sigma }} = \upsilon X_f^{0\sigma } + 2\sigma {\gamma _b}X_f^{\bar \sigma S}$, ${c_{f{p_a}\sigma }} = {\gamma _a}\left( {\sigma \sqrt 2 X_f^{\bar \sigma T0} - X_f^{\sigma T2\sigma }} \right)$, ${c_{f{d_{{z^2}}}\sigma }} = {\gamma _z}\left( {\sigma \sqrt 2 X_f^{\bar \sigma T0} - X_f^{\sigma T2\sigma }} \right)$, ${c_{f{p_z}\sigma }} = {\gamma _p}\left( {\sigma \sqrt 2 X_f^{\bar \sigma T0} - X_f^{\sigma T2\sigma }} \right)$. Here, $\bar \sigma  =  - \sigma $ and ${T2\sigma }$ stands for $T\left( { + 1} \right)$ or $T\left( { - 1} \right)$ depending on the value of the spin label $\sigma  =  \pm {1 \mathord{\left/ {\vphantom {1 2}} \right. \kern-\nulldelimiterspace} 2}$. The explicit form of the TB Hamiltonian Eq.~(\ref{eq1}) in this basis looks like the two-band singlet-triplet Hubbard model:
\begin{eqnarray}
\label{eq12}
{H_{pd}} &=& \sum\limits_f {\left[ {{\varepsilon _1}\sum\limits_\sigma  {X_f^{\sigma \sigma }}  + {\varepsilon _{2S}}X_f^{SS} + {\varepsilon _{2T}}\sum\limits_M {X_f^{TM\,TM}} } \right]} \\ \nonumber
&+&\sum\limits_{f \ne g,\sigma } {\left[ {t_{fg}^{00}X_f^{\sigma 0}X_g^{0\sigma } + t_{fg}^{SS}X_f^{S\bar \sigma }X_g^{\bar \sigma S} + 2\sigma t_{fg}^{0S}\left( {X_f^{\sigma 0}X_g^{\bar \sigma S} + h.c.} \right) } \right.} \\ \nonumber
&+& t_{fg}^{ST}\left\{ {\left( {\sigma \sqrt 2 X_f^{T0\bar \sigma } - X_f^{T2\sigma \sigma }} \right)\left(\upsilon {X_g^{0\sigma } + 2\sigma {\gamma _b}X_g^{\bar \sigma S}} \right) + h.c.} \right\} \\ \nonumber
&+& \left. {t_{fg}^{TT}\left( {\sigma \sqrt 2 X_f^{T0\bar \sigma } - X_f^{T2\sigma \sigma }} \right)\left( {\sigma \sqrt 2 X_g^{\bar \sigma T0} - X_g^{\sigma T2\sigma }} \right)} \right].
\end{eqnarray}

The hopping parameters of the effective Hubbard model are expressed through the microscopic \textit{ab initio} parameters ${t_{pd}}$ and ${t_{pp}}$, $t_{fg}^{00} =  - 2{t_{pd}}{\mu _{fg}}2u\upsilon  - 2{t_{pp}}{v_{fg}}{\upsilon ^2}$, $t_{fg}^{SS} =  - 2{t_{pd}}{\mu _{fg}}2{\gamma _x}{\gamma _b} - 2{t_{pp}}{v_{fg}}\gamma _b^2$, $t_{fg}^{0S} =  - 2{t_{pd}}{\mu _{fg}}\left( {\upsilon {\gamma _x} + u{\gamma _b}} \right) - 2{t_{pp}}{v_{fg}}\upsilon {\gamma _b}$, $t_{fg}^{TT} = \frac{2}{{\sqrt 3 }}{t_{pd}}{\lambda _{fg}}2{\gamma _a}{\gamma _z} + 2{t_{pp}}{v_{fg}}\gamma _a^2 - 2{{t'}_{pp}}{\lambda _{fg}}2{\gamma _p}{\gamma _a}$, $t_{fg}^{ST} = \frac{2}{{\sqrt 3 }}{t_{pd}}{\xi _{fg}}{\gamma _z} + 2{t_{pp}}{\chi _{fg}}{\gamma _a} - 2{{t'}_{pp}}{\xi _{fg}}{\gamma _p}$. Here ${\mu _{fg}}$, ${v_{fg}}$, ${\lambda _{fg}}$, ${\xi _{fg}}$ and ${\chi _{fg}}$ are the coefficients of the oxygen group orbitals construction, and $u$, $\upsilon $, ${\gamma _x}$, ${\gamma _b}$, ${\gamma _a}$, ${\gamma _p}$, and ${\gamma _z}$ are the matrix elements ${\gamma _{\lambda \sigma }}\left( {p,q} \right)$ (see Ref. ~\cite{GavOvchBor} for details).
The QP band structure of La$_2$CuO$_4$ is shown in Fig.~\ref{fig3} for the $\Gamma \left( {0,0} \right) - M\left( {\pi ,\pi } \right) - X\left( {\pi ,0} \right) - \Gamma \left( {0,0} \right)$ and $X\left( {\pi ,0} \right) - Y\left( {0,\pi } \right)$ cuts of the square lattice Brillouin zone. Zero at the energy scale is not the Fermi level but rather fixed by the condition ${\varepsilon _{{d_{{x^2} - {y^2}}}}} = 0$.
The top of the valence band is at the ${\bar M} = \left(\pi/2, \pi/2 \right)$ point, while the bottom of the conductivity band is at the $X$ point. The dispersion of the valence band determined by the hybridization of the two bands which formed by either $X_f^{\bar \sigma S}$ or $X_f^{\bar \sigma TM}$ Hubbard fermions. The hybridization between them is provided by the ${t^{ST}}$ hopping matrix elements in Eq.~(\ref{eq12}). These are fermionic bands, but frequently in the literature terms ``singlet band'' and ``triplet band'' are used. These terms reflect the final two-hole states involved in the QP excitations. The dominant spectral weight in the singlet band stems from the oxygen ${b_{1g}}$ states, while for the bottom of the conductivity band it is from the ${d_{{x^2} - {y^2}}}$ states of Cu.

\subsubsection{Effective Exchange Interaction and its Pressure Dependence}
\label{ES:3.2}

Using  the LDA+GTB approach, we calculate the compression dependence of superexchange constant $J(P)$ in La$_2$CuO$_4$ (La214), where the two-magnon Raman scattering experiments show that $J(P)$ has a substantially weaker pressure dependence~\cite{Aronson_etal1990,Eremets_etal1991,Schilling} unlike the pressure dependencies of superexchange interaction in many other conventional transition-metal oxide~\cite{JohnsonSievers1974, Kaneko_etal1987, KimMoret1988, Massey_etal1990}.

Within the perturbation theory using the atomic orbitals representation~\cite{ZhangRice1988,Eskes_etal1993,Maekawa_etal2004}, a superexchange interaction is obtained at the fourth-order of a perturbation theory and the weak-pressure dependence of $J(P)$ is clearly not consistent with the pressure dependencies of main parameters of the $pd$-model: $t_{pd}\sim a^{-\alpha}$ $\left( {2.5 \leq \alpha \leq 3.0} \right)$~\cite{Smith_1969,Fuchikami_1970,Jough_1975,Shrivastava_etal1976}, $\Delta\sim a^{-\beta}$ $(\beta\approx0.4\pm0.4)$~\cite{Venkateswaran_etal1989}.

Comparison of the results
at the the fourth-order with the calculations in higher orders of perturbation theory~\cite{Eskes_etal1993} and the exact diagonalization of finite clusters~\cite{Eskes_etal1993,Eskes_etal1989,Ohta_etal1991,Stechel_etal1988,Annet_etal1989} show that the in-plane superexchange $J$ depends on the $t_{pd}$ significantly weaker and due to the large value of $t_{pd}/\Delta$ ratio in the CuO$_2$ layer, the fourth order of perturbation theory may be insufficient.

Essentially, there are two acceptable approaches to the superexchange study. The first one is the calculation with the intermediate two-hole states which arise in the hopping from oxygen to oxygen in the perturbation theory of a higher than a fourth order ~\cite{Eskes_etal1993}. Another approach is a cell perturbation theory taking into account all of the excited states. The latter seems more appropriate when it is necessary to work with a large number of excited states. Especially, it is important due to the fact that the energy gap between the singlet and the triplet two-hole cell states involving $2p_z$ apical oxygen orbital can be quite small ~\cite{Kamimura1987,Kamimura_etal1990,Eskes_etal1991,Grant_etal1991,Ohta1_etal1991}.

Hereinafter we calculate the exchange constant $J$ in the LDA+GTB approach and compare our results with the conclusion from the neutron experiments in undeformed La214~\cite{Coldea_etal1990}, the experiments related to the two-magnon Raman scattering in deformed materials~\cite{Aronson_etal1990} at the 0\% and 3\% hydrostatic and uniaxial (along the $c$ axis) compressions, and the fourth-order perturbation theory~\cite{Maekawa_etal2004}. We found that the obtained superexchange parameter $J\approx0.15$ eV in the undeformed La214 is close to the experimental value of 0.146 eV~\cite{Coldea_etal1990} and $J(P)$ increases by about 20\% under the 3\%-hydrostatic compression. At the same time, the superexchange interaction is only slightly reduced by 5\% under the uniaxial compression. According to the available experimental data~\cite{Aronson_etal1990}, the superexchange interaction is increased by 12\% at 2.3\% under the hydrostatic compression. The $J(P)$ correlates with the energy gap $\delta=\varepsilon(^3B_{1g})-\varepsilon (A_{1g})$ between the energy levels of the $^3B_{1g}$ triplet and Zhang-Rice $A_{1g}$ two-hole states. In addition, we also made a GTB calculation of the $A_{1g} \leftrightarrow ^3B_{1g}$ two-hole state crossover hypothetical case. The superexchange interaction $J$ still has the antiferromagnetic character.

The exchange interaction appears in the second order of the cell perturbation theory with respect to the intercluster hoppings~\cite{Jefferson_etal1992}. This corresponds to virtual excitations from the occupied singlet and triplet bands through the insulating gap to the conduction band $m = 0$, ${\alpha _0} = (0,\sigma )$ and back. These perturbations are described by the off-diagonal elements $t_{fg}^{0m}$ with $m \ge 1$ in the expression (\ref{eq6}).
In order to eliminate them, we generalize the projection operator method proposed by Chao \textit{et al.}~\cite{Chao_etal1977} to the multiband Hubbard model. Since the diagonal Hubbard operators are projection operators, the $X$-operator representation allows us to construct this generalization. In our case, the total number of diagonal two-hole operators $X_f^{\mu \mu }$ is $N = N_S + 3N_T$. By neglecting the exponentially low temperature occupation of excited one-hole terms in the absence of doping when none of the two-hole
terms is occupied, we can retain only one lower one-hole term. 
We choose a pair of neighboring sites $(i, j)$ and construct the set of projection operators ${p_\mu }$
\begin{eqnarray}
{p_0} &=& \left( {X_i^{00} + \sum\limits_\sigma  {X_i^{\sigma \sigma }} } \right)\left( {X_j^{00} + \sum\limits_\sigma  {X_j^{\sigma \sigma }} } \right), \label{eq:2.21} \\
{p_\mu } &=& X_i^{\mu \mu } + X_j^{\mu \mu } - X_i^{\mu \mu }\sum\limits_{n'} {X_j^{n'n'}}, \label{eq:2.22}
\end{eqnarray}
where $1 \le \mu \le N$. It is easy to check that each operator ${p_\mu }$ is a projection operator $p_\mu ^2 = {p_\mu }$ and that these operators form a complete system and are orthogonal, ${p_\mu }{p_\nu } = {\delta _{\mu \nu }}{p_\mu }$, $\sum\limits_{\mu  = 0}^N {{p_\mu }}  = 1$. By using the identity
%
$H = \sum\limits_{\mu \nu } {{p_\mu }H{p_\nu }}$
%
we calculate the diagonal and off-diagonal elements of matrices. In this case, the term ${p_0}H{p_0}$ corresponds
to the Hamiltonian component acting in the lower Hubbard band ${\alpha _0} = (0,\sigma )$, etc. It is easy to show that the
equality
%
$\sum\limits_\mu  {{p_\mu }} {H_0}{p_\mu } = {H_0}$
%
is satisfied and that the diagonal elements ${p_\mu }{H_1}{p_\mu }$ describe the hoppings in the band $\mu $ and the off-diagonal
elements ${p_\mu }{H_1}{p_\nu }$ correspond to the hybridization of the bands $\mu $ and $\nu$; that is,
\begin{equation}
{p_\mu }{H_1}{p_\nu } = \sum\limits_{ij} {t_{ij}^{\mu \nu }} X_i^{\mu \dag} X_j^\nu
\label{eq:2.25}
\end{equation}
We choose the off-diagonal matrix elements
\begin{equation}
{\tilde H}(\varepsilon ) = {\tilde H_0} + \varepsilon {\tilde H_1}, \;\;\;
{\tilde H_0} = \sum\limits_\mu  {{p_\mu }H{p_\mu }}, \;\;\;
{\tilde H_1} = \sum\limits_{\mu  \pm \nu } {{p_\mu }H{p_\nu }}
\label{eq:2.26}
\end{equation}
as a perturbation and perform the standard unitary transformation
%
${H'{(\varepsilon )}} = {e^{ - i\varepsilon S}}H(\varepsilon ){e^{i\varepsilon S}}$
%
to eliminate the linear in $\varepsilon$ contributions in ${\tilde H_1}$. If the matrix $\hat S$ satisfies the equation
\begin{equation}
{\tilde H_1} + i{\left[ {{{\tilde H}_0}, S} \right]_ - } = 0
\label{eq:2.28}
\end{equation}
the transformed Hamiltonian has the form
\begin{equation}
H'(\varepsilon ) = {\tilde H_0} + i \varepsilon^2 \left[ {\tilde H}_1, S \right]_- / 2.
\label{eq:2.29}
\end{equation}
In order to solve Eq.~(\ref{eq:2.28}), we multiply each term by ${p_\mu }$ from the left and by ${p_\nu }$ from the right. As a result, we obtain
\begin{equation}
{p_\mu }H{p_\nu }(1 - {\delta _{\mu \nu }}) + i({p_\mu }H{p_\mu })({p_\mu }S{p_\nu }) - i({p_\mu }S{p_\nu })({p_\nu }H{p_\nu }) = 0
\label{eq:2.30}
\end{equation}
This equation coincides in form with the corresponding equation in the work~\cite{Chao_etal1977} and differs from it only in the dimension of matrices.
It follows from (\ref{eq:2.30}) that the diagonal matrix elements have the form ${p_\mu }S{p_\mu } = \eta {p_\mu }$, where $\eta $ is a constant. In order to solve the equation with respect to the off-diagonal elements ${p_\mu }S{p_\nu }$,
we make the approximation ${p_\mu }H{p_\mu } \to {\varepsilon _\mu }$. As a result, the solution has the form
%
${p_\mu }S{p_\nu } = i{p_\mu }H{p_\nu } / \Delta _{\mu \nu }$
%
and the effective Hamiltonian is represented as
\begin{eqnarray}
\label{eq:2.32}
H'(\varepsilon=1) = \sum\limits_\mu {{p_\mu}H} {p_\mu} + \frac{1}{2} \sum\limits_{\nu \ne \mu } {({p_\mu}H} {p_\nu}S - S{p_\mu}H{p_\nu}) = \sum\limits_\mu {{p_\mu}H} {p_\mu} \nonumber\\
-\frac{1}{2} \sum\limits_{\mu  \ne \nu } {\left\{ {\frac{{{{\left[ {{p_\mu }H{p_\nu },{p_\nu }H{p_\mu }} \right]}_ - }}}{{{\Delta _{\mu \nu }}}} + \sum\limits_{\scriptstyle\alpha  \ne \mu ,\hfill\atop\scriptstyle\alpha  \ne \nu \hfill} {\left[ {\frac{{({p_\mu }H{p_\nu })({p_\nu }H{p_\alpha })}}{{{\Delta _{\nu \alpha }}}} - \frac{{({p_\alpha }H{p_\mu })({p_\mu }H{p_\nu })}}{{{\Delta _{\alpha \mu }}}}} \right]}} \right\}}
\end{eqnarray}


The calculation of the terms in the Hamiltonian (\ref{eq:2.32}) for the singlet and triplet bands leads to different results. The interband transitions through the gap are described by the commutator
\begin{equation}
\left[{{p_0}H{p_\nu }, {p_\nu }H{p_0}} \right]_-.
\label{eq:2.33}
\end{equation}
For the $n$-th singlet band ${\alpha _\nu } = (-\sigma, nS)$ the commutator (\ref{eq:2.33}) becomes
%
$\sum\limits_{fgij} {\sum\limits_{\sigma \sigma '} {{{\left[ {X_f^{\sigma 0}X_g^{ - \sigma ,ns},\;X_i^{ns, - \sigma '}X_j^{0\sigma '}} \right]}_-}}}.$
%
The exchange contribution to the Heisenberg part of the Hamiltonian has the form
\begin{equation}
{H_A} = \sum\limits_{ij} {{J_A}\left( {{{\bf{R}}_{ij}}} \right)} \left( {{{\bf{s}}_i}{{\bf{s}}_j} - \frac{1}{4}{n_i}{n_j}} \right)
\label{eq:2.35}
\end{equation}
where $\bf{s}$ and ${n_i}$ are the spin operators for $s = 1/2$ and the number of particles at the $i$-th site, respectively, and
\begin{equation}
{J_A}\left( {{{\bf{R}}_{ij}}} \right) = \sum\limits_n {J_A^{\left( n \right)}} ({{\bf{R}}_{ij}}) = \sum\limits_{n = 1}^{{N_s}} {{{\left| {t_{ij}^{0,ns}} \right|}^2}} /{\Delta _{ns}},\;\;\;{\Delta _{ns}} = {E_{ns}} - 2{\varepsilon _1}.
\label{eq:2.36}
\end{equation}
For the $m$-th triplet band, the commutator (\ref{eq:2.33}) is
%
$\left[ {X_f^{\sigma ;0}\left( {X_g^{\sigma ; - m,2\sigma } + {\textstyle{1 \over {\sqrt 2 }}}X_g^{ - \sigma ;m,0}} \right)\left( {X_i^{m,2\sigma ';\sigma '} + {\textstyle{1 \over {\sqrt 2 }}}X_i^{m,0; - \sigma '}} \right)X_j^{0;\sigma '}} \right].$
%
As a result, the ferromagnetic exchange contribution to the Heisenberg part of the Hamiltonian takes the form
\begin{equation}
{H_F} = \sum\limits_{ij} {{J_B}\left( {{{\bf{R}}_{ij}}} \right)} \left( {{{\bf{s}}_i}{{\bf{s}}_j} + \frac{3}{4}{n_i}{n_j}} \right)
\label{eq:2.38}
\end{equation}
where ${J_B}\left( {{{\bf{R}}_{ij}}} \right) = \sum\limits_m {J_B^{\left( m \right)}\left( {{{\bf{R}}_{ij}}} \right)}  =  - \sum\limits_{m = 1}^{{N_T}} {{{\left| {t_{ij}^{0,m}} \right|}^2}} /2{\Delta _m}$ and ${\Delta _m} = {E_m} - 2{\varepsilon _1}$. By summing up over all singlet and triplet bands, we find the following expression for the effective exchange interaction parameter:
\begin{equation}
J_{ij} = \sum\limits_{n = 1}^{{N_S}} \left| t_{ij}^{0,ns} \right|^2 / \Delta _{ns} - \sum\limits_{m = 1}^{{N_T}} \left| t_{ij}^{0,m} \right|^2 / 2 \Delta_m.
\label{eq:2.39}
\end{equation}
In our case, the excited singlet states also results in the antiferromagnetic contribution, which decreases with an increase in the energy of the excited term at the expense of the denominator. The ferromagnetic contribution of the triplet states is associated with the fact that the spins of two holes upon triplet formation are oriented parallel to each other and then hoppings from site to site carry this parallel spin orientation.
\begin{table}
\caption {Single electron energies, hopping parameters, $J(P)$ and $\delta$ for orthorhombic La214 (all values in eV). Here $x^2$, $z^2$, $p_x$, $p_y$, $p_z$ denote Cu-$d_{x^2-y^2}$, Cu-$d_{3z^2-r^2}$, O$_{p}$-$p_x$, O$_p$-$p_y$, O$_a$-$p_z$ orbitals respectively.}
\label{tab:hopsLa}
\begin{tabular}{c|c|c|c|c}
\footnotesize Parameters &\footnotesize Connecting vectors &\footnotesize 3\%-compr. along the $c$-axis &\footnotesize Undeformed &\footnotesize 3\%-hydrostatic compr.\\
\hline
                                   \footnotesize $\varepsilon_{x^2}$&                        &\footnotesize -2.031     &\footnotesize -1.849   &\footnotesize -1.578\\
                                   \footnotesize $\varepsilon_{x^2}-\varepsilon_{z^2}$&                         &\footnotesize 0.119     &\footnotesize 0.225   &\footnotesize 0.204\\
                                   \footnotesize $\varepsilon_{x^2}-\varepsilon_{p_x}$&	        			          &\footnotesize 0.983    &\footnotesize 0.957    &\footnotesize 1.004\\
                                   \footnotesize $\varepsilon_{x^2}-\varepsilon_{p_y}$&	                        & \footnotesize 0.983    &\footnotesize 0.957    &\footnotesize 1.004\\
                                   \footnotesize $\varepsilon_{x^2}-\varepsilon_{p_z}$&       				          &\footnotesize -0.503    &\footnotesize -0.173    &\footnotesize -0.311\\
\hline
\footnotesize t($x^2$,$x^2$)	 &\footnotesize(-0.493,-0.5)          &\footnotesize-0.173                 &\footnotesize-0.188 &\footnotesize-0.215\\

\hline
\footnotesize t($z^2$,$z^2$)	 &\footnotesize(-0.493,-0.5)           &\footnotesize 0.050                  &\footnotesize 0.054 &\footnotesize 0.062\\

\hline
\footnotesize t($x^2$,$p_x$)	 &\footnotesize(0.246,0.25,-0.02)     &\footnotesize 1.302                 &\footnotesize 1.355 &\footnotesize 1.527\\

\hline
\footnotesize t($z^2$,$p_x$)	 &\footnotesize(0.246,0.25,-0.02)      &\footnotesize-0.547	            &\footnotesize-0.556 &\footnotesize -0.618\\

\hline
\footnotesize t($z^2$,$p_z$)	 &\footnotesize(0,0.04,0.445) 	       &\footnotesize0.851                  &\footnotesize0.773 &\footnotesize 0.875\\

\hline
\footnotesize t($p_x$,$p_y$)	 &\footnotesize(0.493, 0.0) 	       &\footnotesize-0.854                 &\footnotesize-0.858 &\footnotesize-0.935\\
\footnotesize t$'$($p_x$,$p_y$)  &\footnotesize(0.985,0.5,0.041)          &\footnotesize 0.757                 &\footnotesize 0.793 &\footnotesize 0.862\\

\hline
\footnotesize t($p_x$,$p_z$)	 &\footnotesize(-0.246,-0.21,0.465)    &\footnotesize-0.447                 &\footnotesize-0.391 &\footnotesize-0.423\\
\footnotesize t$'$($p_x$,$p_z$)  &\footnotesize(0.246,0.29,-0.425)    &\footnotesize-0.424	            &\footnotesize-0.377 &\footnotesize-0.408\\
\hline
\footnotesize $J(\Delta J\%)$	 &\footnotesize    &\footnotesize 0.14(-5.7\%)                &\footnotesize 0.15 &\footnotesize 0.18(19.9\%)\\
\hline
\footnotesize $\delta$	 &\footnotesize    &\footnotesize 0.82                &\footnotesize 1.33 &\footnotesize 1.45\\
\end{tabular}
\end{table}

Table~\ref{tab:hopsLa} shows the values of hopping parameters and single electron energies for orthorhombic La214 obtained in WF projection procedure for different sets of trial orbitals~\cite{KOGNP} at zero, hydrostatic, and uniaxial (along the $c$ axis) 3\% compressions. Despite the fact that the Table shows the same vector, a system of the connecting vectors varies slightly with the increase of compression. The next to last line of the Table~\ref{tab:hopsLa} shows the values of superexchange constant $J$, calculated via the Eq.~(\ref{eq:2.39}) at the $N_S=15$ and $N_T=10$ in the five-band $pd$-model. The calculations show that the value of the superexchange $J$ increases by $\sim$20\% under hydrostatic 3\% compression. This result can be compared with experimental results~\cite{Aronson_etal1990} for La214. Under the hydrostatic
100 Kbar pressure the $r_{Cu-O}$ reduces by $\sim$ 2.3\%, while the $J$ increases by $\sim$ 12\%.
There are also studies where the linear dependence of the superexchange on the pressure was observed up to a $P=410$ Kbar ~\cite{Eremets_etal1991}.  According to the dependence of the La214 crystal structure on the pressure we can found the pressure $P\sim 205$ Kbar corresponds to the 3\%-deformed material ~\cite{Akhar_etal1988}. The $J(P)$ at this pressure increases by $\sim$ 18\% ~\cite{Eremets_etal1991}. The calculated value $J \approx 0.15$ eV in the undeformed La214 exceeds the value of $0.1-0.13$ eV~\cite{Aronson_etal1990,Eremets_etal1991} obtained in experiments on the two-magnon Raman scattering but agrees well with the value of $J = 0.146$ eV~\cite{Coldea_etal1990} from the neutron experiments. By contrast, under the uniaxial 3\% compression along the $c$-axis $J$ is decreased by 5.7\%, i.e. the superexchange constant changes much weaker. In the both cases of the hydrostatic and the anisotropic compression the superexchange constant $J(P)$ correlates with the inplane hopping parameters and $dd$-excitation energy $\delta=\varepsilon(^3B_{1g})-\varepsilon(A_{1g})$ involving the two-hole states: Zhang-Rice-type singlet state $A_{1g}$ and
${}^3{B_{1g}}$ triplet state (see Table~\ref{tab:hopsLa}). The AFM character of superexchange is maintained even at a hypothetical set of Hamiltonian parameters corresponding the $A_{1g}\leftrightarrow{^3B_{1g}}$ - singlet-triplet crossover.

Number of experimental results on the external and internal pressure influence on the temperature of superconducting state transition in cuprates are quite consistent. These results show that $T_c$ increase with applying pressure along the $a$ and $b$ axes in CuO$_2$ plane and its decrease with pressure along the $c$-axis which is perpendicular to the copper-oxygen layer. The similar effect of pressure on the exchange parameter $J$ is seen from the Table~\ref{tab:hopsLa}. The isotropic pressure results in the increasing $J$ while pressure along the $c$-axis gives decreasing $J$. It is clear that parameters in our calculations reflect this concept wery well: compression along the $c$-axis makes interatomic distance to grow, decrease the wave functions overlapping and thus the hopping t$_{pd}$. Correlation between $J$ and $T_c$ is and additional confirmation for the magnetic mechanism of pairing in cuprates.

\subsubsection{Cascade of Lifshitz Quantum Phase Transitions in La$_{2 - x}$Sr$_x$CuO$_4$}
\label{ES:3.3}

In this section we will discuss doping dependence of the electronic and magnetic structure calculated self-consistently assuming the isotropic spin liquid state with the antiferromagnetic short range order. Contrary to many model calculations within the Hubbard or $t-J$-model we have no empirical fitting parameters. The effective low energy Hamiltonian has been derived within LDA+GTB with \textit{ab initio} calculated parameters \cite{KOGNP}. The low-energy Hamiltonian for La$_{2-x}$Sr$_x$CuO$_4$ (LSCO) is the $t - t' - t'' - {J^*}$-model obtained via exclusion of the interband
excitations. Here ${J^*}$ means that besides the Heisenberg exchange term a three-site correlated hopping ${H_3}$ is also included, ${H_{t - {J^*}}} = {H_{tJ}} + {H_3}$, where
\begin{eqnarray}
\label{eq12tJ}
H_{tJ} &=& \sum\limits_{f, \sigma} (\varepsilon - \mu) X_f^{\sigma \sigma}+\sum\limits_f 2(\varepsilon- \mu) X_f^{SS} \nonumber\\
&+& \sum\limits_{f \ne g, \sigma} \left[ t_{fg} X_f^{S \bar\sigma} X_g^{\bar\sigma S} + \frac{J_{fg}}{4} \left( X_f^{\sigma \bar\sigma} X_g^{\bar\sigma \sigma} - X_f^{\sigma \sigma} X_g^{\bar\sigma \bar\sigma} \right) \right], \\
H_3 &=& \sum\limits_{f \ne m \ne g, \sigma} \frac{\tilde t_{fm} \tilde t_{mg}}{U_{eff}}  \left( X_f^{\sigma S} X_m^{\bar\sigma \sigma} X_g^{S \bar\sigma} - X_f^{\sigma S} X_m^{\bar\sigma \bar\sigma} X_g^{S \sigma} \right).
\end{eqnarray}
Here ${t_{fg}} = t_{fg}^{SS}$ is the hopping in the UHB, $J_{fg} = \tilde{t}_{fg}^2/U_{eff}$ is the exchange interaction due to the interband (UHB $\leftrightarrow$ LHB) hopping ${{\tilde t}_{fg}} = t_{fg}^{0S}$ through the charge-transfer gap ${U_{eff}} = {E_{CT}}$, hole creation operator is now $a_{f\sigma }^\dag  = 2\sigma X_f^{S\bar \sigma }$ and its algebra is different from the bare fermion's one. The spin operators are also easily expressed via $X$-operators, $S_f^ +  = X_f^{\sigma \bar \sigma }$, $S_f^z = \left( X_f^{\sigma \sigma} - X_f^{\bar\sigma \bar\sigma} \right)/2$.
Here we dropped out the triplet state $\left| {TM} \right\rangle $ because the triplet itself and the singlet-triplet excitations do not contribute to the near-Fermi level physics.
The \textit{ab initio} derived parameters for LSCO are (in eV) $t = 0.93$, $t' =  - 0.12$, $t'' = 0.15$, $\tilde t = 0.77$, $\tilde t' =  - 0.08$, $\tilde t'' = 0.12$, $J = 0.29$, $J' = 0.003$, $J'' = 0.007$. The three-site correlated hopping term $H_3$ usually is ignored. Its importance for magnetic mechanism of pairing has been demonstrated by \cite{shn_9}. Here we will show its importance also for the dispersion of electrons in a short range antiferromagnetic background.

Our approach is essentially a perturbation theory with the small parameter $t/U$ contrary to the usual Fermi liquid perturbation expansion in terms of $U$ which is large in cuprates. We use a method of irreducible Green functions which is similar to the Mori-type projection technique, with the zero-order Green function given by the well-known Hubbard I approximation. Beyond it there are spin fluctuations.
To describe them one can calculate the self-energy in the non-crossing approximation by neglecting vertex renormalization that is equivalent to the self-consistent Born approximation (SCBA)~\cite{PlakOud}. Resulting electron self-energy contains the space-time dependent spin correlation function $C\left( {\mathbf{q},\omega } \right)$ and results in the finite quasiparticle lifetime, ${\mathop{\rm Im}\nolimits} \Sigma \left( {\mathbf{k},\omega } \right) \ne 0$. Note that at low temperatures $T \le 10$ K the spin dynamics is much slower than the electron one. A typical spin fluctuation time ${10^{ - 9}}$ sec, is much larger than electronic time ${10^{ - 13}}$ sec~\cite{HarMcDonSing}; that is why we can safely neglect the time dependence of the spin correlation function, $C\left( {\mathbf{q},\omega } \right) \to {C_\mathbf{q}}$. The self-energy becomes static, $\Sigma \left( {\mathbf{k},\omega } \right) \to \Sigma \left( \mathbf{k} \right)$, and ${\mathop{\rm Im}\nolimits} \Sigma  = 0$. This is the essence of our mean-field theory \cite{KorOvch} that is similar to the SCBA. Note that $\Sigma \left( {\mathbf{k},\omega } \right)$ is the object completely different from the one in the Fermi liquid approach because here it is build by the diagrams for the $X$-operators, not the standard Fermionic annihilation-creation operators ${a_{f\sigma }}$. In the usual Fermi liquid expansion dynamical self-energy definitely plays a crucial role in the lightly doped cuprates. Here, our theory starts from a different limit where the lowest order approximation  is represented by the Hubbard I solution. The corrections to the strongly-correlated mean-field approach are small because the starting point is already a reasonable approximation for the Mott-Hubbard insulator. That is proved by the small effect of the frequency dependence of the self-energy in Refs.~\cite{PlakOud,BarHayKov}. Moreover the doping-dependence of the FS is determined by ${\mathop{\rm Re}\nolimits} \Sigma $, and it is qualitatively similar in our approach~\cite{KorOvch} and in the approach which properly takes ${\mathop{\rm Im}\nolimits} \Sigma $ into account~\cite{PlakOud,BarHayKov}.
The vertex corrections to the self-energy are small far from the spin-density wave or the charge-density wave instabilities, that is true for the moderate doping. Our approximation for the self-energy is done in the framework of the mode-coupling approximation which has been proved to be quite reliable even for systems with strong interaction~\cite{Prelovsek,PlakOud2}. As shown in the spin-polaron treatment of the $t - J$ model, the vertex corrections to the non-crossing approximation are small and give only numerical renormalization of the model parameters~\cite{LiuManous}.
Green function ${\left\langle {\left\langle {{X_{\mathbf{k}}^{\bar \sigma S}}} \mathrel{\left | {\vphantom {{X_{\mathbf{k}}^{\bar \sigma S}} {X_{\mathbf{k}}^{S\bar \sigma }}}} \right. \kern-\nulldelimiterspace} {{X_{\mathbf{k}}^{S\bar \sigma }}} \right\rangle } \right\rangle _\omega }$ for a hole moving on the background of the short-range AFM order is
\begin{eqnarray}
\label{eq14}
 G(\mathbf{k},\omega) = \frac{(1+x)/2}{\omega - \varepsilon + \mu - \frac{1 + x}{2} t_\mathbf{k} - \frac{1 - x^2}{4} \frac{\tilde{t}_\mathbf{k}^2}{U_{eff}} + \Sigma(\mathbf{k})},
\end{eqnarray}
where
\begin{eqnarray}
\label{eq15}
 \Sigma(\mathbf{k}) &=& -\frac{2}{1 + x} \frac{1}{N} \sum\limits_\mathbf{q} \left\{ \left[ t_{\mathbf{k} - \mathbf{q}} - \frac{1 - x}{2} J_\mathbf{q}  + \frac{1 - x}{2} \frac{\tilde{t}_{\mathbf{k} - \mathbf{q}}^2}{U_{eff}} \right.\right. \nonumber\\
 &-& \left.\left. \frac{1 + x}{2} \frac{2 \tilde{t}_\mathbf{k} \tilde{t}_{\mathbf{k} - \mathbf{q}}}{U_{eff}} \right] \left( \frac{3}{2}C_\mathbf{q} + K_{\mathbf{k} - \mathbf{q}} \right) - \frac{1 + x}{2} \frac{\tilde{t}_\mathbf{q}^2}{U_{eff}} K_\mathbf{q} \right\}.
\end{eqnarray}
Here, ${t_{\mathbf{k}}}$ and ${{\tilde t}_{\mathbf{k}}}$ are the Fourier transforms of the intraband and interband hopping, respectively. The self-energy is determined by static spin correlation function ${C_{0n}} = \left\langle {S_0^ + S_n^ - } \right\rangle $ and kinematic correlation function ${K_{0n}} = \sum\limits_\sigma  {\left\langle {\tilde a_{0\sigma }^\dag {{\tilde a}_{n\sigma }}} \right\rangle } $ between sites $0$ and $n$. The correlation function ${C_{\mathbf{q}}}$ represents the AFM short-range order. In contrast to approach of Ref.~\cite{PlakOud}, we calculate these correlation functions self-consistently up to $n = 9$ (ninth coordination sphere) together with the chemical potential $\mu $. To get the spin correlation function we also obtain the spin Green function ${\left\langle {\left\langle {{X_{\mathbf{q}}^{\sigma \bar \sigma }}} \mathrel{\left | {\vphantom {{X_{\mathbf{q}}^{\sigma \bar \sigma }} {X_{\mathbf{q}}^{\bar \sigma \sigma }}}} \right. \kern-\nulldelimiterspace} {{X_{\mathbf{q}}^{\bar \sigma \sigma }}} \right\rangle } \right\rangle _\omega }$ in a spherically-symmetric spin liquid state~\cite{ShimTak,ValDzeb} with $\left\langle {{S^z}} \right\rangle  = 0$ and the equal correlation functions for each spin component, $\left\langle {S_0^ + S_n^ - } \right\rangle  = 2\left\langle {S_0^zS_n^z} \right\rangle  = {C_{0n}}$. Both ${C_{0n}}$ and ${K_{0n}}$ are essentially doping-dependent and ${C_{0n}}$ decrease with the doping~\cite{KorOvch}. While the nearest neighbor function ${C_{01}}$ is finite for all studied $x$ up to $x = 0.4$ with a kink at $x = {p^ * } = 0.24$, more distant spin correlations fall down to zero for $x > {p^ * }$.
\begin{figure}
\begin{center}
\includegraphics[angle=0,width=0.75\columnwidth,clip=true]{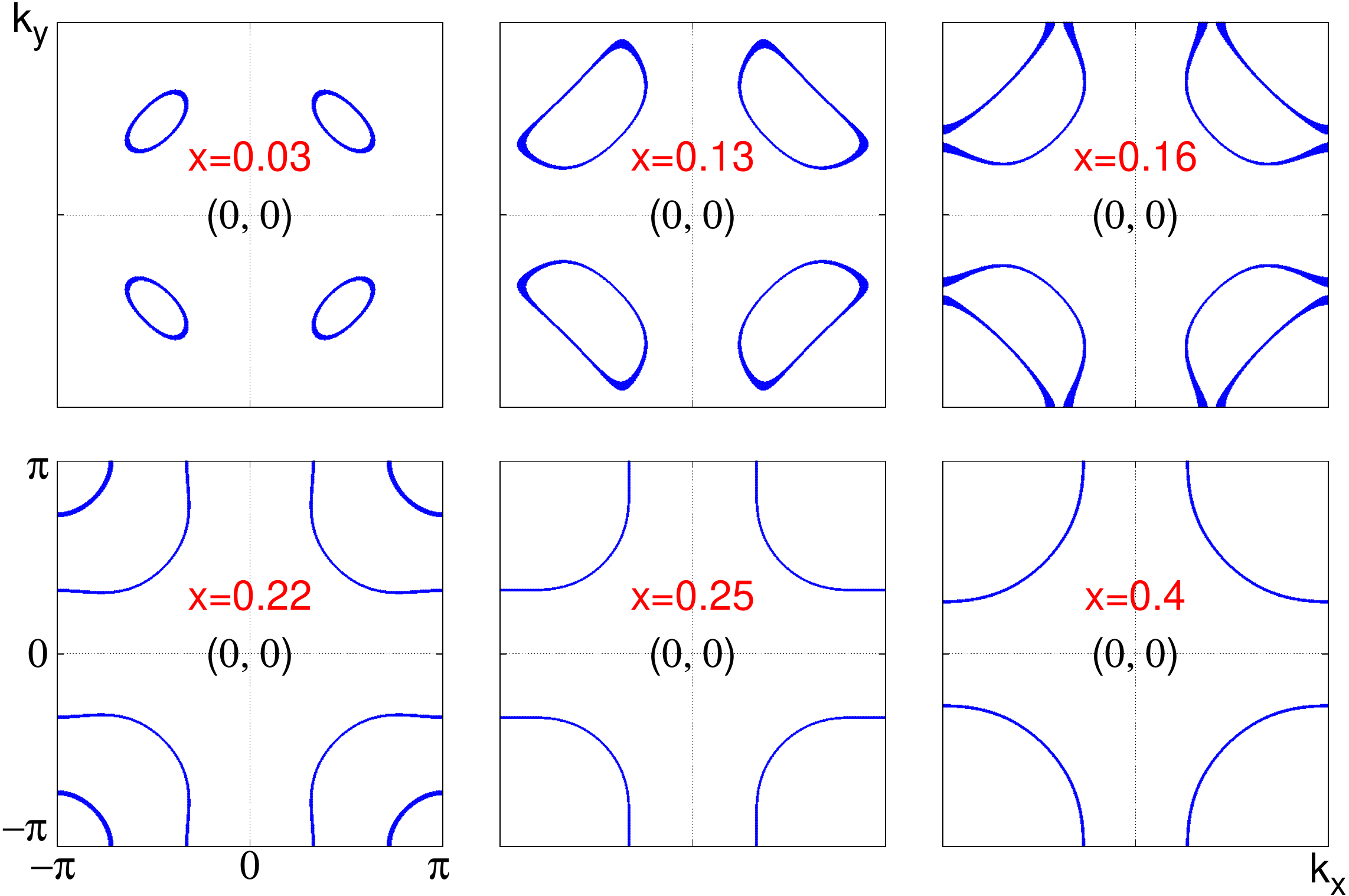}
 \caption{(Color online) Mean-field Fermi surface transformation with doping $x$ as calculated from poles of Eq.~(\ref{eq14}). There are two topological changes: first one is at $x_{c1}=0.15$ and second one is at $x_{c2}=0.24$; see Ref.~\protect\cite{KorOvch} for detailed discussion.}
\label{figFS}
\end{center}
\end{figure}

The calculated FS changes its topology twice with doping~\cite{KorOvch}, see Fig.~\ref{figFS}. Small hole pockets around $\left( { \pm {\pi  \mathord{\left/ {\vphantom {\pi  {2, \pm {\pi  \mathord{\left/ {\vphantom {\pi  2}} \right. \kern-\nulldelimiterspace} 2}}}} \right. \kern-\nulldelimiterspace} {2, \pm {\pi  \mathord{\left/ {\vphantom {\pi  2}} \right. \kern-\nulldelimiterspace} 2}}}} \right)$ points are present at small doping; then they increase in size and touch each other in the non-symmetric points $k =  \pm \pi \left( {1, \pm 0.4} \right)$ at ${x_{c1}} = {p_{opt}} = 0.151$. Above $p_{opt}$, there are two FSs around $\left( {\pi ,\pi } \right)$ with outer being a hole-like and inner being an electron-like. The electron FS collapsed at ${x_{c2}} = {p^ * } = 0.246$, and at $x > {p^ * }$ we have only one large hole surface around $\left( {\pi ,\pi } \right)$. Similar conclusion on the coexistence of hole and electron FS at some intermediate doping have been also drawn recently~\cite{HozLaadFulde,ChakKee}, and earlier for the spin-density wave state of the Hubbard model~\cite{SachChub}.

\begin{figure}[t]
\begin{center}
 \includegraphics*[width=0.8\textwidth]{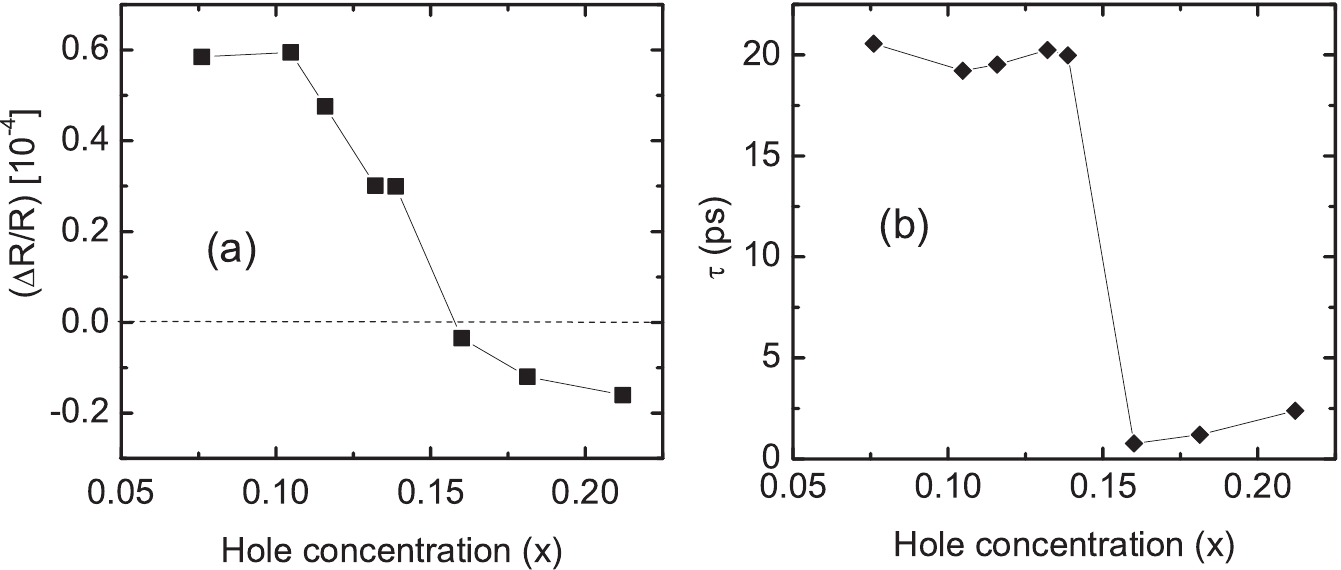}
 \caption{(a) Fractional change in reflectivity and (b) initial lifetime as a function of hole concentration \cite{GedLanOr}.
 The collapse of the quasiparticle lifetime and sign change of $\Delta R$ both occur at optimal doping.}
\label{figrefl}
\end{center}
\end{figure}

Presence of anomalous electronic system properties near ${x_{c1}} = 0.15$ is in agreement with the experimental data on changes in reflectivity and lifetimes. In the work ~\cite{GedLanOr} it was found that change of reflectivity from positive at $x<0.15$ to negative ($x>0.15$) takes place near optimal concentration in the Bi$_2$Sr$_2$Ca$_{1-y}$Dy$_y$Cu$_2$O$_{8+\delta}$, and lifetime abrupt fall down to near zero in the vicinity of $p_{opt}$ (Fig.~\ref{figrefl}). This behavior tells us about significant differences in electronic structure of underdoped and overdoped cuprates.

According to the general Lifshitz analysis~\cite{LifshAzbKag} for the three dimensional (3D) system, a change of topology at the energy $\varepsilon  = {\varepsilon _c}$ either by appearance of a new segment (like we found at ${p^ * }$) or by change of its connectivity (like at ${p_{opt}}$) would result in the additional DOS, $\delta N\left( \varepsilon  \right) \sim {\left( {\varepsilon  - {\varepsilon _c}} \right)^{{1 \mathord{\left/ {\vphantom {1 2}} \right. \kern-\nulldelimiterspace} 2}}}$, and the change in the thermodynamic potential, $\delta \Omega  \sim {\left( {{\varepsilon _F} - {\varepsilon _c}} \right)^{{5 \mathord{\left/  {\vphantom {5 2}} \right. \kern-\nulldelimiterspace} 2}}}$ (the QPT of the 2.5-order), where ${{\varepsilon _F}}$ is the Fermi energy. However, due to the strong anisotropy of electronic and magnetic properties, cuprates are quasi-2D and not isotropic 3D systems. The electron hopping perpendicular to the CuO$_2$ layers in a single-layer cuprates like LSCO is negligibly small. Below we discuss Lifshitz transitions in two dimensional system.

\begin{figure}
\begin{center}
 $\begin{array}{cc}
 (a)\includegraphics[angle=0,width=0.46\columnwidth]{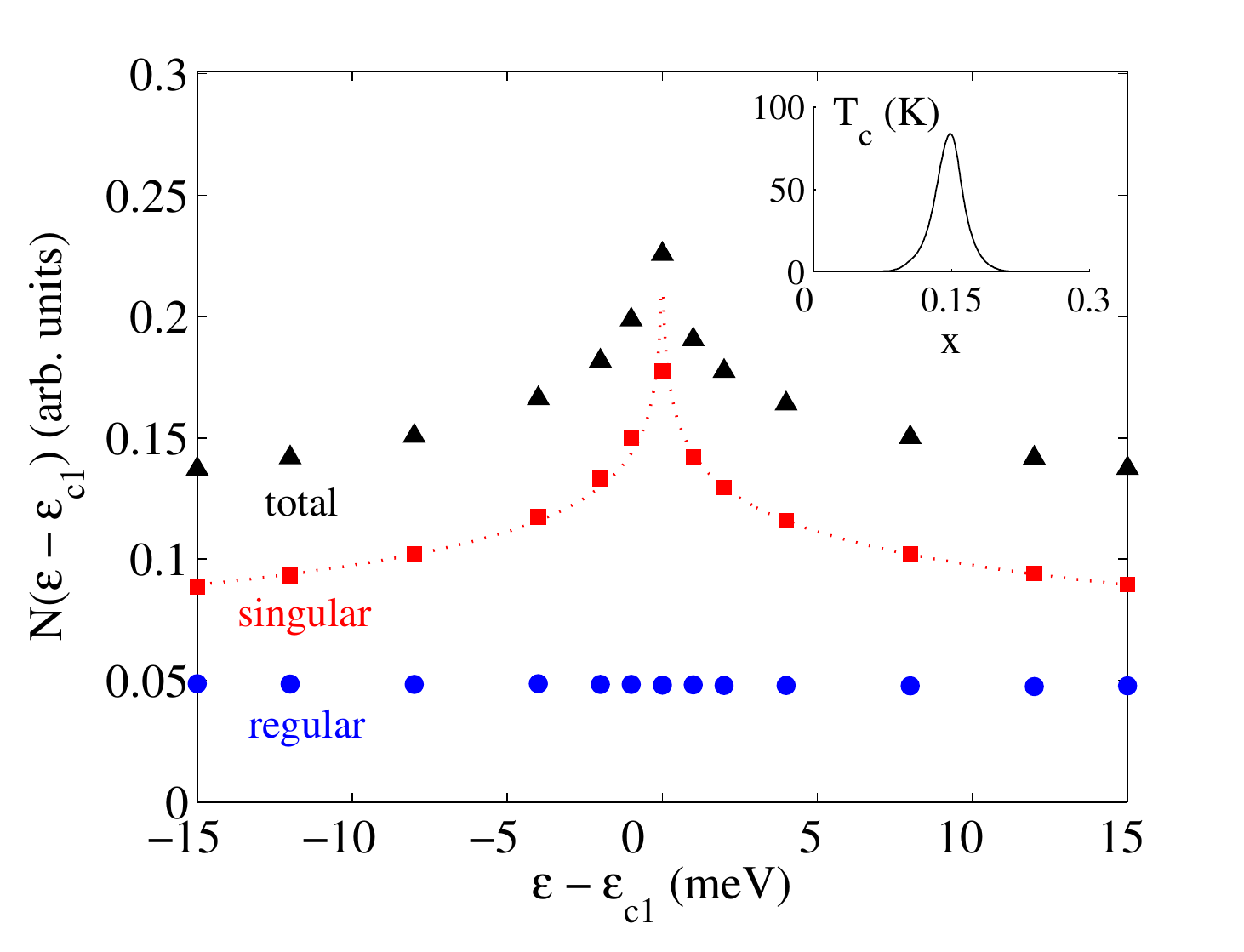}&
 (b)\includegraphics[angle=0,width=0.46\columnwidth]{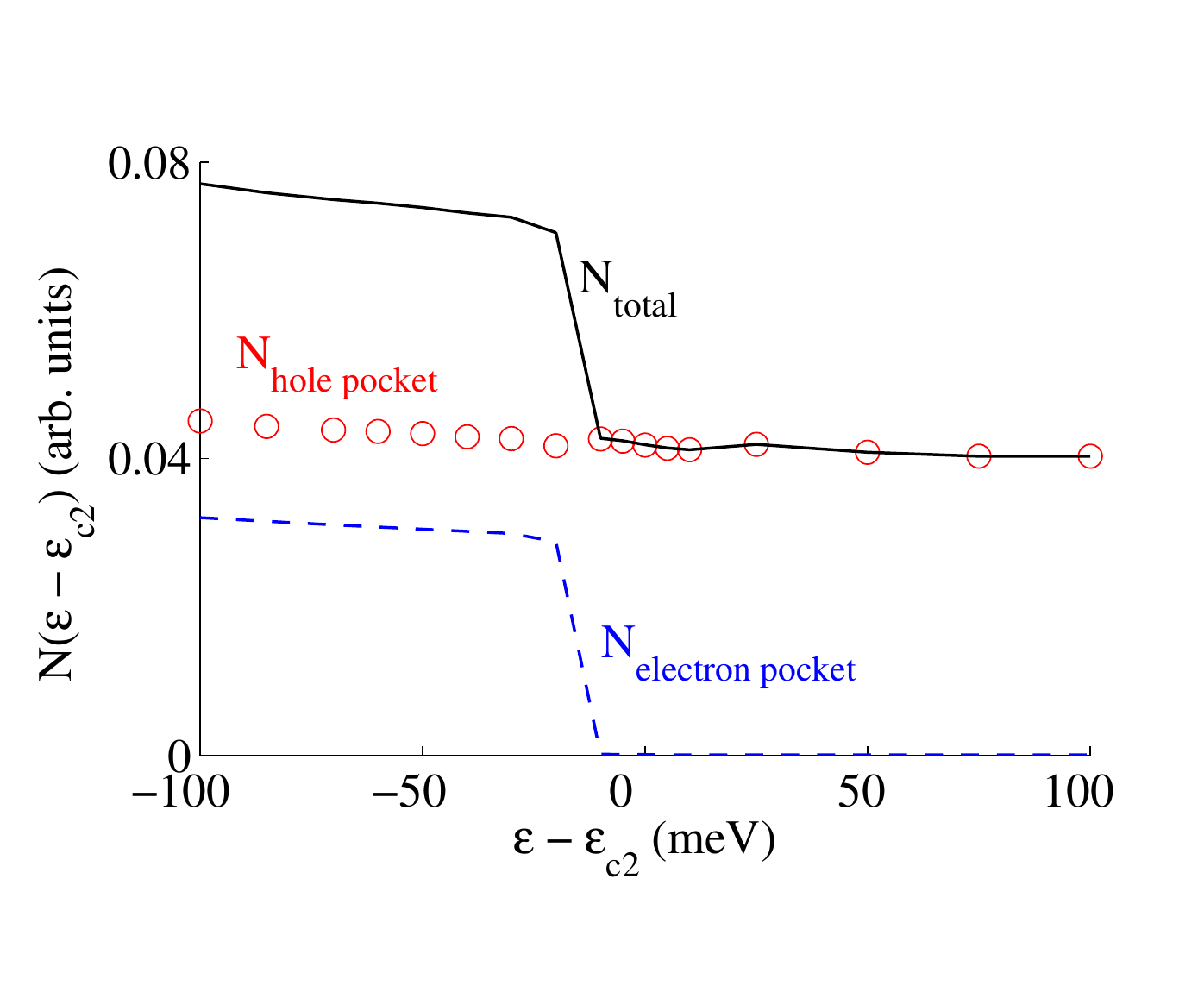}\\
 (c)\includegraphics[angle=0,width=0.46\columnwidth]{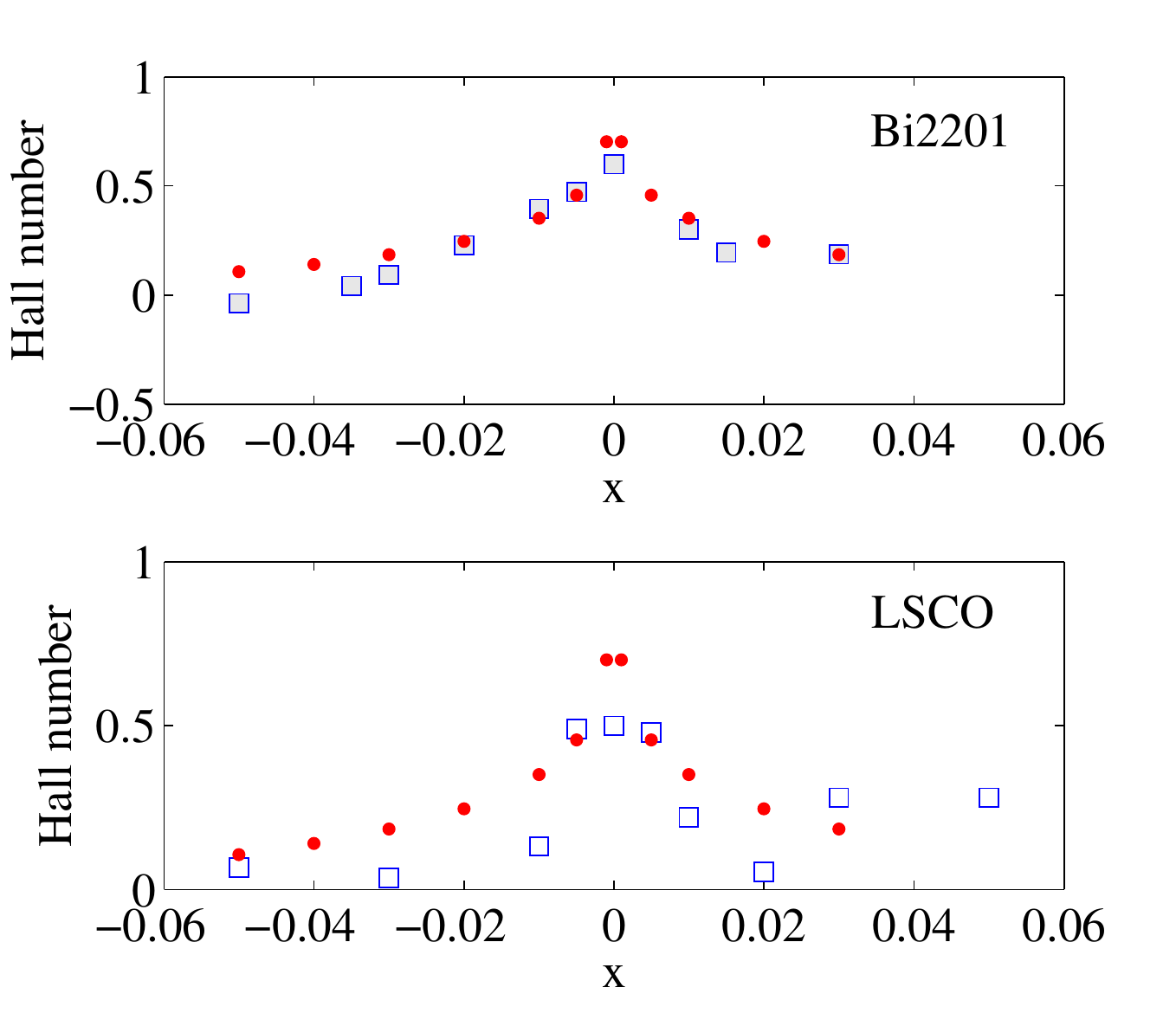}&
 (d)\includegraphics[angle=0,width=0.46\columnwidth]{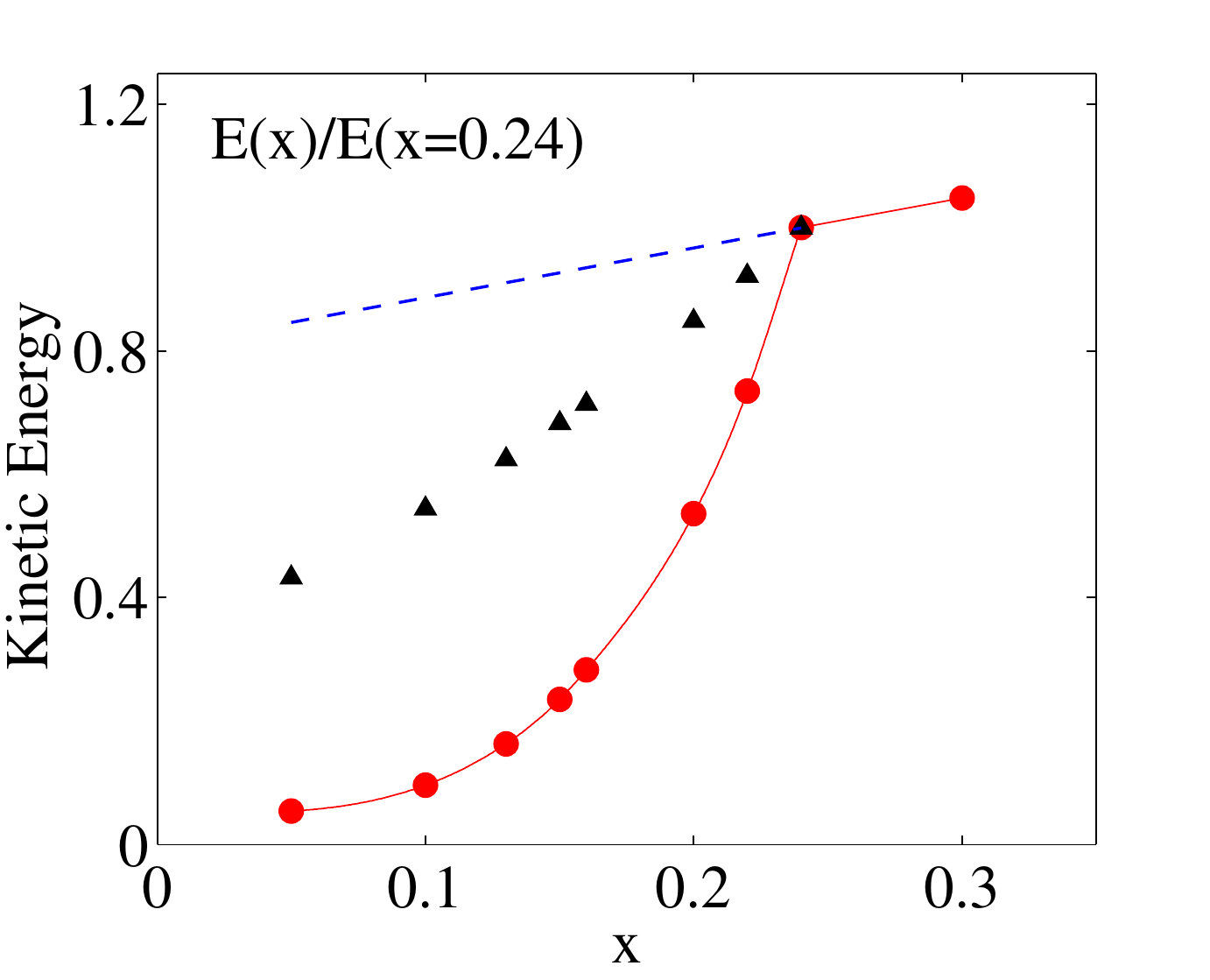}
 \end{array}$
 \caption{Regular, singular, and total density of states $N\left(\varepsilon-\varepsilon_{ci}\right)$ near the optimal doping $\varepsilon_{c1}=\varepsilon_F(p_{opt})$ (a) and near the pseudogap critical point $\varepsilon_{c2} = \varepsilon_F(p^*)$ (b), as calculated from the Green function (\ref{eq14})~\cite{OvchKorshShn,OvchShneyKorsh}. Dotted curve shows the logarithmic fitting. Below $p^* = 0.24$ ($\varepsilon < \varepsilon_{c2}$) a singular step-like contribution to the total DOS appears due to the appearance of the electron pocket. In the inset, the doping dependence of the superconducting critical temperature $T_c(x)$ is shown; the optimal doping is $0.151$. Note that the energy $\varepsilon - \varepsilon_{c1}$ is the energy of holes. (c) Comparison of experimental (blue squares) singular Hall coefficient ~\cite{BalBetMig} for Bi$_2$Sr$_{0.51}$La$_{0.49}$CuO$_{6+\delta}$(top) and for La$_{2-x}$Sr$_x$CuO$_4$(bottom) and our calculated (red filled circles) singular DOS, $N_{sing}(\epsilon_F(x))$, near the optimal doping. Agreement with results on the bulk single crystals(Bi2201) is better than with results on thin films(LSCO). (d) The doping dependence of the dimensionless kinetic energy $E_{kin}(p)/E_{kin}(p^*)$. The calculated dependence is shown by the filled (red) circles. Above ${{p^ * }}$ it obeys a conventional law and is proportional to $1 + p$. The extrapolation of this law to the region $p < p^*$ (blue dashed line) emphasizes the depletion of part of the kinetic energy in the pseudogap region. The calculation for the idealized triangular pseudogap model (\ref{eq16}) is shown by the filled triangles.}
\label{figdos}
\end{center}
\end{figure}

The change of the FS topology at ${x_{c1}} = {p_{opt}}$ results in the logarithmic divergence of DOS, while the emergence of the new electron-like pocket below ${x_{c2}} = {p^ * }$ results in a step in DOS (Fig.~\ref{figdos}). The total DOS is a sum of the singular and regular contributions. We would like to stress that both logarithmic and step DOS singularities are in perfect agreement with the general properties of the van Hove singularities for the 2D electrons~\cite{Ziman}. To continue discussion in terms of the critical points, not the critical energies, we note that near the critical point ${\varepsilon _F}\left( x \right) - {\varepsilon _{c1}} = k\left( {x - {x_{opt}}} \right)$. Contrary to the 3D systems, the thermodynamical potential for the 2D has a singular contribution $\delta \Omega  \sim {\left( {{\varepsilon _F} - {\varepsilon _c}} \right)^2}$ for the step singularity and $\delta \Omega  \sim {\left( {{\varepsilon _F} - {\varepsilon _c}} \right)^2}\ln \left| {{\varepsilon _F} - {\varepsilon _c}} \right|$ for the logarithmic singularity~\cite{Nedorezov}. Thus QPT at ${x_{c2}} = {p^ * }$ is of the second order, while at ${x_{c1}} = {p_{opt}}$ the singularity is stronger. It is immediately follows that the Sommerfeld parameter $\gamma $ in the electronic heat capacity $\gamma  = {{{C_e}} \mathord{\left/  {\vphantom {{{C_e}} T}} \right. \kern-\nulldelimiterspace} T}$ has also a singular step contribution at ${x_{c2}} \le {p^ * }$, and $\delta \gamma  \propto \ln \left| {{\varepsilon _F} - {\varepsilon _c}} \right| \propto \ln \left| {x - {x_{opt}}} \right|$ near ${x_{c1}} = {p_{opt}}$. Similar divergence in the specific heat was found within the dynamical cluster approximation for the Hubbard model~\cite{MikKhatGal}. The coincidence of ${x_{c1}}$ with ${p_{opt}}$ and ${x_{c2}}$ with ${p^ * }$ is not occasional. A singular contribution to the Hall coefficient near the optimal doping has been measured for Bi$_2$Sr$_{0.51}$La$_{0.49}$CuO$_{6+\delta}$ single crystals and for La$_{2-x}$Sr$_x$CuO$_4$ thin films under a strong magnetic field of 60 T ~\cite{BalBetMig}. According to our theory, these extra carriers are induced by the singular DOS. In Fig.~\ref{figdos} we plot the calculated singular DOS ${N_{sing}}\left( z \right)$, $z = x - {x_{opt}}$, together with the singular contribution to the Hall coefficient, ${n_{Hall}}\left( {1.5\,K} \right) - {n_{Hall}}\left( {100\,K} \right)$~\cite{BalBetMig}. The optimal doping in the LSCO thin film $x_{opt} = 0.17$ is shifted from the bulk value $x_{opt} = 0.15$ in the Bi2201 which may be due to the strains in the films. The general agreement of the calculated singular DOS and Hall data provides further support for our analysis.
We compare the superconducting critical temperature dependence ${T_c}\left( x \right)$ in the same model~\cite{ShnOvch} as a function of doping and observe that $T_c(x)$ has a maximum at ${{x_{opt}}}$ (see inset in Fig.~\ref{figdos}a), which indeed equals to ${x_{c1}}$. It is not a coincidence since in the BCS-like theory the maximum in ${T_c}\left( x \right)$ is determined by the maximum DOS, and at ${x_{c1}}$ we have a logarithmic singularity. Kinetic energy, ${E_{kin}} = \sum\limits_n {{t_{0n}}{K_{0n}}} $, reveals a remarkable kink at ${x_{c2}} = {p^ * }$ (Fig.~\ref{figdos}d)  due to the change in DOS. Above ${p^ * }$, ${{{E_{kin}}\left( p \right)} \mathord{\left/ {\vphantom {{{E_{kin}}\left( p \right)} {{E_{kin}}\left( {{p^ * }} \right)}}} \right. \kern-\nulldelimiterspace} {{E_{kin}}\left( {{p^ * }} \right)}} \sim 1 + p$ that is expected for a conventional 2D metal with the hole concentration ${n_h} = 1 + p$ and ${E_{kin}} \sim {\varepsilon _F} \sim {n_h}$. The extrapolation of this law below ${p^ * }$ reveals that actual ${E_{kin}}$ is much smaller. We associate this depletion with the pseudogap formation and fit it with the Loram-Cooper model~\cite{CoopLor,LorLuoCoop} that describes a simple free electron gas with a triangular pseudogap DOS,
\begin{equation}
\label{eq16}
N\left( \varepsilon  \right) = \left\{ \begin{array}{l}
g,\,\,\,\left| {\varepsilon  - {\varepsilon _F}} \right| > {E_g}\\
g\frac{{\left| {\varepsilon  - {\varepsilon _F}} \right|}}{{{E_g}}},\,\,\,\left| {\varepsilon  - {\varepsilon _F}} \right| < {E_g}
\end{array} \right..
\end{equation}
Here ${E_g} = {{J\left( {{p^ * } - p} \right)} \mathord{\left/ {\vphantom {{J\left( {{p^ * } - p} \right)} {{p^ * }}}} \right. \kern-\nulldelimiterspace} {{p^ * }}}$ is a doping dependent pseudogap and $J$ is the nearest-neighbor exchange parameter. Thus observed depletion of the kinetic energy together with the jump in DOS relates the QPT at ${x_{c2}}$ to the pseudogap.

The energy dependence of the electron self-energy is crucial and determines the Mott-Hubbard transition in the Hubbard model as was convincingly demonstrated by the dynamical mean-field theory (DMFT)~\cite{GeoKotKra}. Cluster generalization of DMFT~\cite{HetTahzad,KotSavPal,Potthoff,MaiJarPru} is necessary to study electron correlations in a two-dimensional CuO$_2$ layer where the nearest neighbor spin correlations require the momentum dependent self-energy. The cellular DMFT (CDMFT) method provides $\bf{k}$-dependent self-energy and results in the phase diagrams that have features similar to the ones experimentally observed in cuprates~\cite{SenLavMar,KanKyu,HauKot,MacJarMai,KyuSenTrem}. Recently, the exact diagonalization version of CDMFT (CDMFT+ED) was used to study the electronic structure of the doped Mott-Hubbard insulator~\cite{SakMotIma,SakMotIma2}. The sequence of the FS transformations with doping in Refs.~\cite{SakMotIma,SakMotIma2} is very similar to ours.

Another fruitful approach to study the two-dimensional Hubbard model is the composite operator method~\cite{Avella2007,Avella2014}. It results in the FS evolution similar to what is seen in Fig.~\ref{figFS} and the quasiparticle dispersion qualitatively similar to Ref.~\cite{KorOvch}.

\begin{figure}
\begin{center}
 \includegraphics[angle=0,width=0.7\columnwidth]{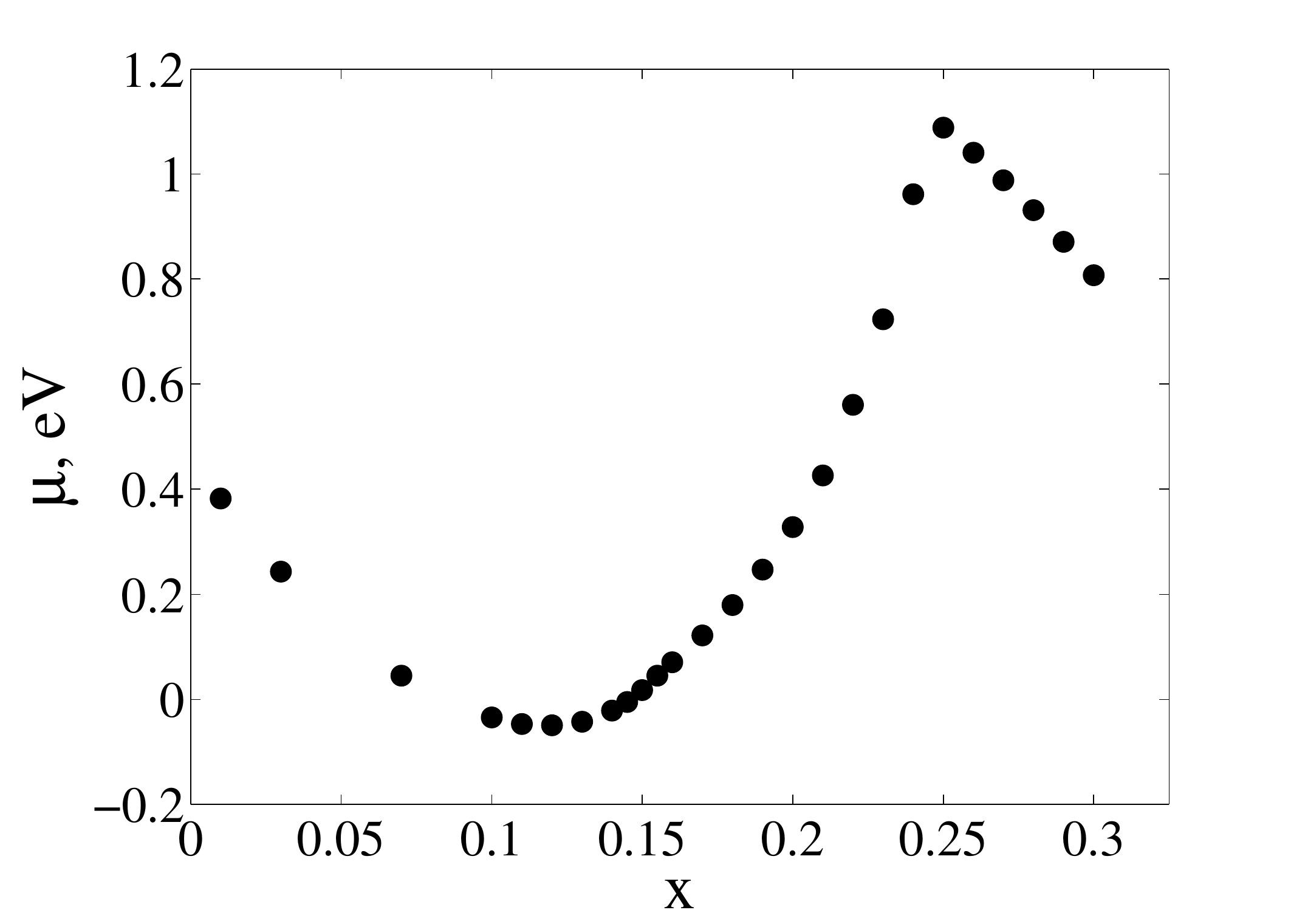}
 \caption{Dependence of the chemical potential on the doping level $x$.}
\label{Chem_pot_bilayer}
\end{center}
\end{figure}

It is seen from the DOS that the chemical potential changes nonlinearly upon the increase of doping. Fig.~\ref{Chem_pot_bilayer} shows this dependence in a greater detail. The chemical potential first decreases and attains its minimal value near $x = 0.13$, and then increases with the doping. This increase continues to a level of $x =0.25$, after which a decrease in the chemical potential is observed again.
This implies the possibility of phase separation in the crystal, that is often discussed in HTSC cuprates. We want to emphasize that in our calculation the negative electronic compressibility and tendency to the phase separation occurs in the concentration region between critical points.

\subsubsection{Electronic Structure of Bilayer Cuprates}
\label{ES:3.4}
It is well known that all compounds belonging to the class of HTSC cuprates have one or several CuO$_2$ layers in their structure. It is generally accepted at present that superconductivity and the electronic structure in the vicinity of the Fermi level are determined by the copper-oxygen planes, while the remaining atoms supply charges to these planes. However the physical properties of YBa$_2$Cu$_3$O$_7$ and other cuprates with two identical CuO$_2$ layers are different. The main  proof of this statement is dependence of T$_c$ on the number of the CuO$_2$ layers in the unit cell. The superconducting transition temperature increases with the number n of CuO$_2$ layers up to $n = 3$, and then decreases ~\cite{ChenLin}. The decrease in T$_c$ for $n > 3$ is traditionally explained by the deficit of carriers in the plane. However, the increase in T$_c$ has not yet been adequately explained. In addition, some other effects such as nonuniform distribution of doped carriers between the planes ~\cite{KarpYama1,KarpYama2,MoriTohMae,TrokLeNoc,StatSong,MagKit,JulCarHor,ZhKitAsa,TokIshKit,KotTok} and the coexistence of the antiferromagnetic and superconducting phases in the same unit cell, but in different planes ~\cite{MoriMae}, take place only in multilayer HTSC materials. Therefore, it would be interesting to study the effect of coupling between CuO$_2$ layers in the form of interlayer hoppings on the electronic structure of cuprates.
The simplest effect of coupling between the layers has been known for a long time and is associated with the formation of bonding and antibonding bands, which leads to splitting of the Fermi surface manifested most strongly in the vicinity of antinodal points $\left( {\pi ,0} \right)$. In the present Chapter, the main focus is on the influence of interlayer hoppings on the evolution of the Fermi surface upon an increase in doping level. We will demonstrate the splitting of Lifshitz transitions due to interlayer hoppings.

The unit cell of YBa$_2$Cu$_3$O$_7$ consists of two CuO$_5$ pyramids formed by a plane CuO$_4$ cluster and an apical oxygen atom, CuO chains, Y atoms between CuO$_5$ planes, and Ba atoms. In our model unit cell, we retain only a bilayer block of two CuO$_5$ pyramids. Unit cell of a bilayer cuprate is symmetric relative to the reflection in the plane passing through the yttrium atom parallel to the plane of the CuO$_2$ layer. The wavefunctions for the upper and lower CuO$_5$ pyramids are identical; therefore, we can consider an individual pyramid as elementary cluster. This statement could have been absolutely valid in the absence of coupling between CuO$_2$ planes; the hoppings between copper oxide layers brings to the formation of common wavefunctions for the two layers. Electronic structure of individual CuO$_2$ layer is defined by ${d_{{x^2} - {y^2}}}$, $p_x$, and $p_y$-orbitals which are characterized by a high density of probability only in a plane. In contrast, the density of probability of these orbitals in the direction perpendicular to the layers is very low. Therefore overlap of ${d_{{x^2} - {y^2}}}$, $p_x$, and $p_y$-orbitals from neighbor CuO$_2$ layers is small. The overlap of $p_z$-oxygen orbitals in which the electron density is mainly concentrated along the $z$ axis is much stronger ~\cite{OKALJ}. The strongest interlayer overlap is observed for $s$-orbitals of copper and $s$-orbitals with $p$-orbitals of oxygen atoms in the neighboring plane.
The influence of $s-s$ overlapping on the electronic structure is realized via effective overlapping of $d_x$-orbitals which is made possible by interlayer $s-p$ overlapping and intralayer $p-d$ overlapping. This result was demonstrated in ~\cite{OKALJ} using the one-band model. The symmetry of overlapping of the $s$ and $d$-orbitals is taken into account in the form of the hopping integral between the layers, which has the following form for YBa$_2$Cu$_3$O$_7$~\cite{OKALJ}:

\begin{equation}
\label{eq17}
{t_ \bot }\left( {\bf k} \right) = {t_ \bot }{\left( {\cos \left( {{k_x}a} \right) - \cos \left( {{k_y}b} \right)} \right)^2},
\end{equation}
In further analysis, when writing the expressions for various points in the ${\bf k}$-space, we will assume that the equality $a = b = 1$ holds for lattice constants.

The characteristic values of the intracell hopping integral $t_\bot$ that can be extracted from LDA calculations ~\cite{OKALJ} (${t_ \bot } = 0.06$ eV) or from ARPES data ~\cite{FengArmLu,ChuGromFed} (${t_ \bot } \approx0.029$ eV) are two orders of magnitude lower than the maximum integral $t_{pd}$ of intralayer hopping between $d_x$-orbitals of copper and $p$ orbitals of oxygen. Therefore, the interlayer hoppings can be treated as a small perturbation. This allows us to describe the coupling between CuO$_2$ layers in the framework of the cluster perturbation theory.

The Hamiltonian of the $t - t' - t'' - {J^ * }$-model for a bilayer has the form
\begin{equation}
\label{eq18}
H_{t - t' - t'' - {t_ \bot } - {J^*}} = \sum\limits_\alpha  {H_{t - t' - t'' - {J^*}}^\alpha }  + \sum\limits_{f\sigma } {{t_ \bot }\left( {{\rm X}_{fu}^{S\sigma }{\rm X}_{fd}^{\sigma S} + {\rm X}_{fd}^{S\sigma }{\rm X}_{fu}^{\sigma S}} \right)},
\end{equation}
in which subscript $\alpha $ indicates the upper (u) or lower (d) CuO$_2$ plane.
To determine the quasiparticle excitation spectrum, we use the same GTB approach as was given above for single CuO$_2$ layer and treat the interlayer hopping as the perturbation. Thus we have found electronic Green's function
\begin{equation}
\label{eq19}
\left\langle {\left\langle {{{\rm X}_{\mathbf{k}}^{\sigma S}}}
 \mathrel{\left | {\vphantom {{{\rm X}_{\mathbf{k}}^{\sigma S}} {{\rm X}_{\mathbf{k}}^{S\sigma }}}}
 \right. \kern-\nulldelimiterspace}
 {{{\rm X}_{\mathbf{k}}^{S\sigma }}} \right\rangle } \right\rangle  = \frac{{\left( {{p_\sigma } + x} \right)}}{2}\left( {\frac{1}{{E - {E_ + }\left( {\mathbf{k}} \right)}} + \frac{1}{{E - {E_ - }\left( {\mathbf{k}} \right)}}} \right)
 \end{equation}
where ${p_\sigma }=\left\langle {{X^{\sigma\sigma}}}\right\rangle $ is the filling factor and poles ${E_ + }\left( {\mathbf{k}} \right)$ and ${E_ - }\left( {\mathbf{k}} \right)$ determine the dispersion dependences of antibonding and bonding bands formed
due to the interlayer coupling. The values of the poles are defined as
\begin{equation}
\label{eq20}
{E_ \pm }\left( {\bf{k}} \right) = {\varepsilon _2} - {\varepsilon _1} - \mu  + \left( {{p_\sigma } + x} \right){t_{\bf{k}}} + {p_{\bar \sigma }}{J_0} + {p_{\bar \sigma }}\left( {{p_\sigma } + x} \right)\frac{{{{\tilde t}_{\bf{k}}}^2}}{{{E_{ct}}}} + {\Sigma _ \bot }\left( {\bf{k}} \right) \pm {\tilde t_{ \bot {\bf{k}}}}
\end{equation}
where ${\Sigma _ \bot}\left( \bf{k} \right)$ is the self-energy that is obtained from Eq.~(\ref{eq15}) by addition of terms with interlayer kinetic correlator ${K_ \bot }$ under sum over $\bf{q}$, and the expression

\begin{equation}
\label{eq21}
{\tilde t_{ \bot {\bf{k}}}} = \left( {{p_\sigma } + x} \right){t_ \bot }{\left( {\cos {k_x} - \cos {k_y}} \right)^2}\left( {1 + {3{C_ \bot }}/2{{\left( {{p_\sigma } + x} \right)}^2}} \right)
\end{equation}
characterizes the energy of band splitting due to interlayer coupling. A band splitting equal to $\tilde t_{ \bot {\bf{k}}}$, which is known as the bilayer splitting, was confirmed by many experiments ~\cite{FengArmLu,ChuGromFed,SchPark,ChGromFed1,KordBorGol,AsAvRoc,KamRosFre1,GrFedChu,BorKordZab,KamRosFre2,KoKhKa,KoKhSa,InBorEr,OkIshUch}. It is seen that bilayer splitting largely depends on the wavevector. This splitting is zero in the nodal direction and attains its maximal value in the antinodal direction.

In the expression (\ref{eq20}), ${C_ \bot }$ are interlayer spin correlation functions.
They were found by the exact diagonalization of a bilayer cluster Cu$_2$O$_{10}$. It was found that ${C_ \bot } \approx  - 0.1$. The ground state of the unit cell Cu$_2$O$_{10}$ with one hole per layer is a singlet, spins of holes in the neighbor layers are antiparallel. This is a reason for the negative sign of ${C_ \bot }$. Antiferromagnetic ordering of spins in neighbor layers was confirmed experimentally in ~\cite{TranCoxKunn} with the help of neutron diffraction.

\begin{figure}
\begin{center}
\includegraphics[angle=0,width=0.9\columnwidth]{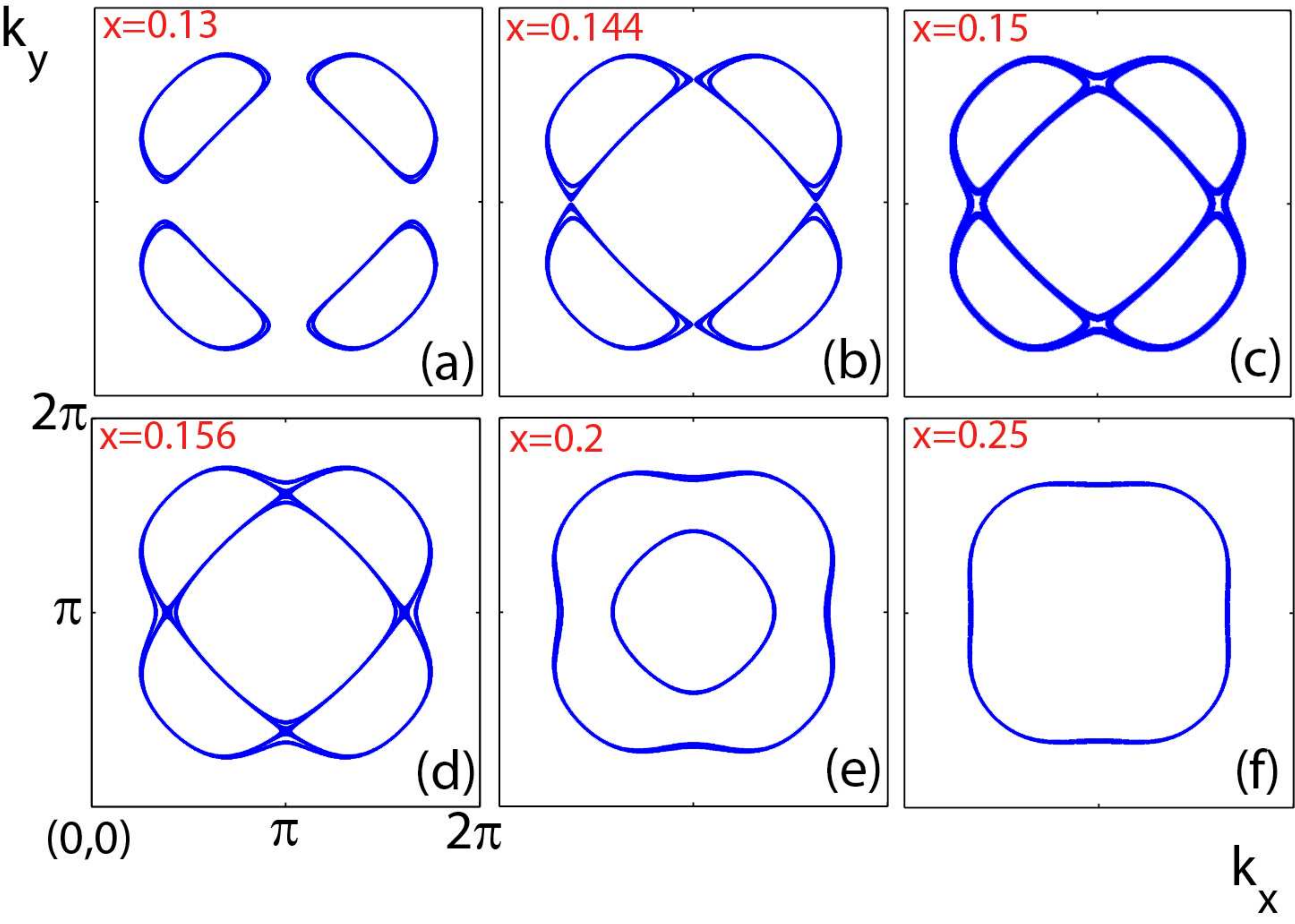}
 \caption{Bilayer cuprate Fermi surface evolution with doping.}
\label{FS_disp_bilayer}
\end{center}
\end{figure}

Note that we assume equal number of doping carriers ($x$) in each of the two layers. In the range of weakly doped compounds, the Fermi surface consists of four elements each of which is formed by two hole pockets inserted into each other (see Fig.~\ref{FS_disp_bilayer}a). The first quantum phase transition occurs for ${x_{c1}} = 0.144$ (Fig.~\ref{FS_disp_bilayer}b). At this point, wider pockets merge into one, which gives rise to an outer hole pocket and the inner electron pocket inserted in it. Along with the newly formed pockets, four growing pockets also exists; they merge for ${x_{c2}} = 0.156$. As a result, we are left with two hole and two electron pockets, each of which exhibit interlayer splitting at antinodal points (Fig.~\ref{FS_disp_bilayer}e). Upon a further increase in the doping level, the inner electron pockets shrink and become less and less distinguishable from  each other (Fig.~\ref{FS_disp_bilayer}f). In the vicinity of ${x_{c3}} = 0.25$, the electron pockets disappear upon a subsequent quantum phase transition, and only two large hole pockets are left (Fig.~\ref{FS_disp_bilayer}g). Naturally, it should be two critical points, but the electron pockets are split so weakly that these critical points cannot be distinguished. Upon an increase in $x$, antiferromagnetic correlations become weaker, which is manifested in an increase in energy at point $\left( {\pi ,\pi } \right)$, indicating the return to the paramagnetic state. It should be noted that the pattern of topological variations of the Fermi surface of a double-layer cuprate is almost the same as for a single-layer cuprate. The main difference is associated with the splitting of the Lifshitz transition near the optimal doping level into two transitions at $x_{c1}$ and $x_{c2}$.

\begin{figure}
\begin{center}
 \includegraphics[angle=0,width=0.7\columnwidth]{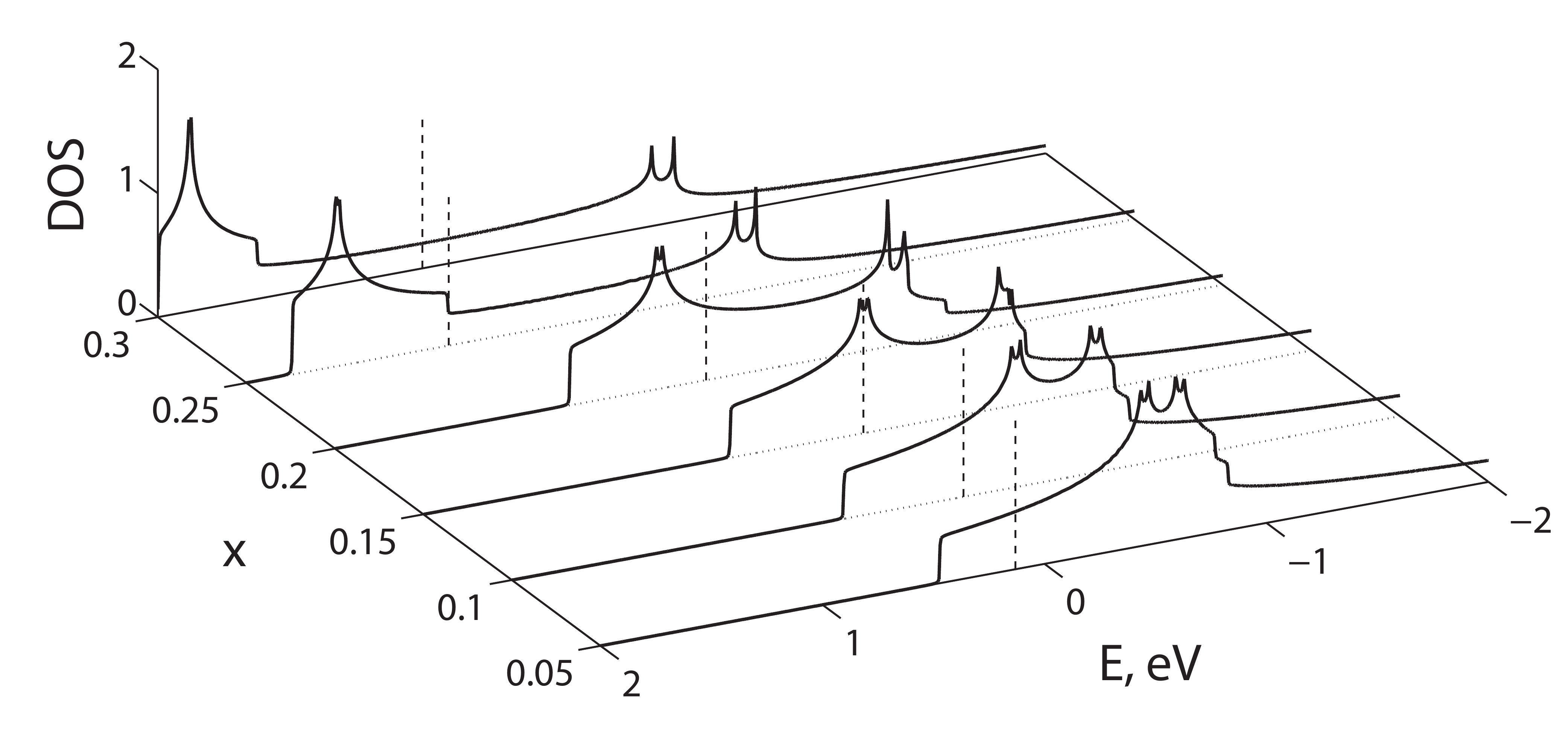}
 \caption{Density of states of bilayer cuprate at $t_ \bot=0.03$ eV. Dash lines denote chemical potential level.}
\label{DOS_bilayer}
\end{center}
\end{figure}

Fig.~\ref{DOS_bilayer} shows DOS for different hole concentrations. For low concentrations $x = 0.05$ and 0.1 in Fig.~\ref{DOS_bilayer}, as well as for $x = 0.13$ in Fig.~\ref{FS_disp_bilayer}a, the interlayer splitting corresponds to the filled part of the band deep beneath the Fermi level. At point $x_{c1} = 0.144$, carriers at the boundary of the Brillouin zone fill the neck between the pockets, which appears as the formation of a bridge between the neighboring domain on the Fermi surface. First, a bridge is formed at the antibonding band from the two bands split as a result of tunneling. This antibonding band forms a sheet of the open Fermi surface. Soon (at $x_{c2} = 0.156$), the bridge is formed for the bonding band. Both sharp increments of the hole density correspond to two van Hove singularities in Fig.~\ref{DOS_bilayer}.

DOS obtained in our analysis for $x = 0.25$ can be compared with the results of ~\cite{BarMaksZhu}, where the hole spectrum of two CuO$_2$ planes was investigated using the two-band model for a magnetic polaron. Bilayer splitting of the DOS observed in ~\cite{BarMaksZhu} for $x = 0.24$ due to the presence of bonding and antibonding bands is qualitatively identical to that calculated in our analysis.

\section{Magnetic Mechanism of Pairing}

One of the most striking problem in cuprates is that almost every single experimental fact has several equally good explanations in completely different theories. Therefore, a theory that runs for an election to be an ``ultimate theory'' have to explain number of observations. Here we concentrate on one particular class of theories that involves spin-fluctuation or, more generally, magnetic mechanism of pairing. Even so, there are too many different approaches and methods to review all of them here and we apologise before those whose theories we missed to mention.

Here we first present general ideas and the history of the spin-fluctuation mediated pairing in the context of the Fermi-liquid, then briefly discuss applications of such theories to cuprates, and after that we concentrate on the limit of strong electronic correlations.

\subsection{Spin Fluctuation Pairing}
\label{subsec:sf}

When a new class of superconductors is discovered it is common to discuss electronic pairing mechanisms as soon as there is some evidence that the electron-phonon mechanism is not strong enough to produce observed critical temperatures; this was the case in the cuprates, sodium cobaltates, and Fe-based superconductors. Berk-Schrieffer \cite{BerkSchrieffer} type spin fluctuation theories are quite popular because they are relatively simple and give some qualitatively correct results in the well-known cases of $^3$He and the cuprates. It is important that this type of description cannot be regarded as the complete answer even in superfluid $^3$He, where the true pairing interaction contains a significant density fluctuation component, while in the cuprates it is controversial whether the full pairing interaction can be described by a simple boson exchange theory at all. Nevertheless spin fluctuation theories can explain the symmetry of the order parameter in both systems quite well, in part because other interaction channels are projected out in the ground state.
Below we illustrate the basic equations leading to the canonical $d$-wave case within the one-band Berk-Schrieffer approach, history of which has been reviewed by Scalapino \cite{ScalapinoSFhistory}.


The original proposal of superconducting pairing arising from magnetic interactions was put forward by Emery \cite{Emery} and by Berk and Schrieffer \cite{BerkSchrieffer}, who were interested primarily in transition metal elements and nearly ferromagnetic metals. Such systems are considered to be close to a ferromagnetic ordering transition in the Stoner sense, i.e. their susceptibility may be approximated by $\chi=\chi_0/(1-U\chi_0)$, where $U$ is a local Hubbard-like Coulomb matrix element assumed to be large since $U\chi_0\simeq 1$ ($\chi_0$ is the susceptibility in the absence of interactions). Physically this means a spin up electron traveling through the medium polarizes the spins around it ferromagnetically lowering the system's energy. The spin triplet pairing interaction for such a correlated electron gas is therefore  attractive, while the singlet interaction  turns out to be repulsive, which can overcompensate the attractive electron-phonon interaction in this system \cite{BerkSchrieffer}.

The excitations being ``exchanged'' in such a picture are not well-defined collective modes of the system such as phonons or magnons,
but rather ``paramagnons'', defined by the existence of a peak-like structure in the the imaginary part of the small $\q$ susceptibility.
Effective pairing vertex in the equal- and opposite-spin channels, $\Gamma_{\uparrow \uparrow}$ and $\Gamma_{\uparrow \downarrow}$, is
\begin{eqnarray}
{\Gamma _{ \uparrow  \uparrow }} &=& \frac{{ - {U^2}{\chi _0}({{\bf{k}}^\prime } - {\bf{k}})}}{{1 - {U^2}{\chi _0}^2({{\bf{k}}^\prime } - {\bf{k}})}},\label{eq:Gamma1}\\
\Gamma_{\uparrow \downarrow} &=&
U^2\left({3\over 2}\chi^s -{1\over 2}\chi^c\right)+U, \label{eq:Gamma3}
\end{eqnarray}
where we have defined $\chi^s\equiv \chi_0/(1-U\chi_0)$ and $\chi^c=\chi_0/(1+U\chi_0)$, and in the last step we have changed $-\k$ to $\k$ in the second term of $\Gamma_{\uparrow\downarrow}$ because we assume we work in the even parity (singlet pairing) channel.
The total pairing vertex in the triplet (singlet) channel is $\Gamma_t={1\over 2}\Gamma_{\uparrow\uparrow}$ ($\Gamma_s={1\over 2} (2 \Gamma_{\uparrow\downarrow} -\Gamma_{\uparrow\uparrow} )$). In the original paramagnon theory, $\chi_0(\q)$ is the  noninteracting
susceptibility of the (continuum) Fermi gas, i.e. the Lindhard function. This function at small frequency  has a maximum at $\q=0$,
meaning correlations are indeed ferromagnetic.  Thus due to the negative sign in the equation for $\Gamma_{\uparrow \uparrow}$ (note $\chi_0>0$ and $U\chi_0<1$ to prevent a magnetic instability), pairing is attractive in the triplet channel and singlet superconductivity is suppressed.

\subsubsection{Antiferromagnetic Spin Fluctuations}

In the context of heavy fermion systems it was realized \cite{ScalapinoHF,VarmaHF} that strong AFM spin fluctuations in either the weak or strong coupling limit lead naturally to the spin singlet, $d$-wave pairing. The weak coupling argument has been reviewed by Scalapino \cite{ScalapinoPhysRep}. Suppose the susceptibility is strongly peaked near some wave vector $\Q$. The form of the singlet interaction is
%
\begin{eqnarray}
\Gamma_s(\k,\k') = \frac{3}{2} U^2 \frac{\chi_0(\q)}{1-U\chi_0(\q)} \label{eq:singlet_approx}
\end{eqnarray}
if we neglect terms which are small near the random phase approximation (RPA) instability $U\chi_0(\q) \rightarrow 1$ \cite{Nakajima}. This now implies that $\Gamma_s(\q)$ is also peaked at this wavevector, but is always repulsive. Nevertheless, if one examines the BCS gap equation for this interaction
\begin{equation}
\Delta_\k= -{\sum_{\k^\prime}}^\prime \Gamma_s(\k,\k^\prime) {\Delta_\k^\prime \over 2E_\k^\prime } {\rm tanh} {E_\k^\prime \over 2T},
\label{eq:gapeqn}
\end{equation}
one sees immediately that an isotropic state cannot be a solution, but that
if the state changes sign,
\begin{equation}
\Delta_\k= - \Delta_{\k+\Q},
\label{eq:changesign}
\end{equation}
a solution will be allowed.

In the cuprates, $\chi$ is peaked at $\Q\simeq (\pi,\pi)$, and the two possible states of this type which involve pairing on nearest neighbor bonds only are
\begin{equation}
\Delta_\k^{d,s} = \Delta_0 (\cos k_xa \mp \cos k_y a ).
\end{equation}
Which state will be stabilized then depends on the Fermi surface in question.
So we need to use the fact that the states close to the Fermi surface are the most important in Eq.~(\ref{eq:gapeqn}), and examine the pairing kernel for these momenta.  For example, for a $(\pi/a,0)\rightarrow (0,\pi/a)$ scattering, $\Delta_\k^s$ satisfies
Eq.~(\ref{eq:changesign}) by being zero, whereas $\Delta_\k^d$ is nonzero and changes sign, contributing to the condensation energy. It should therefore not be surprising that the end result of a complete numerical evaluation of Eq.~(\ref{eq:gapeqn}) over a cuprate Fermi surface gives $d$-wave pairing.


\subsubsection{Applications of the Spin-Fluctuation Pairing to the Cuprates}

\begin{figure}
\begin{center}
\includegraphics[angle=0,width=0.99\columnwidth]{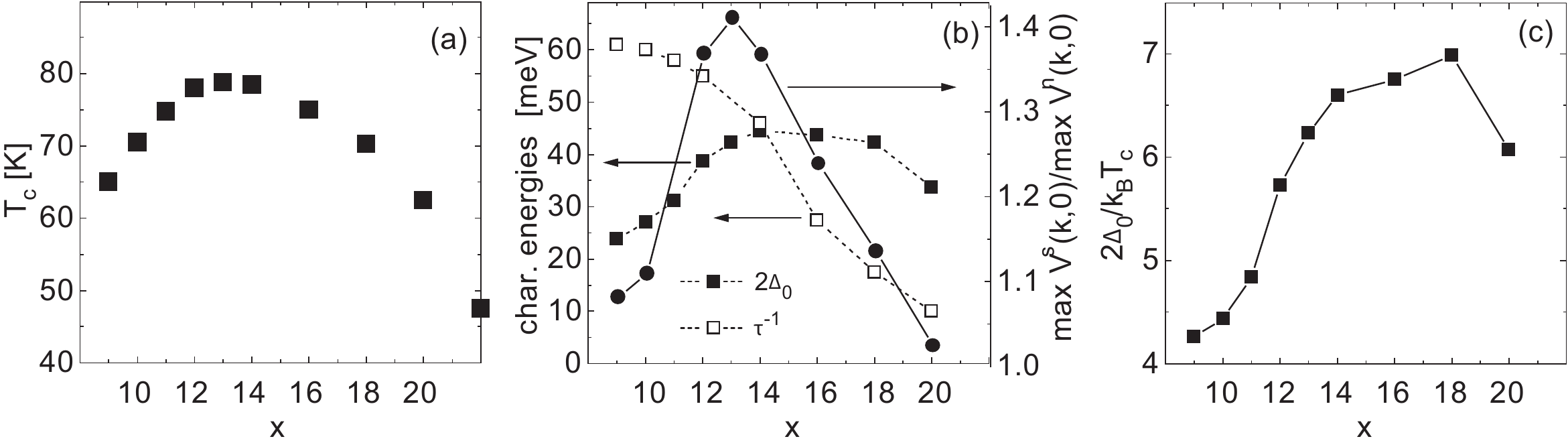}
\caption{Exemplary doping dependence of the superconducting state in FLEX \cite{Grabowski}: (a) $T_c$ obtained from $\Delta(T=T_{c})=0$, (b) ratio of the effective interactions in the superconducting and normal states $max(V_s ({\bf k},0))/max(V_n ({\bf k},0))$, $2\Delta_{0}$, and the inverse quasi particle lifetime $\tau^{-1}$, (c) ratio $2 \Delta_{0}/ k_B T_c$. Doping is given in \%.}
\label{flexTc}
\end{center}
\end{figure}
Eq.~(\ref{eq:singlet_approx}) for the pairing vertex in the singlet channel $\Gamma_s(\q)$ can be solved together with equations for the renormalization of the electronic band structure due to the scattering on the spin fluctuations. If this is done self-consistently and spin fluctuations are treated in the RPA, it is called fluctuation-exchange approximation (FLEX).
In particular, single-band FLEX \cite{flex,flex1} employs sum of all
particle-hole(particle) ladder graphs for the generating functional self-consistently valid for intermediate strength of the
correlations. The FLEX equations for the single-particle Green function $G$, the self-energy $\Sigma$, the effective interaction
$V$, the bare ($\chi^0$) and renormalized spin ($\chi^s$) and charge ($\chi^c$) susceptibilities read
\begin{eqnarray}
G_{\bf k}(\omega_n) &=& \left[ \omega_n - \varepsilon_{\bf k} + \mu - \Sigma_{\bf k}(\omega_n) \right]^{-1}, \\
\Sigma_{\bf k}(\omega_n) &=& \frac{T}{N} \sum\limits_{{\bf p}, m} V_{\bf k-p}(\omega_n - \omega_m) G_{\bf p}(\omega_m), \\
V_{\bf q}(\nu_m) &=& U^2 \left[ \frac{3}{2} \chi^s_{\bf q}(\nu_m) +
\frac{1}{2} \chi^c_{\bf q}(\nu_m)
- \chi^0_{\bf q}(\nu_m) \right], \\
\chi^0_{\bf q}(\nu_m) &=& - \frac{T}{N} \sum\limits_{{\bf k}, n} G_{\bf k+q}(\omega_n + \nu_m) G_{\bf k}(\omega_n), \\
\chi^{s,c}_{\bf q}(\nu_m) &=& \frac{\chi^0_{\bf q}(\nu_m)}{1 \mp U \chi^0_{\bf q}(\nu_m)},
\end{eqnarray}
where $\omega_n = i \pi T (2n+1)$, $\nu_m = i \pi T (2m)$, and $\varepsilon_{\bf k}$ is the bare band dispersion.
In the last equation the '$-$' sign in the denominator corresponds to the $\chi^{s}_{\bf q}(\nu_m)$, while the '$+$' sign corresponds
to the $\chi^{c}_{\bf q}(\nu_m)$.
Then solution of the FLEX equations performed numerically for a particular lattice partition and number of $\omega$-points in the wide energy range.

FLEX was applied to the case of the one-band Hubbard model for cuprates \cite{Lenck,Monthoux,Dahm,Langer,Grabowski,Altmann,Manske}, as well as generalized for the multiband case, see, e.g., \cite{Esirgen,Ueda_etal}. The theory even can explain the shape of the superconducting dome, see Fig.~\ref{flexTc} and Ref.~\cite{Grabowski}. Other advantages include description of the low-energy kink feature seen in ARPES as having pure electronic origin and demonstration of the spin-resonance asymmetry for electron and hole doped cuprates \cite{Manske}. The main disadvantage, if one discuss the case of cuprates, is the lack of strong electronic correlations in the whole scheme of FLEX. As we have shown in the previous Section, the Mott-Hubbard physics significantly affects normal state properties. Considering this, while FLEX results looks sometimes very convenient, they have to be regarded as some (maybe crude) approximation to the theory of superconductivity in cuprates. With this in mind, we move further to the discussion of a more exotic pairing mechanism with the large total momentum and then to the magnetic mechanisms of Cooper pairing in the limit of strong electronic correlations.

\subsection{Pairing With the Large Total Momentum}

Natural extension of the spin fluctuation theory in the spirit of the Fermi-liquid approach is to add the screened Coulomb interaction. The history goes back to Yang \cite{Yang}, who have shown that states formed from the localized on one lattice site singlet pairs ($\eta$ pairing) are metastable eigenstates of the simple Hubbard model with the total pair momentum of $\K=0$ and $\K=(\pi,\pi)$. In the extended Hubbard model allowing for the hopping of the repulsive pair between neighboring sites \cite{Penson} the state with the superconducting condensate of $\eta_{\pi}$ pairs with the large momenta $\K=(\pi,\pi)$ can become ground state competing with the condensate of $\eta_0$ pairs ($\K=0$) \cite{Bulka,Japaridze}. Spatially inhomogeneous state appearing when there is a superconducting pairing with the finite total momentum is similar to the Fulde-Ferrell-Larkin-Ovchinnikov (FFLO) state in weakly ferromagnetic superconductors \cite{Fulde,Larkin}. For pairs with $\K \neq 0$, domains of the momentum space where the pairing of electrons or holes is allowed are significantly reduced compared to the $\K=0$ case. Thus, FFLO state corresponds to the condensate of pairs with the small total momentum. However, the screened Coulomb repulsion can result in the singlet superconducting pairing with the large total pair momentum $\K$ \cite{fflo7}.

If besides the dominant screened Coulomb repulsive pairing interaction \cite{fflo7} one considers a conventional phonon-mediated pairing \cite{fflo4,fflo5,fflo5t} and the antiferromagnetic (AFM) magnon-mediated pairing (spin fluctuations due to the short-range AFM order) \cite{fflo6}, then the competition between these mechanisms may lead to phase transitions inside the superconducting dome with the symmetry change from the $d$-wave at extremely low doping through the extended $s$-wave to the conventional $s$-wave symmetry in the overdoped regime \cite{Belyavsky2008}.

\begin{figure}
\begin{center}
 $\begin{array}{cc}
 (a)\includegraphics[angle=0,width=0.46\columnwidth]{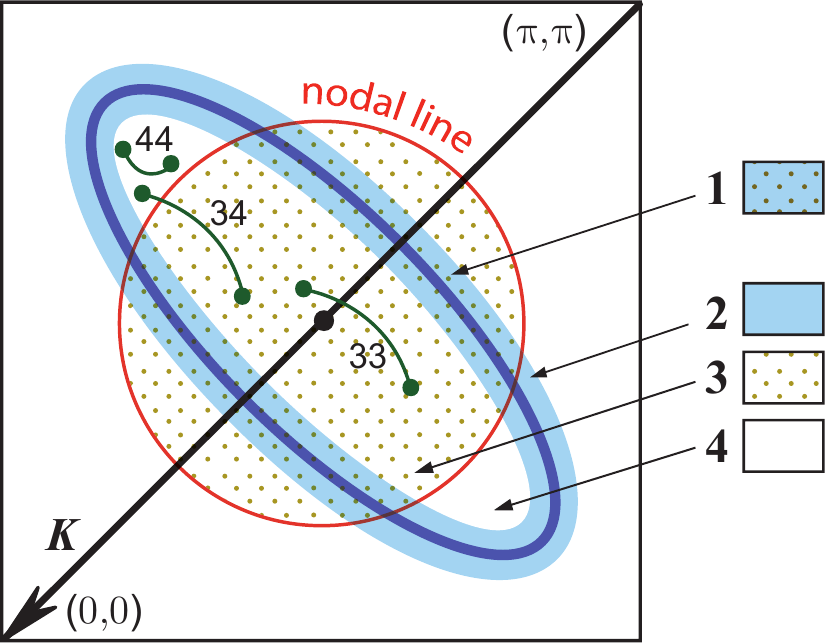}&
 (b)\includegraphics[angle=0,width=0.38\columnwidth]{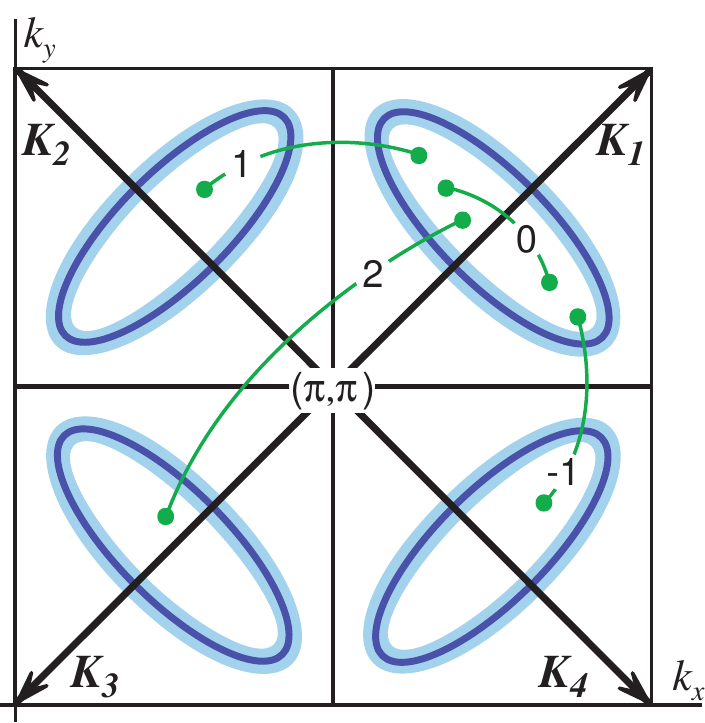}
 \end{array}$
 \caption{(a) Pocket-like Fermi contour (bold line) and nodal line inside a quarter of the Brillouin zone, schematically. The shaded elliptical layer enveloping the FC corresponds to the domain of definition of phonon-mediated pairing attraction. Subdomains 1 and 2 (3 and 4) correspond to different sign of the order parameter inside (outside) the elliptical layer. Transitions under scattering inside subdomains of equal signs (33 and 44) and between these subdomains (34) are shown schematically. (b) Domains of kinematic constraint corresponding to four crystal equivalent total pair momenta. Transitions under scattering inside a domain of kinematic constraint (0), between neighbor (1 and -1), and opposite domains (2) are shown schematically.}
\label{fig:domains}
\end{center}
\end{figure}

One can consider the large total pair momentum $\K$ as a remnant of the AFM wave vector of the parent compound ground state. In such a case, one-particle states composing a pair should belong to a domain of kinematic constraint ${\Xi}$ inside the Brillouin zone, Fig.~\ref{fig:domains}a. The logarithmic singularity of the self-consistency equation survives in the case of mirror nested Fermi contour (FC) typical for the cuprates \cite{fflo7} since the filled and vacant parts of this domain turn out to be separated by finite pieces of the FC forming a ``pair'' Fermi contour (PFC) on which kinetic energy of the pair with momentum $\K$ ($\K$-pair) vanishes.

Coulomb pairing results in the SC order parameter that changes sign inside $\Xi$. Two subdomains of ${\Xi}$ with the order parameter of constant sign should be separated by a nodal line crossing the PFC. Since phonon-mediated pairing exists only inside a narrow layer enveloping the FC, ${\Xi}$ can be divided into four different parts. Two of them correspond to taking into account both Coulomb and phonon-mediated pairing in this layer inside and outside the nodal line, whereas in the other two parts, exterior to the layer (inside and outside the nodal line as well), phonon-mediated pairing can be neglected. The nodal line disposition is determined by the total pairing interaction. We suppose that Coulomb pairing inside ${\Xi}$ dominates all of the rest pairing interactions. Therefore, we consider a fixed nodal line as it were due to Coulomb pairing only. All above considered subdomains of the domain of kinematic constraint are presented in Fig.~\ref{fig:domains}a which corresponds to the case of the FC in the form of small hole pockets \cite{fflo6}. In this case, PFC coincides with the FC and the domain of kinematic constraint $\Xi$ is a quarter of the Brillouin zone, but it should be noted that a half of the pocket belonging to the first magnetic Brillouin zone of the parent compound forms the \textit{main band} with high spectral weight whereas the complementary half of the pocket in the second magnetic zone forms the so-called \textit{shadow band} with considerably lower spectral weight \cite{fflo8}. Therefore, the extension of the pocket with doping leads to a progressive decrease of the spectral weight of the shadow band that results in an effective restriction of the part of the domain of kinematic constraint contributing into the logarithmic singularity of the self-consistency equation.

There are four crystallographically equivalent large momenta $\K_j$ ($j = 1, 2, 3, 4$ enumerating quarters of the Brillouin zone) corresponding to their own domains of kinematic constraint $\Xi_j$, see Fig.~\ref{fig:domains}b. Superconducting order parameter is degenerate with respect to the symmetry transformations of the copper plane. Thus its components corresponding to equivalent momenta can have different signs. Here we consider electron-phonon attraction and AFM interaction to be weak compared to the dominant screened Coulomb repulsion. If the electron-phonon attraction dominates over the AFM magnon-mediated pairing one can expect the $s$-wave or the extended $s$-wave order parameter symmetry. That may correspond to the overdoped cuprates. In the opposite case corresponding to the underdoped materials the order parameter is of $d$-wave type with four nodal lines along the Brillouin zone diagonals as follows from the spin-fermion pairing \cite{fflo9}. We note, however, that while $s$-wave symmetry order parameter does not change sign upon rotation on $\pi/2$ in the momentum space, it has nodal lines inside each domain of the kinematic constraint connected to the Coulomb pairing \cite{fflo7}.

\begin{figure}
\begin{center}
 $\begin{array}{cc}
 (a) \includegraphics[angle=0,width=0.41\columnwidth]{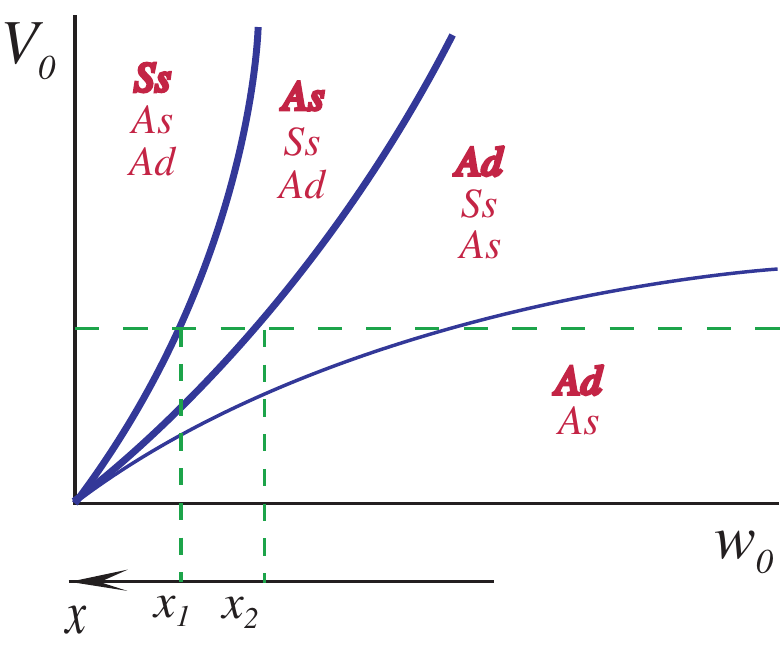}&
 (b) \includegraphics[angle=0,width=0.47\columnwidth,height=1.8in]{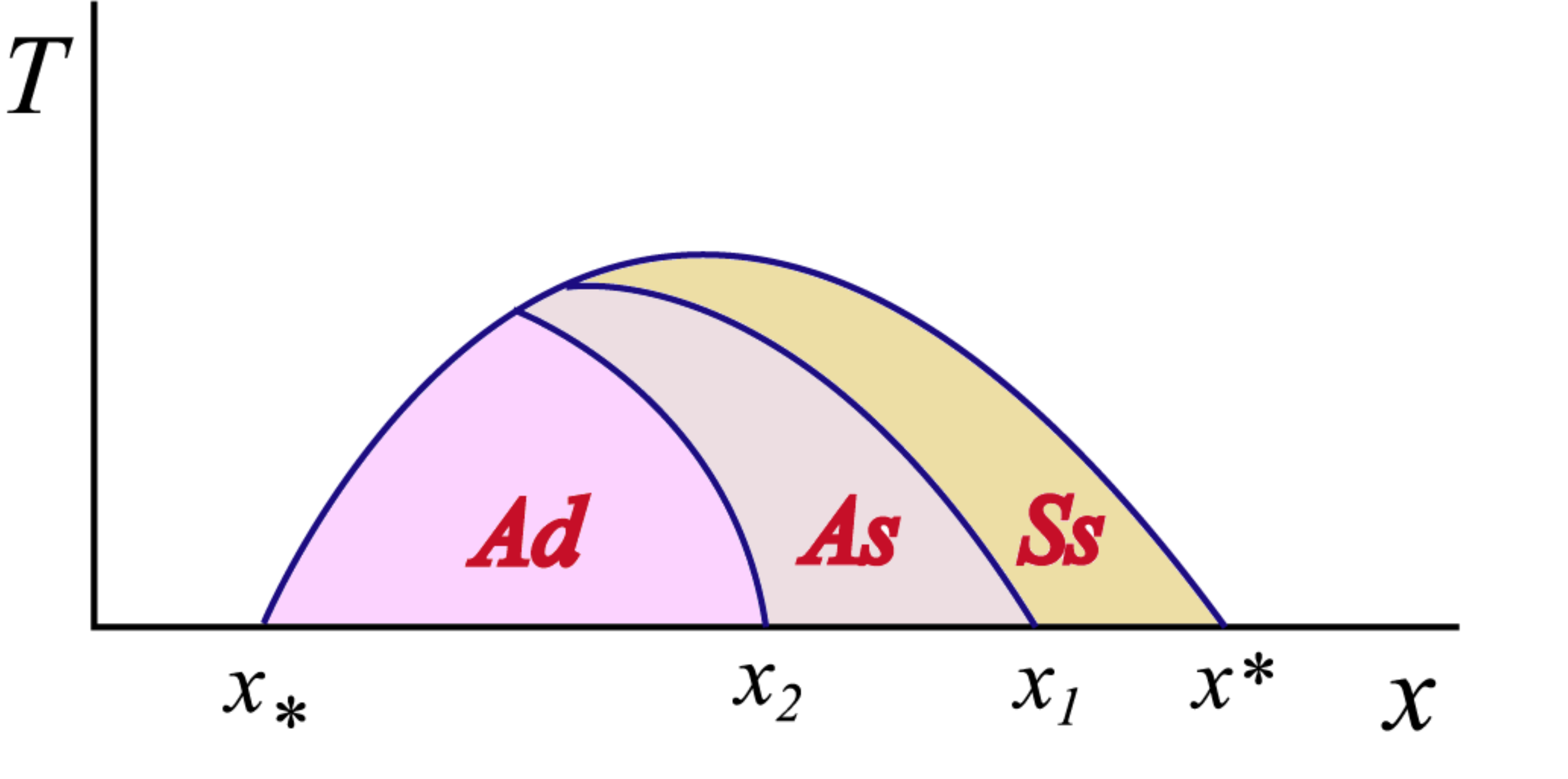}
 \end{array}$
 \caption{(a) Dependence of the solutions of the self-consistency equation on the the repulsive contribution to the pairing interaction $w^{}_0$ and phonon-mediated attraction $V_0$. Symbols of the dominating symmetry are in bold face. Relation between $w^{}_0$ and doping $x$ is shown schematically in the bottom of the figure. (b) Possible distribution of the order parameter symmetry inside the SC dome. Curves separating domains with different symmetries correspond to phase transitions between different SC phases.}
\label{fig:V0Tdoping}
\end{center}
\end{figure}

Order parameter $\Delta_j(\bm{k})$ defined inside ${\Xi}_j$ is determined by self-consistency equation system
\begin{equation}\label{SCE}
{\Delta}^{}_j({\bm{k}})=-{\frac{1}{2}}\sum_{i=1}^4 \sum_{{\bm{k}}_{}^{\prime}\in{\Xi}^{}_i} {\frac{U^{}_{ji}({\bm{k}}-{\bm{k}}_{}^{\prime})\; {\Delta}^{}_i({\bm{k}}_{}^{\prime})} {{\sqrt{{\xi}_{}^2({\bm{k}}_{}^{\prime})+ {\Delta}^{2}_i({\bm{k}}_{}^{\prime})}}}},
\end{equation}
where ${\bm{k}}$ is the relative motion momentum defined in ${\Xi}^{}_j$, $U^{}_{ji}({\bm{k}}-{\bm{k}}_{}^{\prime})$ is a
matrix element of the pairing interaction (subscripts $j$ and $i$ show that ${\bm{k}}\in {\Xi}^{}_j$ and ${\bm{k}}_{}^{\prime} \in
{\Xi}^{}_i$),
\begin{equation}
2{\xi}({\bm{k}})={\varepsilon}({\bm{K}}/2+{\bm{k}}) + {\varepsilon}({\bm{K}}/2-{\bm{k}})
\end{equation}
is the kinetic energy of the ${\bm{K}}$~-~pair, ${\varepsilon}({\bm{k}})$ is electron dispersion with respect to the chemical potential. The outer sum in Eq.~(\ref{SCE}) includes contributions of all of the crystal equivalent domains ${\Xi}^{}_i$ and the inner one is taken over one of them.

To solve the self-consistency equation (\ref{SCE}) we use a step-wise approximation for the pairing interaction. We present the matrix element as a set of constants corresponding to scattering inside each domain ${\Xi}^{}_j$ and between the neighbor and opposite (along a diagonal of the Brillouin zone) domains. Thus, we have three constants describing a scattering due to the phonon-mediated pairing (defined in the vicinity of the Fermi contour) and also three constants related to a scattering due to the AFM magnon-mediated pairing. Screened Coulomb repulsion is analogously approximated by the constants corresponding to scattering between neighbor and opposite domains. Such an approximation is similar to the Tolmachov model \cite{fflo10}, where the electron-phonon pairing is considered along with the screened Coulomb repulsion. In addition, the scattering inside each of the domains should be described by two constants that correspond to scattering inside the subdomains of ${\Xi}^{}_j$ in which the order parameter is positive and negative and another one constant corresponding to the scattering between these subdomains \cite{fflo11}. All of the cases of the scattering discussed here are presented in Fig.~\ref{fig:domains}.

Crystal symmetry allows to reduce the self-consistency equation to two different systems of four algebraic equations for the average values of the order parameter inside and outside the domain of the electron-phonon attraction on both sides of nodal line \cite{Belyavsky2008}. Each system corresponds to its own type of the order parameter orbital symmetry: $s$-wave and $d$-wave. Inside the whole Brillouin zone there are four types of solutions with the different orbital symmetry depending on the relation between pairing constants. Inside each domain of kinematic constraint corresponding to one of the crystallographically equivalent total pair momentum $\K$ there are either antisymmetric (A) or symmetric (S) solution reflecting the symmetry of the pairing interaction. These solutions form $s$- or $d$-wave total symmetry in the whole Brillouin zone defining the order parameter symmetry with respect to the rotation in the momentum space.

The symmetric solution corresponding to ${\Delta}^{}_{1}={\Delta}^{}_{2}\equiv {\Delta}$ has the form
\begin{equation}\label{S}
{|{\Delta}|}^{}_l = 2{\varepsilon}^{}_D\cdot \exp{\left (-{\frac{1}{g\, V_{l}^{\ast}}}\right )}
\end{equation}
where index $l=0$ ($l=1$) corresponds to $s$-wave ($d$-wave) order parameter symmetry, ${\varepsilon}^{}_D$ is the Debye energy scale, $g$ is the averaged density of states at the PFC, and $V_{l}^{\ast}$ is the effective coupling constant that determined by all three pairing interactions and depends on the orbital symmetry of the order parameter.

Antisymmetric solution with the opposite signs of $\Delta_1$ and $\Delta_2$ is
\begin{eqnarray}
{|{\Delta}^{}_1|}^{}_l &=& |{\Delta}^{\prime}_{}| [1+{\sigma}a^{}_l], \label{A1}\\
{|{\Delta}^{}_2|}^{}_l &=& |{\Delta}^{\prime}_{}| [1-{\sigma}a^{}_l], \label{A2}
\end{eqnarray}
where
\begin{equation}\label{A0}
|{\Delta}_{}^{\prime}| = 2{\varepsilon}^{}_0\cdot \exp{\left (-{\frac{1}{g\, w^{}_0}}\right )},
\end{equation}
${\varepsilon}^{}_0$ is the energy scale of the kinematic constraint domain $\Xi$, $w^{}_0$ is the parameter describing qualitatively the repulsive (Coulomb and AFM magnon-mediated) contributions to the pairing interaction. Parameter ${\sigma}$ describes the asymmetry of the order parameter sign distribution (the case of $\sigma \ll 1$ was considered here). Parameter $a^{}_l$ depends on all coupling constants and energy scales ${\varepsilon}^{}_0$ and ${\varepsilon}^{}_D$ in a complicated way.

Note that the form of solutions (\ref{S}) and (\ref{A0}) is the same in the cases of both $s$- and $d$-orbital symmetry but effective coupling constant in the $s$-wave solution differs from the $d$-wave one.

To show the relation between basic pairing interactions one can introduce two parameters: $w^{}_0$ and $V^{}_0$. Latter describes the attractive phonon-mediated contribution. Fig.~\ref{fig:V0Tdoping}a, corresponding to relatively weak AFM magnon-mediated pairing, represents the domains of the phase plane ($w^{}_0$, $V^{}_0$) in which each of the solutions exists and the domains in which one of the solutions dominates the others. In this phase diagram, one can distinct three regions with different symmetry of the order parameter: a region of strong correlations with dominating antisymmetric $d$-wave ($Ad$) solution, a region of intermediate correlations with dominating antisymmetric $s$-wave ($As$) solution, and a region of weak correlations with symmetric $s$-wave ($Ss$) solution. Bold lines separating these regions correspond to phase transitions between phases with different symmetry of the SC state. It should be noted that the $Ad$ and $As$ solutions exist in the whole phase diagram, whereas the $Ss$ solution exists only above the thin line shown in Fig.~\ref{fig:V0Tdoping}a. Weak AFM magnon-mediated pairing considered here cannot result in the symmetric $d$-wave ($Sd$) solution.

Doping results in a considerable decrease of the AFM contribution into the pairing interaction but weakly affects the Coulomb and phonon contributions. Therefore, one can assume that $w^{}_0$ decreases with doping whereas $V^{}_0$ remains unaffected by doping. Then, one can expect that different SC states can exist in the SC dome in the doping~--~temperature ($x,T$) phase diagram as shown schematically in Fig.~\ref{fig:V0Tdoping}b. At extremely low doping, the AFM contribution is large enough
and the SC state corresponds to the $Ad$ symmetry (extended $d$-wave symmetry: four nodal lines along the diagonals of the Brillouin zone and generic nodal lines due to repulsive pairing in each domain of kinematic constraint). An increase of doping can lead to the phase transition from the $Ad$ into $As$ state (extended $s$-wave symmetry: only generic nodes exist). A further increase of the doping can result in more phase transition from the $As$ into the $Ss$ state (anisotropic $s$-wave symmetry without any nodal lines).

While the common believe is that the order parameter symmetry in hole-doped cuprates is always $d$-wave, there are discussions that the order parameter in the volume and surface layer of the same compound can manifest different symmetries: extended $s$- and $d$-waves, respectively \cite{fflo1,fflo2}. Doping dependence of the pairing interaction within the theory of the large total pair momentum can in principle give an explanation of the mentioned interpretation of the experimental data.



The problem with the present approach is similar to the one in FLEX: it is strictly applicable only for the case of weak/moderate correlations. In the case of strong correlations and band splitting into the upper and lower Hubbard subbands the theory have to include effect of spectral weight redistribution between these subbands. There is no theory for pairing with the large total momentum in strong correlation limit yet.

\subsection{Pairing in the Limit of Strong Electronic Correlations}

In this section we consider microscopic approaches to magnetic mechanisms of Cooper pairing in the limit of strong electronic correlations. Discussing the normal state properties we demonstrated that the effects of strong electronic correlations significantly change the band structure, Fermi surface topology, and the thermodynamics of the cuprates, especially the underdoped ones. Convenient way of describing such systems is to use Hubbard $X$-operators.
Since these operators describe electron as the superposition of quasiparticle excitations in the limited Hilbert space where double occupied states are forbidden our consideration thoroughly differs from the Fermi-liquid approach. The same no-double occupancy constraint is kept at all steps of the calculations also in the Monte-Carlo variational method for projection wavefunctions \cite{shn_1}, in the theories of resonating valence bonds \cite{shn_2} and of renormalized mean field constructed in the Gutzwiller approximation \cite{shn_3,shn_4}.
In the simplest way magnetic mechanism can be understood within the $t-J$-model. Then for hole-doped cuprates spin singlet pairing of the Hubbard fermions from upper Hubbard band is given by the anomalous average ${B_{{\bf{k}}\sigma }} = \left\langle {X_{ - {\bf{k}}}^{\sigma S}X_{\bf{k}}^{\bar \sigma S}} \right\rangle$. Applying the projection technique \cite{shn_5,shn_6} for the normal ${G_{{\bf{k}}\sigma }} = \left\langle {\left\langle {\left. {X_{\bf{k}}^{\bar \sigma S}} \right|X_{\bf{k}}^{S\bar \sigma }} \right\rangle } \right\rangle $ and superconducting $F_{{\bf{k}}\sigma }^ {\dag}  = \left\langle {\left\langle {\left. {X_{ - {\bf{k}}}^{S\sigma }} \right|X_{\bf{k}}^{S\bar \sigma }} \right\rangle } \right\rangle $ Green functions we get in generalized Hartree-Fock approximation the BCS-type equation for the order parameter which in the case of the $t-J^*$-model (\ref{eq12tJ}) reads
\begin{eqnarray}
\label{shn_eq1}
\Delta _{\bf{k}}^{tJ^*} &=& \frac{1}{N}{\varphi _{\bf{k}}}\sum\limits_{\bf{q}} {\left( {2{t_{\bf{q}}} + \frac{{1 - x}}{2}\left( {{J_{{\bf{k}} + {\bf{q}}}} + {J_{{\bf{k}} - {\bf{q}}}}} \right) + \frac{{2\left( {1 + x} \right){{\tilde t}_{\bf{k}}}{{\tilde t}_{\bf{q}}}}}{{{E_{ct}}}} - \frac{{1 - x}}{2}\cdot\frac{{{{\tilde t}_{\bf{q}}}^2}}{{{E_{ct}}}}} \right)} \nonumber \\
&\times& \frac{{{\Delta _{\bf{q}}}{\varphi _{\bf{q}}}}}{{2{E_{\bf{q}}}}}\tanh \frac{{{E_{\bf{q}}}}}{{2{k_B}T}}.
\end{eqnarray}
Here ${\Delta _{\bf{q}}} = {\Delta _0}{\varphi _{\bf{q}}}$, where ${\varphi _{\bf{q}}} = (\cos {q_x}a - \cos {q_y}a)$ is the angle-dependent part of the ${d_{{x^2} - {y^2}}}$-gap. The first term in brackets of (\ref{shn_eq1}) describes intraband kinematical interaction and formally it originates from projected character of Hubbard operators \cite{shn_6}. In the mean-field approximation kinematical interaction produces only $s$-wave pairing since its contribution to the gap is ${\bf k}$-independent \cite{shn_7}. Other terms in brackets of (\ref{shn_eq1}) are caused by virtual interband transitions. The second one corresponds to the Heisenberg exchange interaction renormalized by three-site correlated hoppings \cite{shn_9,shn_8} while the third and the fourth terms describe contributions from correlated hoppings on three different sites.To proceed we use the nearest-neighbor approximation for the exchange parameter. In the case of singlet pairing ${\Delta _{\bf{k}}} = {\Delta _{ - {\bf{k}}}}$ only symmetric contribution $\left( {{J_{{\bf{k + q}}}} + {J_{{\bf{k}} - {\bf{q}}}}} \right)$ enters the T$_c$-equation

\begin{equation}
\label{shn_eq2}
1 = \frac{{1 - x}}{2}J\frac{1}{N}\sum\limits_{\bf{p}} {\frac{{{{\left( {\cos p{}_x - \cos p{}_y} \right)}^2}}}{{{\xi _{\bf{p}}}}}} \tanh \frac{{{\xi _{\bf{p}}}}}{{2{k_B}T}},
\end{equation}
where $k_B$ is the Boltzmann constant, $T$ is temperature, and ${\xi _{\bf{q}}} = \sqrt {\varepsilon _{\bf{q}}^2 + {{\left| {{\Delta _{\bf{q}}}} \right|}^2}}$ is quasiparticle excitation spectrum. System of equations on T$_c$ (\ref{shn_eq2}), normal phase dispersion ${\varepsilon _{\bf{q}}}$ (see expression (\ref{eq15})), and chemical potential $\mu$ self-consistently defines dependence of critical temperature on doping, ${T_c}\left( x \right)$. For \textit{ab initio} derived parameters of LSCO, computations reproduce the general structure of the superconducting dome (inset in Fig.~\ref{figdos}a) with the optimal doping at $x = 0.15$ and disappearance of superconductivity in the underdoped region below $x = 0.05$.
Dynamics of discussed above pairing via superexchange interaction $J \sim {\tilde t}^2/E_{ct}$ involves virtual excitations $\tilde t$ above the charge transfer gap, which is set by ${E_{ct}}$ \cite{shn_10,shn_11}, so the pairing interaction is essentially instantaneous. To consider the spin-fluctuation exchange pairing we should go beyond the mean field theory (see, for example \cite{PlakOud2,shn_12,shn_13,shn_14}). In the spin-fluctuation exchange picture, the pairing is viewed as arising from the exchange of particle-hole spin fluctuations whose dynamics reflect the frequency spectrum seen in inelastic magnetic neutron scattering. This spectrum covers an energy range which is small compared with ${E_{ct}}$ or the bare bandwidth. In this case, the pairing interaction is retarded.
A microscopic approach to the description of the spin-fluctuation mechanism of pairing in HTSC implies self-consistent solution for the self-energy of the electrons and the magnetic susceptibility in the related model, i.e. Hubbard model, $t-J$ model or multiband $pd$-model. The equation for the superconducting phase is the Eliashberg equation \cite{shn_16} where gap is determined by a complex, frequency-dependent interaction due to spin-fluctuations. In the general case that is extremely difficult task. The problem can be simplified in the weak or strong coupling limit and/or by applying a phenomenological form of susceptibility (see, for example \cite{shn_17}) with parameters determined from the experiment. The success of the spin-fluctuation theory of high-temperature superconductivity is connected primarily with the fact that it explains \cite{shn_18,shn_19,shn_20,shn_21,shn_22}  the highly recognized ${d_{{x^2} - {y^2}}}$-symmetry of the order parameter in cuprates \cite{shn_23}. However there are arguments against the magnetically mediated pairing scenario in these compounds, the main one is a weakness of antiferromagnetic fluctuations in optimally doped compounds seen in inelastic magnetic neutron scattering experiments \cite{shn_24,shn_25}.
We believe that lack of consensus about microscopic theory of HTSC is strictly related with the fact that different strong interactions persist in cuprates and therefore can drive superconductivity. One of them is electron-phonon interaction (EPI) that is present in all substances and can be essential in layered cuprates due to specific features of their crystal structure. Accumulated experimental data give convincing evidence for strong EPI in cuprates \cite{shn_26}, but they do not indicate that this interaction plays a leading role in superconducting pairing.
For normal metals such unambiguous evidence was isotope effect on the superconducting transition temperature. In HTSC cuprates, for which the index of oxygen-isotope effect on T$_c$ strongly depends on the doping level \cite{shn_27,shn_28,shn_29}, the situation appears to be more complicated. The isotope exponent ${\alpha _O}\left( x \right)$ is minimal at optimal doping; moreover, its value is an order of magnitude smaller than the value predicted via the BCS theory \cite{shn_30}. Some authors treated this behavior as evidence of a smallness of the electron-phonon contribution to the pairing potential. At the same time, in weakly doped and overdoped samples an increase in the value of ${\alpha _O}\left( x \right)$ is observed; in some systems, the isotope index in the underdoped region considerably exceeds the BCS theory value (i.e., $\alpha \left( x \right) = 0.5$).
To analyze specific isotope effect dependence on doping the interaction of correlated electrons with phonons should be included in the previous scheme. In the strong electronic correlations regime the EPI is the interaction of phonon with wave number ${\bf q}$, frequency ${\omega _{{\bf{q}}\nu }}$, and mode $\nu$ and Hubbard fermions. Then the total Hamiltonian \cite{shn_31} is given by

\begin{equation}
\label{shn_eq3}
{H_{eff}} = {H_{t - J^*}} + {H_{el - ph - el}},                                                                                                                           \end{equation}
where effective electron-electron interaction  mediated by phonons is ${H_{el - ph - el}} = \sum\limits_{{\bf{kk'q}}} {\sum\limits_{\sigma \sigma '} {{V_{{\bf{kk'q}}}}X_{{\bf{k}} + {\bf{q}}}^{2\bar \sigma }X_{{\bf{k}}' - {\bf{q}}}^{2\bar \sigma '}X_{{\bf{k}}'}^{\bar \sigma '2}X_{\bf{k}}^{\bar \sigma 2}} }$ with the effective interaction neglecting retardation effects  ${V_{{\bf{kk'q}}}} =  - \sum\limits_v {{{{g_v}\left( {{\bf{k}}{\bf{,q}}} \right){g_v}\left( {{\bf{k}}'{\bf{,}} - {\bf{q}}} \right)} \mathord{\left/ {\vphantom {{{g_v}\left( {{\bf{k}}{\bf{,q}}} \right){g_v}\left( {{\bf{k}}'{\bf{,}} - {\bf{q}}} \right)} {{\omega_{{\bf{q}},v}}}}} \right. \kern-\nulldelimiterspace} {{\omega _{{\bf{q}},v}}}}}$. In the generalized mean field approximation the gap and T$_c$-equations look like equations (\ref{shn_eq1}) and (\ref{shn_eq2}) respectively but with renormalized coupling constant describing the sum of exchange and phonon pairing mechanisms \cite{shn_32}

\begin{equation}
\label{shn_eq4}
{\lambda _{tot}} = \frac{{1 - x}}{2}J + {\lambda _{EPI}}\theta \left( {\left| {{{\xi'}_{\bf{q}}}} \right| - {\omega _D}} \right).                                                                                         \end{equation}

The $\theta$-function as usually means that phonon pairing occurs in a narrow energy window of the $\omega _D$ width near the Fermi energy. In accordance with definition of isotope effect exponent ${\alpha _O} =  - \frac{{{\rm{d}}\ln \left( {{T_c}} \right)}}{{{\rm{d}}\ln \left( {{M_O}} \right)}} =  - \frac{{{M_O}}}{{{T_c}}}\frac{{dT}}{{d{M_O}}}$ we get the analytical expression for ${\alpha _O}\left( x \right)$ which reads

\begin{equation}
\label{shn_eq5}
{\alpha _O}\left( x \right) = \frac{{\frac{{{\omega _D}}}{N}\sum\limits_{\bf{q}} {{\lambda _{EPI}}\delta \left( {\left| {{\xi _{\bf{q}}}} \right| - {\omega _D}} \right)} \frac{{\varphi _{\bf{q}}^2}}{{{\xi _{\bf{q}}}}}\tanh \left( {\frac{{{\xi _{\bf{q}}}}}{{2{k_B}T}}} \right)}}{{\frac{1}{N}\sum\limits_{\bf{q}} {{\lambda _{tot}}\left( {\bf{q}} \right)} \frac{{\varphi _{\bf{q}}^2}}{{{T_c}}}{{\coth }^{ - 2}}\left( {\frac{{{\xi _{\bf{q}}}}}{{2{k_B}T}}} \right)}}.
\end{equation}

\begin{figure}
 \includegraphics*[width=0.95\textwidth]{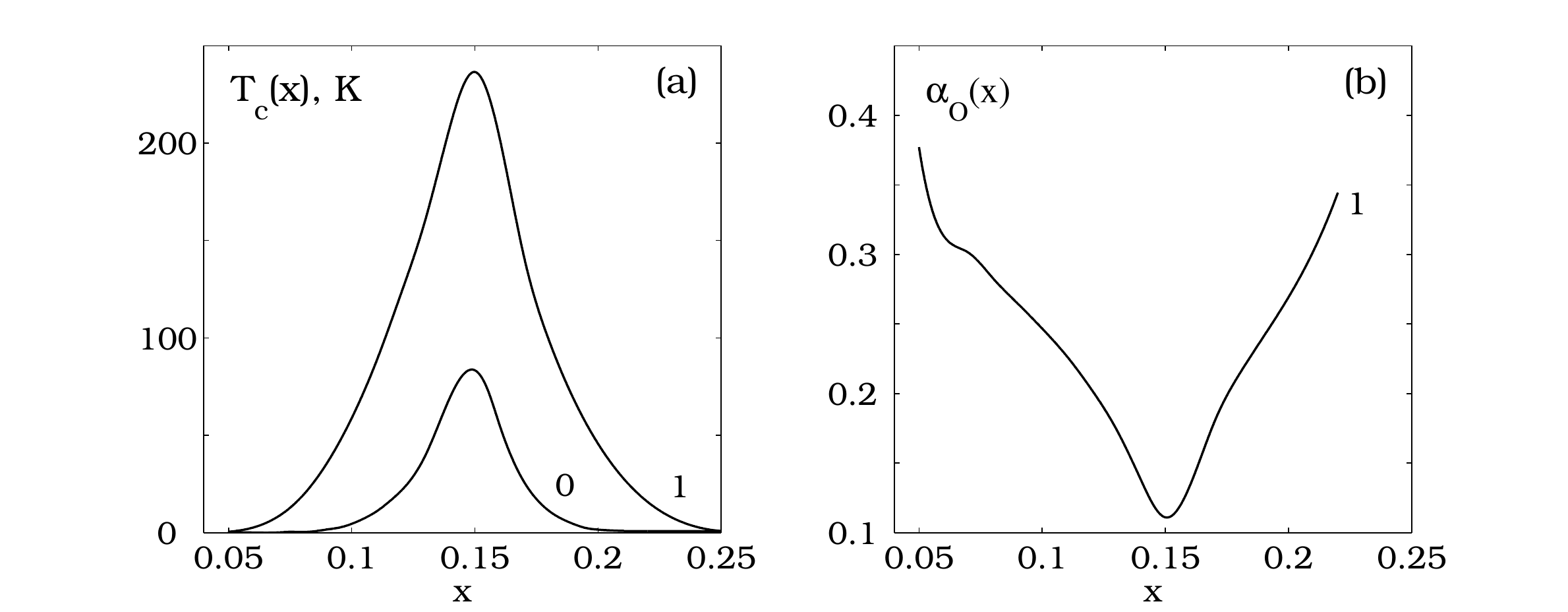}
 \caption{The doping dependence of the critical temperature (a) and the oxygen isotope exponent (b) for the effective electron-phonon coupling indicated near the curve.}
\label{Isotope}
\end{figure}

Eq.~(\ref{shn_eq5}) can be written in a simplified way, namely ${\alpha _O} = {{{\eta _{ph}}} \mathord{\left/ {\vphantom {{{\eta _{ph}}} {\left( {{\eta _J} + {\eta _{ph}}} \right)}}} \right. \kern-\nulldelimiterspace} {\left( {{\eta _J} + {\eta _{ph}}} \right)}}$ where ${\eta _{ph}}$ and ${\eta _J}$ characterize phonon and magnetic contributions to Cooper pairing respectively. Obviously, the combination of any unconventional mechanism of superconductivity with the EPI mediated pairing leads to the reduction of the isotope effect exponent \cite{shn_33}. The only unknown parameters in expression (\ref{shn_eq5}) is the electron-phonon coupling constant ${\lambda _{EPI}}$; thus we can consider it as the fitting parameter. Calculated oxygen isotope effect exponent ${\alpha _O}$ as the function of hole concentration is shown in Fig.~\ref{Isotope}b.  The value ${G \mathord{\left/ {\vphantom {G {J = 1}}} \right. \kern-\nulldelimiterspace} {J = 1}}$ provides ${\alpha _O} = 0.11$ close to the La$_{2 - x}$Sr$_x$CuO$_4$ experimental data at the optimal doping. Increase of the isotope exponent away from the optimal doping is obtained. In our picture concentration dependence of the function ${\alpha _O}\left( x \right)$ generally stems from the strong concentration dependence of the correlated electrons band structure. We cannot rigorously describe the dependence of coupling constant ${\lambda _{EPI}}$ on doping but it is obvious that doped carriers screen electron-phonon interaction therefore ${\lambda _{EPI}}$ decreases with increasing doping. For ${\alpha _O} = {{{\eta _{ph}}} \mathord{\left/  {\vphantom {{{\eta _{ph}}} {\left( {{\eta _J} + {\eta _{ph}}} \right)}}} \right. \kern-\nulldelimiterspace} {\left( {{\eta _J} + {\eta _{ph}}} \right)}}$ this effect should lead to a large value of isotope exponent in underdoped compounds compared with overdoped ones.
With this value of EPI coupling the magnetic and phonon contributions to the $T_c$ are approximately equal, see Fig.~\ref{Isotope}a.
Summarizing our discussion, we want to emphasize that present analysis allows us to find a compromise in the struggle between the adepts of magnetic and phonon mechanisms of high temperature superconductivity: we claim that both mechanisms are important and work together. Therefore both magnetic and phonon mechanisms of ${d_{{x^2} - {y^2}}}$-pairing should be considered in realistic theory of superconductivity in cuprates.

\subsubsection{Influence of Interlayer Hoppings on T$_c$ of Bilayer Cuprates}

The importance of coupling between CuO$_2$ layers was established shortly after the discovery of high-temperature superconductivity, since it was found that
T$_c$ increase in each cuprate family (homologous series of Bi, Tl, and Hg-based superconductors) when the number of CuO$_2$ planes was changed from $n = 1$ to 3. Based on these findings, it was expected that, by increasing the number of CuO$_2$ planes to $n \sim 10$ (which is allowed by crystal chemistry), it would be possible to reach T$_c$ = 300 K. The creation of artificial multilayer structures of the Bi-22(n - 1) system with T$_c$ = 250 K for Bi$_2$Sr$_2$Ca$_7$Cu$_8$O$_{20 + x}$ (Bi-2278) \cite{Lagues} seemed to confirm this hypothesis. However, these structures turned out to be unstable. Later, stable Bi-2278 structures were created using molecular beam epitaxy (MBE), but their T$_c$ values did not exceed 60 K \cite{LogGB}. Moreover, it was also demonstrated \cite{BozEV} that a Bi-2212 film of unit cell thickness (with $n=2$ CuO$_2$ planes) had T$_c$ = 70 K. The subsequent progress in MBE technology revealed the role of individual LaSrCuO layers, as parts of a multilayer La$_{1.55}$Sr$_{0.45}$CuO$_4$/La$_2$CuO$_4$ metal-dielectric heterostructure, in the formation of the superconducting state of the whole sample \cite{LogGB}. It was found that suppression of the superconductivity in a single La$_{2 - x}$Sr$_x$CuO$_4$ layer (assumed to be doped to a
nearly optimum level) significantly influences T$_c$ of the whole structure.
The nature of the T$_c$ dependence on the number of CuO$_2$ planes in the unit cell is still not completely clear. In this study, we will not consider the role of tunneling between  superconducting layers through dielectric interlayers in the unit cell (according to the ``high-temperature cuprate superconductor as heterostructure'' model \cite{KapKop}). We restrict ourselves to the influence of the presence of two CuO$_2$ planes in a single superconducting metallic layer. Generally speaking, the addition of a second CuO$_2$ plane permits growth of T$_c$ by two types of mechanisms. The first type, which can be called dynamic, is related to interplanar interactions and hopping. The second type is based on specific features of the structure and the distribution of impurities and, hence, can be called impurity mechanisms. The present study is devoted to the dynamic mechanisms of single-electron hopping, interplanar exchange interaction, and interplanar hopping of quasiparticle pairs.
Experimental NMR data revealed the inequivalence of two external and $n - 2$ internal CuO$_2$ planes in the structures with $n \ge 3$ \cite{Han}. For this reason, we will restrict ourselves to two-layer structures of the YBa$_2$Cu$_3$O$_7$ (YBCO) and Bi-2212 types with equivalent CuO$_2$ planes. Chakravarty et al.~\cite{ChakSAS} proposed an effective theoretical mechanism of the increase in T$_c$ in these structures based on the postulated tunneling of Cooper pairs between CuO$_2$ planes with a tunneling probability amplitude of ${T_J} \sim 10{\Delta _0}$, where ${\Delta _0}$ is the superconducting gap in one plane. However, Schneider and Singer \cite{SchnS} demonstrated the inapplicability of this approach to real high-temperature cuprate superconductors by estimating from experimental data the ratio ($\eta $) of the free energy due to interplanar coupling to the total energy of a superconductor for various families of cuprates. In particular, this ratio was $\eta  \sim {10^{ - 3}}$ for YBCO and $\eta  \sim {10^{ - 4}}$ for Hg cuprates. The smallness of $\eta$ is evidence in favor of the model of a Josephson structure with weak coupling that does not influence the superconducting properties of separate CuO$_2$ planes. Above in the Sect 3.4 we have studied the role of hopping between CuO$_2$ planes in the normal phase
A superconducting phase with a magnetic mechanism of pairing is described in the same manner as for single layer cuprate in a previous Sect. with the interplayer hopping as the perturbation. It will be shown that the interplanar hopping does not increase the critical temperature.

To describe superconducting phase of bilayer cuprate we add to the Hamiltonian (\ref{eq18}) two more terms to get the
\begin{eqnarray}
{H_{t - t' - t'' - {t_ \bot } - {J^*} - {J_ \bot }}} = H_{t - t' - t'' - {t_ \bot } - {J^*}} + {H_{{J_ \bot }}} + {H_{{T_ \bot }}},
\label{eq30} \\
{H_{{J_ \bot }}} = \sum\limits_{{m_u}{n_d}\sigma } {{J_ \bot }\left( {{m_u},{n_d}} \right)} \left( {X_{{m_u}}^{\sigma \bar \sigma }X_{{n_d}}^{\bar \sigma \sigma } - X_{{m_u}}^{\sigma \sigma }X_{{n_d}}^{\bar \sigma \bar \sigma }} \right),
\label{eq32} \\
{H_{{T_ \bot }}} = \sum\limits_{u_1,u_2,d_1,d_2,\sigma} {{T_ \bot }\left( {{u_1},{u_2},{d_1},{d_2}} \right)\left( {X_{{u_1}}^{S\bar \sigma }X_{{u_2}}^{S\sigma }X_{{d_1}}^{\sigma S}X_{{d_2}}^{\bar \sigma S} + h.c.} \right)}.
\label{eq33}
\end{eqnarray}
Here, subscript $\alpha $ refers to CuO$_2$ planes facing up ($u$) and down ($d$). Hamiltonians ${H_{{J_ \bot }}}$, and ${H_{{T_ \bot }}}$ describe interlayer exchange interaction, and interlayer pair hopping, respectively. In the latter Hamiltonian, ${u_1}$, ${u_2}$ and ${d_1}$, ${d_2}$ refer to pairs of sites in the upper and lower CuO$_2$ planes, respectively. The dependence of the interlayer exchange parameter ${J_ \bot }\left( {{m_u},{n_d}} \right)$ on the distance between sites in the upper ($m_u$) and lower ($n_d$) CuO$_2$ planes is determined by the form of this function in the $\bf{k}$ space:

\begin{equation}
\label{eq34}
{J_ \bot }\left( {\bf{k}} \right) = \frac{{t_ \bot ^2\left( {\bf{k}} \right)}}{{{E_{ct}}}} = \frac{{t_ \bot ^2{{\left( {\cos {k_x} - \cos {k_y}} \right)}^4}}}{{{E_{ct}}}}.
\end{equation}

The interlayer pair hopping in the reciprocal space is described by the following Hamiltonian:

\begin{equation}
\label{eq35}
{H_{{T_ \bot }}} = \sum\limits_{\bf{k}} {{T_ \bot }\left( {\bf{k}} \right)\left( {X_{\bf{k}}^{S\sigma }X_{ - {\bf{k}}}^{S\bar \sigma }X_{ - {\bf{k}}}^{\bar \sigma S}X_{\bf{k}}^{\sigma S} + h.c.} \right)},
\end{equation}
where the hoping integrals ${T_ \bot }\left( {\bf{k}} \right)$ are determined by the ${t_ \bot }{\left( {\bf{k}} \right)^2}/{t_{01}}$ values \cite{ChakSAS} as

\begin{equation}
\label{eq36}
{T_ \bot }\left( {\bf{k}} \right) = \frac{{{t_ \bot }{{\left( {\bf{k}} \right)}^2}}}{{{t_{01}}}}{\left( {\cos {k_x} - \cos {k_y}} \right)^4}.
\end{equation}

A generalization of the BCS-like mean field theory of the d-pairing presented above for single layer cuprate is straightforward, see paper \cite{MakOvch,MakOvchShney}. As a result, we obtain the anomalous intralayer averages ${B_{\bf{q}}} = \left\langle {X_{\bf{q}}^{\sigma S}X_{ - {\bf{q}}}^{\bar \sigma S}} \right\rangle $, anomalous interlayer averages ${B_{ \bot {\bf{q}}}} = \left\langle {X_{\left( u \right){\bf{q}}}^{\sigma S}X_{\left( d \right) - {\bf{q}}}^{\bar \sigma S}} \right\rangle $ and two energy gaps, which include all possible interactions leading to the potential pairing of particles:
\begin{eqnarray}
{\Delta _{\bf{k}}} &=& - \frac{1}{N}\frac{1}{{{p_\sigma } + x}} \sum\limits_{\bf{q}} \left[ \left( 2{t_{\bf{q}}} - {p_\sigma }\left( {{J_{{\bf{k}} + {\bf{q}}}} + {J_{{\bf{k}} - {\bf{q}}}}} \right) + 2{p_\sigma }\frac{{\tilde t_{\bf{q}}^2}}{{{E_{ct}}}} - 4\left( {{p_\sigma } + x} \right)\frac{{{{\tilde t}_{\bf{k}}}{{\tilde t}_{\bf{q}}}}}{{{E_{ct}}}} \right){B_{\bf{q}}} \right. \nonumber\\
&-& \left. 2{{\tilde t}_{ \bot {\bf{q}}}}{B_{ \bot {\bf{q}}}} \right], \label{eq37} \\
{\Delta _{ \bot {\bf{k}}}} &=& \frac{1}{N}\frac{1}{{{p_\sigma } + x}}\sum\limits_q \left( {p_\sigma }\left( {{J_{ \bot {\bf{k}} + {\bf{q}}}} + {J_{ \bot {\bf{k}} - {\bf{q}}}}} \right)+\left( {\left( {{p_\sigma } + x} \right){T_{jos}}} \right){B_{ \bot {\bf{q}}}} \right), \label{eq38}
\end{eqnarray}
where $p=(1-x)/2$ and $x$ is the concentration of doped holes. The first term on the right-hand side of Eq.~(\ref{eq37}) reflects the kinematic mechanism of pairing~\cite{ZaitsIv}, the second term is related to the exchange, the  third and fourth are due to three-center interactions in one CuO$_2$ plane, and the last term allows for the hoping between CuO$_2$ layers. The appearance of superconducting gap (\ref{eq38}) is due to the possible pairing of quasiparticles from different planes by means of interlayer exchange interaction. The symmetry of the gap will be considered below. In the general case, the superconducting gap in each CuO$_2$ plane has the form of ${\Delta ^{u\left( d \right)}} = \Delta _0^{u\left( d \right)}{e^{i{\theta ^{u\left( d \right)}}}}$, where phase $\theta $ is the sum of the mean phase ${\theta _0}$ and phase fluctuation $\delta \theta $. The self-consistent calculation using a system of equations for each CuO$_2$ plane of the multilayer compound with allowance for the interaction between neighboring layers gave the following expression \cite{BycSpa}:
\begin{equation}
\label{eq39}
{\Delta _{j,k}} = 2i{\Delta _k}\sin \left( {\frac{{\pi j}}{{n + 1}}} \right),
\end{equation}
where $j$ runs through the numbers of CuO$_2$ layers and $n$ is the total number of these layers. For example, in our case $n = 2$ and, hence, ${\Delta _2} = {\Delta _1}$. Below we will rely on this result and assume mean phases in the neighboring layers to be the same and equal to zero. Since the region of 3D superconductivity (dome-shaped for both single- and multilayer cuprates) is qualitatively the same as the concentration dependence of T$_c$ obtained in our calculations and only exhibits quantitative differences, we use the mean field approximation without taking into account fluctuations of the order parameter. The real parts of superconducting gaps for the upper and lower layer are assumed to be the same, while the phase is assumed to be fixed and equal to zero. The system of the equations of motion for both normal and anomalous intralayer and interlayer Green's functions is as follows:
\begin{equation}
\label{eq40}
\left( {\begin{array}{*{20}{c}}
{E - {\xi _{\bf{k}}}}&{ - {{\tilde t}_{ \bot {\bf{k}}}}}&{ - {\Delta _{\bf{k}}}}&{ - {\Delta _{ \bot {\bf{k}}}}}\\
{ - {{\tilde t}_{ \bot {\bf{k}}}}}&{E - {\xi _{\bf{k}}}}&{ - {\Delta _{ \bot {\bf{k}}}}}&{ - {\Delta _{\bf{k}}}}\\
{ - \Delta _{\bf{k}}^ * }&{ - \Delta _{ \bot {\bf{k}}}^ * }&{E + {\xi _{\bf{k}}}}&{{{\tilde t}_{ \bot {\bf{k}}}}}\\
{ - \Delta _{ \bot {\bf{k}}}^ * }&{ - \Delta _{\bf{k}}^ * }&{{{\tilde t}_{ \bot {\bf{k}}}}}&{E + {\xi _{\bf{k}}}}
\end{array}} \right)\left( {\begin{array}{*{20}{c}}
{G_{{\bf{k}}\sigma }^u}\\
{G_{{\bf{k}}\sigma }^{du}}\\
{F_{{\bf{k}}\sigma }^u}\\
{F_{{\bf{k}}\sigma }^{du}}
\end{array}} \right) = \left( {\begin{array}{*{20}{c}}
{{p_\sigma } + x}\\
0\\
0\\
0
\end{array}} \right).
\end{equation}
where $\xi _{\bf{k}}$ is the law of dispersion in the normal phase (expression (\ref{eq20}) without $t_{ \bot {\bf{k}}}$). Solving system of equations (\ref{eq40}) gives the dispersion of quasiparticle bands in the superconducting state
\begin{equation}
E_{\bf{k}}^{1,2} =  \pm \sqrt {{{\left( {{\xi _{\bf{k}}} + {{\tilde t}_{ \bot {\bf{k}}}}} \right)}^2} + \Delta {{_{\bf{k}}^{\left(  +  \right)}}^2}}, \;\;\;
E_{\bf{k}}^{3,4} =  \pm \sqrt {{{\left( {{\xi _{\bf{k}}} - {{\tilde t}_{ \bot {\bf{k}}}}} \right)}^2} + \Delta {{_{\bf{k}}^{\left(  -  \right)}}^2}}. \label{eq41}
\end{equation}
where $\Delta _{\bf{k}}^{\left(  +  \right)} = {\Delta _{\bf{k}}} + {\Delta _{ \bot {\bf{k}}}}$ and $\Delta _{\bf{k}}^{\left(  -  \right)} = {\Delta _{\bf{k}}} - {\Delta _{ \bot {\bf{k}}}}$. As a result, the general expression for the superconducting gap takes the following form:
\begin{eqnarray}
\label{eq42}
{\Delta _{\bf{k}}} &=& - \frac{1}{N}\frac{1}{{{p_\sigma } + x}}\sum\limits_{\bf{q}} \left( {t_{\bf{q}}} - {p_\sigma }\left( {{J_{{\bf{k}} + {\bf{q}}}} + {J_{{\bf{k}} - {\bf{q}}}}} \right) + 2{p_\sigma }\frac{{\tilde t_{\bf{q}}^2}}{{{E_{ct}}}} 4\left( {{p_\sigma } + x} \right)\frac{{{{\tilde t}_{\bf{k}}}{{\tilde t}_{\bf{q}}}}}{{{E_{ct}}}} \right) \nonumber\\
&\times& \left[ {\frac{{\Delta _{\bf{q}}^{\left(  +  \right)}}}{{4E_{\bf{q}}^1}}\tanh \frac{{E_{\bf{q}}^1}}{{2\tau }} + \frac{{\Delta _{\bf{q}}^{\left(  -  \right)}}}{{4E_{\bf{q}}^3}}\tanh \frac{{E_{\bf{q}}^3}}{{2\tau }}} \right] - \frac{1}{N}\frac{1}{{{p_\sigma } + x}}\sum\limits_{\bf{q}} {{\tilde t}_{ \bot {\bf{q}}}} \Pi(\bf{q},\tau) \label{gap}, \\
{\Delta _{ \bot {\bf{k}}}} &=& \frac{1}{N}\frac{1}{{{p_\sigma } + x}}\sum\limits_{\bf{q}} \left( {{p_\sigma }\left( {{J_{ \bot {\bf{k}} + {\bf{q}}}} + {J_{ \bot {\bf{k}} - {\bf{q}}}}} \right) } + \left( {{p_\sigma } + x} \right){T_{jos}} \right) \Pi(\bf{q},\tau),
\label{gapper}
\end{eqnarray}
where $\Pi(\bf{q},\tau) = \left[ {\frac{{\Delta _{\bf{q}}^{\left(  +  \right)}}}{{4E_{\bf{q}}^1}}\tanh \frac{{E_{\bf{q}}^1}}{{2\tau }} - \frac{{\Delta _{\bf{q}}^{\left(  -  \right)}}}{{4E_{\bf{q}}^3}}\tanh \frac{{E_{\bf{q}}^3}}{{2\tau }}} \right]$, $\tau  = {k_B}T$ and $k_B$ is the Boltzmann constant. It is known that the kinematic mechanism does not obey the condition of symmetry of the superconducting gap ${d_{{x^2} - {y^2}}}$ observed in the ARPES experiments \cite{OkawaIU} and scanning tunnelling spectroscopy \cite{HowFK,HoogBP}. Expression (\ref{eq42}) shows that the pairing of separate quasiparticles via interlayer hopping is impossible for the ${d_{{x^2} - {y^2}}}$-symmetry of the gap, since the product of ${\tilde t_{ \bot {\bf q}}}$ and ${\Delta _{\bf q}}$ in this sum gives a cubic power of the difference of cosines and this term disappears upon summation. In the case of a bilayer cuprate, in contrast to the self-consistent equation for the superconducting gap in a single-layer cuprate, the ${\Delta _{\bf k}}$ gap related to the intralayer pairing is supplemented by the ${\Delta _{ \bot {\bf k}}}$ gap, which reflects the interlayer pairing via exchange interaction. In the case of pairing in the CuO$_2$ plane, consideration is usually restricted to the exchange between nearest neighbors. For the  interlayer exchange, the nearest neighbors are CuO$_2$ layers in the unit cell with $\left( {{R_x},{R_y}} \right) = \left( {0,0} \right)$, for which the magnitude of exchange interaction is ${J_{ \bot 00}} = 0.011$ eV. In this case, terms ${J_{ \bot 00}}$ in the expansion of ${J_{ \bot {\bf{k}} \pm {\bf{q}}}}$ will not contribute to the equation for T$_c$, since they cancel the dependence on ${\bf k}$ and ${\bf q}$ and, because of the ${d_{{x^2} - {y^2}}}$-symmetry of the gap, the summation over ${\bf q}$ eventually yields zero. The dependence of ${J_ \bot }\left( {\bf{k}} \right)$ on the wave vector resulted in the fact that ${J_{ \bot 01}}$  was zero. Therefore, it can be seen that, in the adopted nearest-neighbor approximation, the mechanism of pairing via interlayer exchange interaction can be rejected.

Taking into account the ${d_{{x^2} - {y^2}}}$-symmetry of the gap, the contribution from the tunneling of quasiparticle pairs to the self-consistent equation for the gap disappears for the same reason.
Indeed, the product of an even power (second for single-particle hopping and fourth for pair hopping) of the difference of cosines in the hopping integral and the $\left( {\cos {k_x} - \cos {k_y}} \right)$ factor due to the gap symmetry leads to vanishing of the sum over {\bf q}.
Eventually, with allowance for the ${d_{{x^2} - {y^2}}}$-symmetry of the gap, ${\Delta _{\bf{k}}} = \frac{{{\Delta _0}}}{2}\left( {\cos {k_x} - \cos {k_y}} \right) = {\Delta _0}{\varphi _{\bf{k}}}$ in the approximation of ${J_{01}}$ exchange between nearest neighbors in the CuO$_2$ layer, the equation for T$_c$ can be transformed as follows:
\begin{equation}
\label{eq43}
1 = \frac{1}{N}{p_\sigma }{J_{01}}\sum\limits_{\bf{q}} {\frac{{\varphi _{\bf{q}}^2}}{2}\left( {\frac{1}{{E_{\bf{q}}^ + }}\tanh \frac{{E_{\bf{q}}^ + }}{{2\tau }} + \frac{1}{{E_{\bf{q}}^ - }}\tanh \frac{{E_{\bf{q}}^ - }}{{2\tau }}} \right)}, \end{equation}
where $E_{\bf{q}}^ +  = {\xi _{\bf{q}}} + {\tilde t_{ \bot {\bf{q}}}}$ and $E_{\bf{q}}^ -  = {\xi _{\bf{q}}} - {\tilde t_{ \bot {\bf{q}}}}$. It is clearly seen that the only result of the inclusion of interlayer interactions into the initial model is the appearance of a sum of two terms corresponding to the presence of bonding and antibonding bands.

\begin{figure}
\begin{center}
$\begin{array}{cc}
\includegraphics[angle=0,width=0.47\columnwidth]{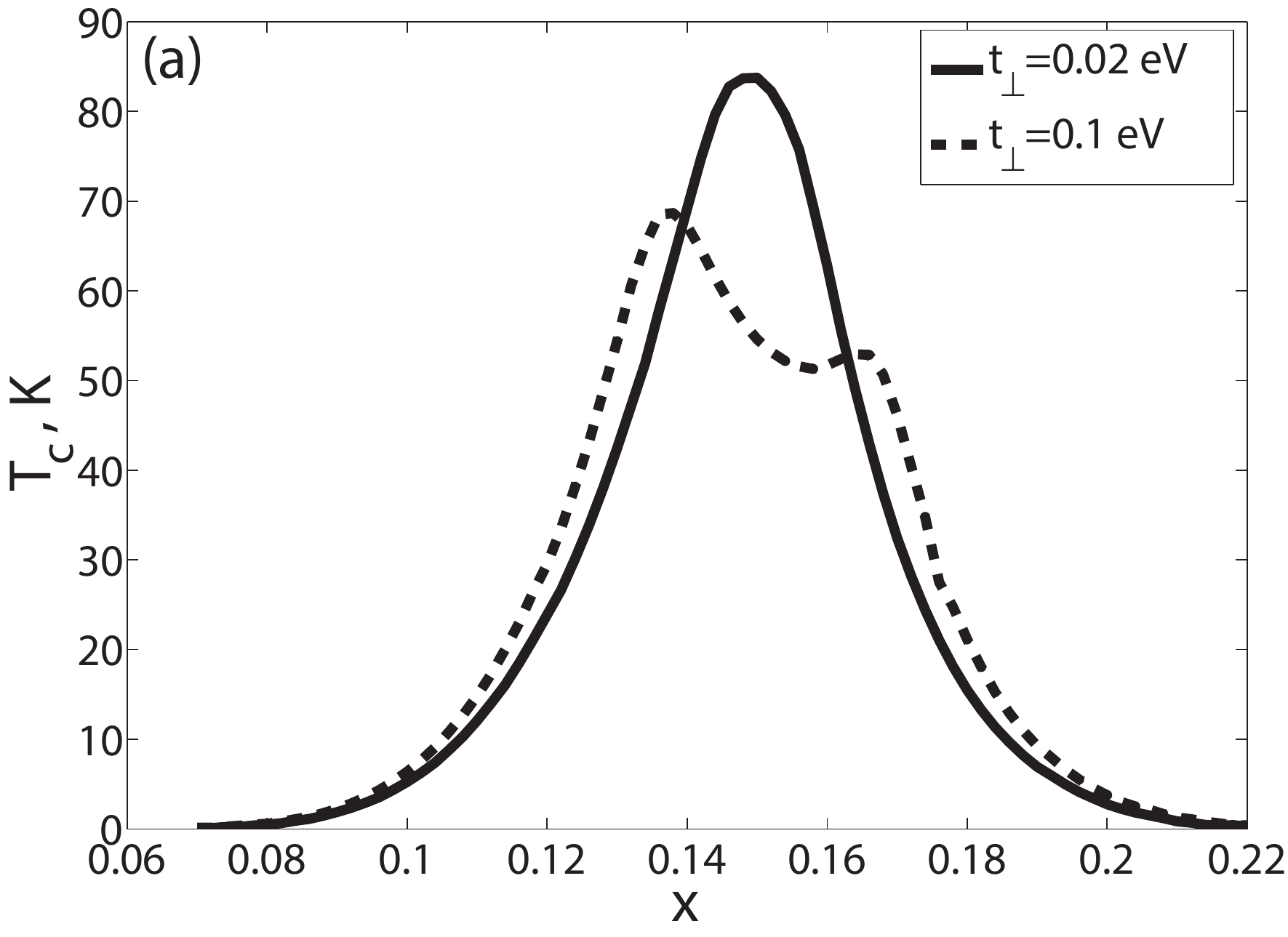} &
\includegraphics[angle=0,width=0.47\columnwidth]{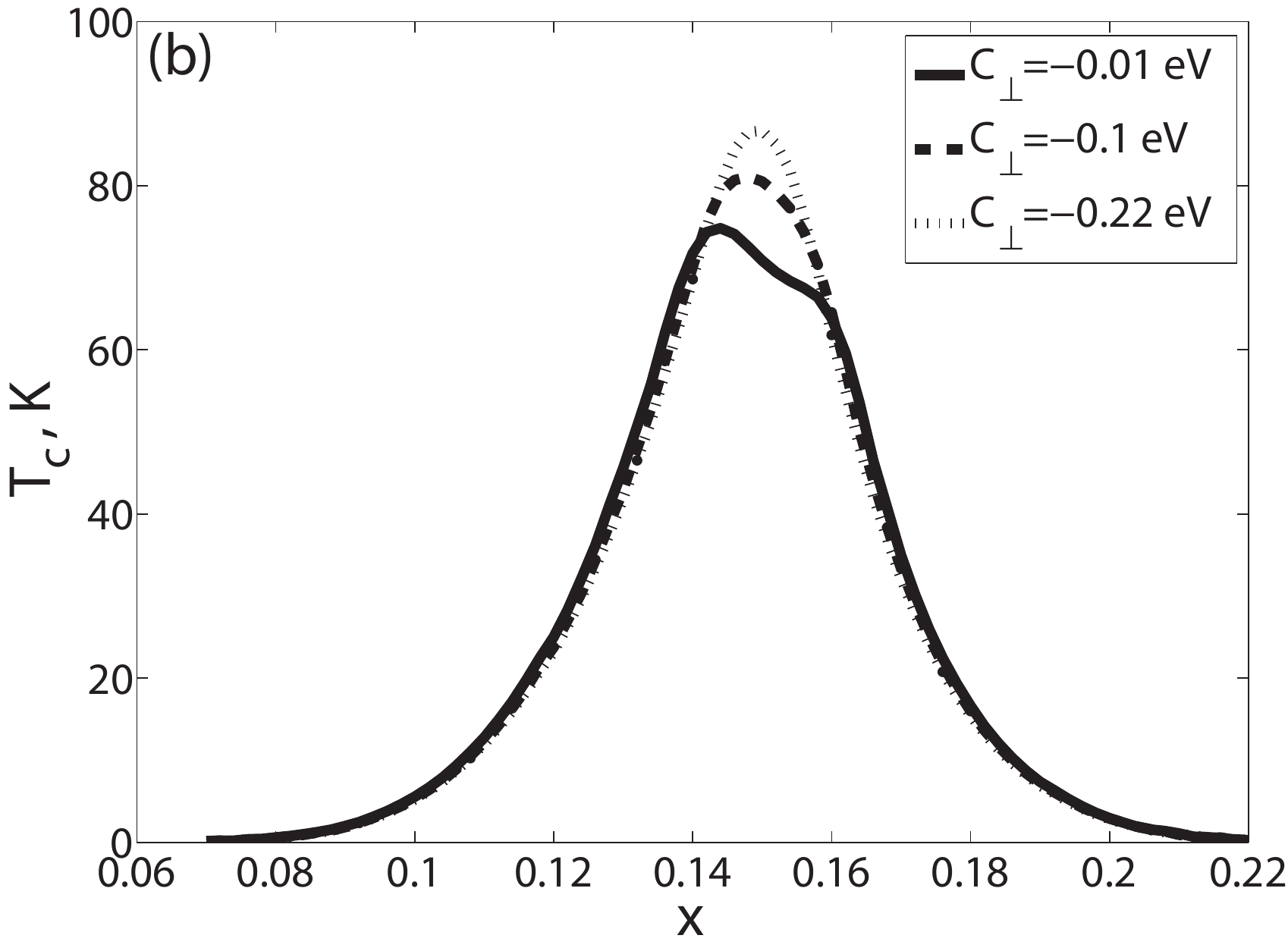}
\end{array}$
\caption{(a) $T_c$ versus doping $x$ for ${t_ \bot } = 0.02$ and ${t_ \bot } = 0.1$ eV.
All calculations were performed for an interlayer spin correlator ${C_ \bot } = -0.1$.
(b) Dependence of T$_c(x)$ on intracell interlayer spin correlators (${C_ \bot } = -0.01, -0.1, - 0.22$).}
\label{Tc_x_bilayer}
\end{center}
\end{figure}

Fig.~\ref{Tc_x_bilayer}a shows the concentration dependences of $T_c$ for different values of single-particle interlayer hopping integral ${t_ \bot }$. For realistic values of ${t_ \bot }= 0.02$ eV and ${C_ \bot } =  - 0.1$, the $T_c(x)$ is almost the same as that for single-layer cuprates, the only difference is a small decrease in T$_c$ at all concentrations of doping carriers. The main factor that prevents the increase in T$_c$ with an increasing number of layers is redistribution of the density of states (DOS) of a single band (for single-layer cuprates) between two bands in the case of bilayer cuprates, which is manifested by the appearance of coefficient 1/2 in Eq.~(\ref{eq43}), while there are no additional mechanisms of pairing in bilayer system as compared to the single-layer case. Thus, in the framework of the generalized mean field approximation, it is impossible to speak of an increase in T$_c$ due to the interlayer one-particle hopping.

An increase in the hopping integral ${t_ \bot }$ is accompanied by an increase in bilayer splitting between the bonding and antibonding bands and, hence, in the distance between two peaks in the DOS (Fig.~\ref{Tc_x_bilayer}a). Each peak corresponds to a certain van Hove singularity. With a change in the level of doping, the DOS peaks exhibit shifts and their intensities vary, but the two-peak structure is retained. Coincidence of the chemical potential with each of the van Hove singularities corresponds to a maximum in T$_c$. A rather large intracell interplanar hopping integral (${t_ \bot } = 0.1$ eV) leads to significant splitting of bands and the resulting difference of maxima with respect to the doping level (Fig.~\ref{Tc_x_bilayer}a). Indeed, the first maximum corresponds to $x = 0.144$, while the second corresponds to $x = 0.156$. No such two-peak structure in the concentration dependence of T$_c$ has been observed in experiments.

Spin correlations enter the energy of coupling between CuO$_2$ layers via formula (\ref{eq21}) and, at first glance, it might seem that their presence would increase this energy. However, because of the negative sign of the correlator, this contribution will decrease the binding energy so that, in fact, the antiferromagnetic correlations suppress bilayer splitting. This, in turn, poses a limitation on the possibility of hopping between layers: a quasiparticle with spin up cannot pass to the neighboring plane because the one-particle state with this spin is occupied, so that only intralayer transitions of quasiparticles with opposite spins are possible. Fig.~\ref{Tc_x_bilayer}b shows how strongly the antiferromagnetic correlations can decrease bilayer splitting.
The effect of magnetic correlations on the magnitude of bilayer splitting is naturally manifested in the concentration dependence of T$_c$. An increase in the level of magnetic correlations to ${C_ \bot } =  - 0.22$ leads to disappearance of the two-peak structure in T$_c(x)$.

A two-peak concentration dependence was obtained earlier in~\cite{ValKr}
assuming that ${t_ \bot }$ was nonzero only for the interlayer hopping inside the unit cell, i.e., that the hopping integral was independent of the 2D wave vector. The two-peak concentration dependence of T$_c$ disappears for $t_{\bot}/t_{01} \approx 0.03$, replaced by a usual parabolic single-peak curve. The absence of an increase in the maximum of T$_c$ due to the interlayer hopping suggests that the hopping of quasiparticles between CuO$_2$ layers cannot play a significant role in the dependence of $T_c$ on the number of CuO$_2$ layers in the unit cell.
This agrees with the results of experiments on uniaxial compression ~\cite{Meingast,WelpGF,MeinKW,KundA}, which showed that a decrease in the spacing of CuO$_2$ layers under uniaxial compression along the $c$ axis influences T$_c$ only via an increase in the concentration of carriers in the CuO$_2$ layers. In other words, an increase in the interlayer coupling weakly influences T$_{c(max)}$; i.e., $\frac{{d{T_c}}}{{d{P_c}}}$ is decreased by an order of magnitude lower than the rate of change in $T_{c(max)}$ for compression in the plane: $\frac{{d{T_c}}}{{d{P_a}}} \approx  - 1.9\,K \cdot GP{a^{ - 1}}$ and $\frac{{d{T_c}}}{{d{P_b}}} \approx  + 2.2\,K \cdot GP{a^{ - 1}}$. This is additional evidence for the decisive role of the CuO$_2$ layer in the formation of superconductivity. Compression along the a and b axes leads to a decrease in the lattice parameters, which enhances coupling between the orbitals of copper and oxygen atoms and, hence, increases the hopping integral and exchange interactions. The exchange interactions between nearest neighbors in the plane, according to Eq.~\ref{eq43}, directly influence T$_c$. It should be noted that the values of derivatives with respect to pressure reflect the so-called ``internal'' properties of cuprates, i.e., changes in the atomic and electron structures without affecting the doping level in the CuO$_2$ plane. Schilling \cite{Schilling} used the terms of ``physiological'' and ``pathological'' cuprates, the latter being represented by the LaSrCuO and YBaCuO systems, in which the doping changes not only the hole concentration, but the structure as well. In contrast, the structures of physiological cuprates based on Tl and Hg are stable and doping only influences the concentration of holes. Experiments with compression are performed in the latter families of cuprates-i.e., on compounds with the most planar and least curved CuO$_2$ planes. This choice is related to the absence of effects related to pressure-induced structural phase transitions. In addition, these compounds are characterized by the maximal T$_c$ values among all cuprates. These facts suggest that defects such as the curved CuO$_2$ planes are not favorable for superconductivity. From the standpoint of microscopic theory, this can be explained by decreasing overlap (and, hence, interaction constants) for orbitals in this plane in the case of its deformation. However, this by no means implies that all other possible inhomogeneities are also unfavorable. It is quite possible that inhomogeneities and disorder in fact cause an increase in T$_c$ upon the addition of CuO$_2$ layers to the unit cell with rearrangement of the atomic and electron structure of the pool of charge carriers and the introduction of new atoms. Indeed, an additional atom (e.g., Ca or Y) appears between copper-oxygen planes in bilayer cuprates in comparison to single-layer ones. A disorder that arises when divalent Ca$^{2+}$ cation is replaced by trivalent Y$^{3+}$ stabilizes the general cell structure due to the presence of an additional positive charge. Moreover, this substitution affects the electron structure of CuO$_2$ layers much more weakly as compared to the influence of disorder brought about by atomic substitutions near the apical oxygen \cite{EisKF}. In single-layer cuprates, the local substitutions of atoms situated near the apical oxygen may displace this oxygen and deform the CuO$_6$ octahedron, since the opposite apical oxygen in this octahedron is fixed. Naturally, this deformation would modify the wavefunction and energy of the entire octahedron and, hence, affect the superconductivity. It was shown \cite{EisKF} for the family of bismuth cuprates that the partial replacement of Bi$^{3+}$ by various lanthanide atoms (Ln-La, Pr, Sm, Eu. Gd) of the same valence significantly influenced T$_c$. The greater the ion radius of substituted atoms, the higher the critical temperature. Reasons for the observed increase in T$_c$ under these conditions are not completely clear, but it is evident that the presence of impurities and inhomogeneities in the atomic surrounding of CuO$_2$ layers influences the superconducting state either directly or indirectly (via apical oxygen). In multilayer cuprates, the influence of such inhomogeneities has an apparently softer character. In bilayer cuprates, there are two CuO$_5$ pyramids spaced by a relative large distance (0.3 nm) instead of one CuO$_6$ octahedron in single-layer compounds. The space between layers serves a kind of buffer that smoothens possible distortions in one of the two pyramids thus ensuring a more stable superconducting state and higher T$_c$ values.

\section{Conclusions}

Problem of describing the high-$T_c$ superconductivity is tightly related to the limited knowledge of how to deal with the strongly correlated systems. Here we show how the correlations affect normal state that sometimes behave quite unusual when compared to the intuitive Fermi-liquid picture. One of the significant feature is the set of quantum phase transitions with the Fermi surface topology change that is as we believe related to the widely discussed pseudogap phenomena. Another aspect of correlations is that the superconductivity in cuprates can not be attributed to a single pairing mechanism. Spin fluctuations and electron-phonon mechanisms are tightly coupled and the effects of correlations make this binding tighter. While in some sense we are only at the beginning of the way toward the ``ultimate'' theory of high-$T_c$ superconductivity, we are speeding up with the development of both numerical and analytical theoretical tools.

\subsection{Acknowledgements}

We would like to thank V.I. Belyavsky, P.J. Hirschfeld, I. Eremin, M.V. Eremin, Yu.V. Kopaev, I.A. Nekrasov, Z.V. Pchelkina, N.M. Plakida, and V.V. Val'kov for the discussions of the topics reviewed here. The authors acknowledge support from FCP Scientific and Research-and-Educational Personnel of Innovative Russia for 2009-2013 (GK 16.740.12.0731 and GK 16.740.11.0740), RFBR (Grants 12-02-31534 and 12-02-31597), Siberian Federal University (Theme \#F-11), Governmental support of leading scientific schools of Russia (NSh-1044.2012.2), Program of SB RAS \#44, Presidium of RAS program \#20.7. EIS, IAM, and MMK are grateful to The Dynasty Foundation and ICFPM for the financial support.

\end{document}